\documentclass[12pt]{article}
\usepackage{graphicx}
\usepackage{slashed}
\usepackage{axodraw4j}
\usepackage{amsmath}

\begin{document}
\title{{\bf TASI Lectures on Effective Field Theory \\ and \\ Precision Electroweak Measurements} }
\author{ {\bf Witold Skiba} \bigskip \\ {\it  Department of Physics, Yale University, New Haven, CT 06520}}
\date{}
\maketitle
\begin{abstract}
The first part of these lectures provides a brief introduction to the concepts and techniques of effective field theory.  The second part reviews precision electroweak constraints using effective theory methods. Several simple extensions of the Standard Model are considered as illustrations. The appendix contains some new results on the one-loop contributions of electroweak triplet scalars to the $T$ parameter and contains a discussion of decoupling in that case.
\end{abstract}
\newpage

\tableofcontents

\section{Introduction}
 
Phenomena involving distinct energy, or length, scales can often be analyzed by considering one relevant scale at a time. In most branches of physics, this is such an obvious statement that it does not require any justification. The multipole expansion in electrodynamics is useful because the short-distance details of charge distribution are not important when observed from far away. One does not worry about the sizes of planets, or their geography, when studying orbital motions in the Solar System. Similarly, the hydrogen spectrum can be calculated quite precisely without knowing that there are quarks and gluons inside the proton.  

Taking advantage of scale separation in quantum field theories leads to effective field theories (EFTs)~\cite{Weinberg}.  Fundamentally, there is no difference in how scale separation manifests itself in classical mechanics, electrodynamics, quantum mechanics, or quantum field theory. The effects of large energy scales, or short distance scales, are suppressed by powers of the ratio of scales in the problem. This observation follows from the equations of mechanics, electrodynamics, or quantum mechanics. Calculations in field theory require extra care to ensure that large energy scales decouple~\cite{WilsonKogut,A-C}. 

Decoupling of large energy scales in field theory seems to be complicated by the fact that integration over loop momenta involves all scales. However, this is only a superficial obstacle which is straightforward to deal with in a convenient regularization scheme, for example dimensional regularization. The decoupling of large energy scales takes place in renormalizable quantum field theories  whether or not EFT techniques are used. There are many precision calculations that agree with experiments despite neglecting the effects of heavy particles. For instance, the original calculation of the anomalous magnetic moment of the electron, by Schwinger, neglected the one-loop effects arising from weak interactions. Since the weak interactions were not understood at the time, Schwinger's calculation included only the photon contribution, yet it agreed with the experiment within a few percent~\cite{gminus2}. Without decoupling, the weak gauge boson contribution would be of the same order as the photon contribution. This would result in a significant discrepancy between theory and experiment and QED would likely  have never been established as the correct low-energy theory. 

The decoupling of heavy states is, of course, the reason for building high-energy accelerators. If quantum field theories were sensitive to all energy scales, it would be much more useful to increase the precision of low-energy experiments instead of building large colliders. By now, the anomalous magnetic moment of the electron is known to more than ten significant digits. Calculations agree with measurements despite that the theory used for these calculations does not incorporate any TeV-scale dynamics, grand unification, or any notions of quantum gravity. 

If decoupling of heavy scales is a generic feature of field theory, why would one consider EFTs? That depends on whether the dynamics at high energy is known and calculable or else the dynamics is either non-perturbative or unknown. If the full theory is known and perturbative, EFTs often simplify calculations. Complex computations can be broken into several easier tasks. If the full theory is not known, EFTs allow one to parameterize the unknown interactions, to estimate the magnitudes of these interactions, and to classify their relative importance. EFTs are applicable to both cases with the known and with the unknown high-energy dynamics  because in an effective description only the relevant degrees of freedom are used. The high-energy physics is encoded indirectly though interactions among the light states. 

The first part of these lecture notes introduces the techniques of EFT\@. Examples of EFTs are constructed explicitly starting from theories with heavy states and perturbative interactions. Perturbative examples teach us how things work: how to organize power counting, how to estimate the magnitudes of terms, and how to stop worrying about non-renormalizable interactions. Readers familiar with the concepts of EFTs are encouraged to go directly to the discussion of precision electroweak measurements.  

The second part of these notes is devoted to precision electroweak measurements using an EFT approach. This is a good illustration of why using EFTs saves time. The large body of precision electroweak measurements can be summarized in terms of constraints on coefficients of effective operators. In turn, one can use these coefficients to constrain extensions of the Standard Model (SM) without any need for detailed calculations of cross sections, decay widths, etc.

The topic of precision electroweak measurements  consists of two major branches. One branch engages in comparisons of experimental data with accurate calculations in the SM~\cite{EW}. It provides important tests of the SM and serves as a starting point for work on extensions of the SM\@. This subject is not covered in these notes. Another branch is concerned with extensions of the SM and their viability when compared with measurements. This subject is discussed here using EFT techniques.  The effective theory applicable to precision electroweak measurements is well known: it is the Standard Model with higher-dimensional interactions. Since we are not yet sure if the Higgs boson exists, one could formulate an effective description with or without the Higgs boson. Only the EFT that includes the Higgs boson is discussed here. While there are differences between the theories with and without the Higgs boson, these differences are technical instead of conceptual.

These notes describe how effective theories are constructed and constrained and how to use EFT for learning about extensions of the SM\@. The best known example of the application of EFT to precision electroweak measurements are the $S$ and $T$ parameters. The $S$ and $T$ parameters capture only a subset of available precision measurements. The set of effective parameters can be systematically enlarged depending on the assumptions about the underlying theory.  Finally, several toy extensions of the SM are presented as an illustration of how to constrain the masses and couplings of heavy states using the constraints on EFTs.  A more complicated example of loop matching of electroweak scalar triplets is presented in the Appendix. The one-loop results in the Appendix have not been published elsewhere. 

\section{Effective Field Theories}

The first step in constructing EFTs is identifying the relevant degrees of freedom for the measurements of interest. In the simplest EFTs that will be considered here, that means that light particles are included in the effective theory, while the heavy ones are not. The dividing line between the light and heavy states is based on whether or not the particles can be produced on shell at the available energies. Of course, all field theories must be effective since we do not know all the heavy states, for example at the Plank scale. There are many more sophisticated uses of EFTs, for instance to heavy quark systems, non-relativistic QED and QCD, nuclear interactions, gravitational radiation, etc. Some of these applications of EFTs are described in the lecture notes in Refs.~\cite{Aneesh, Ira, DBKaplan, Walter}.

Formally, the heavy particles are ``integrated out'' of the action by performing a path integral over the heavy states only
\begin{equation}
	\int {\mathcal D} \, \varphi_H \, e^{i \int {\mathcal L}(\varphi_L,\, \varphi_H)} = e^{i \int {\mathcal L}_{eff}(\varphi_L)},
\end{equation}
where $\varphi_L$, $\varphi_H$ denote the light and the heavy states. Like most calculations done in practice, integrating out is performed using Feynman diagram methods instead of path integrals. The effective Lagrangian can be expanded into a finite number of terms of dimension four or less, and a tower of  ``higher dimensional'' terms, that is terms of  dimension more than four
\begin{equation}
\label{eq:Leff}
 {\mathcal L}_{eff}(\varphi_L)= {\mathcal L}_{d\leq 4} + \sum_i \frac{O_i}{\Lambda^{dim(O_i)-4}},
\end{equation}
where $\Lambda$ is an energy scale and $dim(O_i)$ are the dimensions of operators $O_i$. What is crucial for the EFT program is that the expansion of the effective Lagrangian in Eq.~(\ref{eq:Leff}) is into local terms in space-time. This can be done when the effective Lagrangian is applied to processes at energies lower than the masses of the heavy states $\varphi_H$.

Because we are dealing with weakly interacting theories, the dimension of terms in the Lagrangian is determined by adding the dimensions of all fields making up a term and the dimensions of derivatives. The field dimensions are determined from the kinetic energy terms and this is often referred to as the engineering dimension.  In strongly interacting theories, the dimensions of operators often differ significantly from the sum of the constituent field dimensions determined in free theory. In weakly interacting theories considered here, by definition, the effects of interactions are small and can be treated order by order in perturbation theory. 

The sum over higher dimensional operators in Eq.~(\ref{eq:Leff}) is in principle an infinite sum. In practice, just a few terms are pertinent. Only a finite number of terms needs to be kept because the theory needs to reproduce experiments to finite accuracy and also because the theory can be tailored to specific processes of interest. The higher the dimension of an operator, the smaller its contribution to low-energy observables. Hence, obtaining results to a given accuracy requires a finite number of terms.~\footnote{Not all terms of a given dimension need to be kept. For example, one may be studying $2\rightarrow 2 $ scattering. Some operators may contribute only  to other scattering processes, for example  $2\rightarrow 4 $, and may not contribute indirectly through loops to the processes of interest at a given loop order.} This is the reason why non-renormalizable theories are as good as renormalizable theories. An infinite tower of operators is truncated and a finite number of parameters is needed for making predictions, which is exactly the same situation as in renormalizable theories. 

It is a simplification to assume that different higher dimensional operators in Eq.~(\ref{eq:Leff}) are suppressed by the same scale $\Lambda$. Different operators can arise from exchanges of distinct heavy states that are not part of the effective theory. The scale $\Lambda$ is often referred to as the cutoff of the EFT. This is a somewhat misleading term that is not to be confused with a regulator used in loop calculations, for example a momentum cutoff. $\Lambda$ is related to the scale where the effective theory breaks down. However, dimensionless coefficients do matter. One could redefine $\Lambda$ by absorbing dimensionless numbers into the definitions of operators. The breakdown scale of an EFT is a physical scale that does not depend on the convention chosen for $\Lambda$. This scale could be estimated experimentally by measuring the energy dependence of amplitudes at small momentum. In EFTs, amplitudes grow at high energies and exceed the limits from unitarity at the breakdown scale. It is clear that the breakdown scale is physical since it corresponds to on-shell contributions from heavy states. 

The last remark regarding Eq.~(\ref{eq:Leff}) is that terms in the ${\mathcal L}_{d\leq 4}$ Lagrangian also receive contributions from the heavy fields.  Such contributions may not lead to observable consequences as the coefficients of interactions in  ${\mathcal L}_{d\leq 4}$ are determined from low-energy observables. In some cases, the heavy fields violate symmetries that would have been present in the full Lagrangian ${\mathcal L}(\varphi_L,\, \varphi_H=0)$ if the heavy fields are neglected. Symmetry-violating effects of heavy fields are certainly observable in ${\mathcal L}_{d\leq 4}$.

\subsection{Power counting and tree-level matching}

EFTs are based on several systematic expansions. In addition to the usual loop expansion in quantum field theory, one expands in the ratios of energy scales. There can be several scales in the problem: the masses of heavy particles, the energy at which the experiment is done, the momentum transfer, and so on. In an EFT, one can independently keep track of powers of the ratio of scales and of  the logarithms of scale ratios. This could be useful, especially when logarithms are large. Ratios of different scales can be kept to different orders depending on the numerical values, which is something that is nearly impossible to do without using EFTs.

When constructing an EFT one needs to be able to formally predict the magnitudes of different $O_i$ terms in the effective Lagrangian. This is referred to as power counting the terms in the Lagrangian and it allows one to predict how different terms scale with energy. In the simple EFTs discussed here, power counting is the same as dimensional analysis using the natural $\hbar=c=1$ units, in which $[mass]=[length]^{-1}$. From now on, dimensions will be expressed in the units of $[mass]$, so that energy has dimension 1, while length has dimension $-1$. The Lagrangian density has dimension 4 since $\int {\mathcal L}\,  d^4 x$ must be dimensionless. 

The dimensions of fields are determined from their kinetic energies because in weakly interacting theories these terms always dominate. The kinetic energy term for a scalar field, $\partial_\mu \phi\, \partial^\mu \phi$, implies that $\phi$ has dimension 1, while that of a fermion, $i\,  \overline{\psi} \, \slashed{\partial} \, \psi$, implies that $\psi$ has dimension $\frac{3}{2}$ in 4 space-time dimensions. 

A Yukawa theory consisting of a massless fermion interacting with with two real scalar fields: a light one and a heavy one will serve as our working example. The Lagrangian of the high-energy theory is taken to be 
\begin{equation}
\label{eq:full}
 {\mathcal L} = i \overline{\psi} \,\slashed{\partial}  \, \psi + \frac{1}{2} (\partial_\mu \Phi)^2 - \frac{M^2}{2} \Phi^2 + \frac{1}{2} (\partial_\mu \varphi)^2 - \frac{m^2}{2} \varphi^2 
                           - \lambda\, \overline{\psi} \psi \Phi  - \eta \, \overline{\psi} \psi \varphi.
 \end{equation} 
Let us assume that $M>>m$. As this is a toy example, we do not worry whether it is natural to have a hierarchy between $m$ and $M$. The Yukawa couplings are denoted as $\lambda$ and $\eta$. We neglect the potential for $\Phi$ and $\varphi$ as it is unimportant for now. 

As our first example of an EFT, we will consider tree-level effects. We want to find an effective theory with only the light fields present: the fermion $\psi$ and scalar $\varphi$. The interactions generated by the exchanges of the heavy field $\Psi$ will be mocked up by new interactions involving the light fields. 

We want to examine the $\psi \psi \rightarrow \psi \psi$ scattering process to order $\lambda^2$ in the coupling constants, that is to the zeroth order in $\eta$, and keep terms to the second order in the external momenta. 

\begin{figure}[htb]
\begin{center}
  \begin{picture}(353,68) (39,-39)
    \SetWidth{2.0}
    \Line[arrow,arrowpos=0.5,arrowlength=6.667,arrowwidth=2.667,arrowinset=0.2](40,26)(110,26)
    \Line[arrow,arrowpos=0.5,arrowlength=6.667,arrowwidth=2.667,arrowinset=0.2](110,26)(180,26)
    \Line[arrow,arrowpos=0.5,arrowlength=6.667,arrowwidth=2.667,arrowinset=0.2](40,-24)(110,-24)
    \Line[arrow,arrowpos=0.5,arrowlength=6.667,arrowwidth=2.667,arrowinset=0.2](110,-24)(180,-24)
    \Line[arrow,arrowpos=0.5,arrowlength=6.667,arrowwidth=2.667,arrowinset=0.2](250,26)(320,26)
    \Line[arrow,arrowpos=0.365,arrowlength=6.667,arrowwidth=2.667,arrowinset=0.2](320,26)(390,-24)
    \Line[arrow,arrowpos=0.5,arrowlength=6.667,arrowwidth=2.667,arrowinset=0.2](250,-24)(320,-24)
    \Line[arrow,arrowpos=0.365,arrowlength=6.667,arrowwidth=2.667,arrowinset=0.2](320,-24)(390,26)
    \Line[dash,dashsize=5](110,26)(110,-24)
    \Line[dash,dashsize=5](320,26)(320,-24)
    \Text(210,-6)[lb]{$-$}
    \Text(95,-3)[lb]{$\Phi$}
    \Text(45,9)[lb]{$\psi$}
    \Text(45,-41)[lb]{$\psi$}
    \Text(255,-41)[lb]{$\psi$}
    \Text(255,9)[lb]{$\psi$}
    \Text(305,-3)[lb]{$\Phi$}
     \Text(33,22)[lb]{$1$}
     \Text(33,-28)[lb]{$2$}
     \Text(184,22)[lb]{$3$}
     \Text(184,-28)[lb]{$4$}
      \Text(243,22)[lb]{$1$}
     \Text(243,-28)[lb]{$2$}
     \Text(394,22)[lb]{$3$}
     \Text(394,-28)[lb]{$4$}
  \end{picture}
\end{center}
\caption{\label{fig:tree} Tree-level diagrams proportional to $\lambda^2$ that contribute to $\psi \psi \rightarrow \psi \psi$ scattering.}
\end{figure}
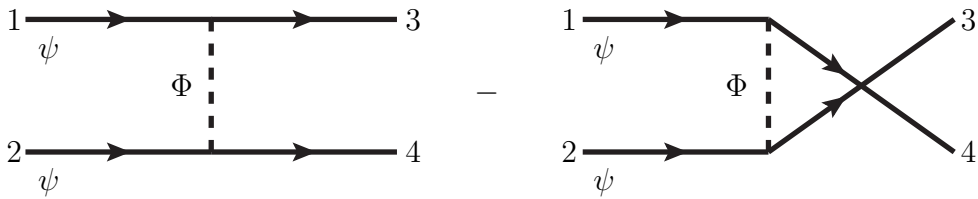

Integrating out fields is accomplished by comparing amplitudes in the full and effective theories. In this example, the only amplitude we need to worry about is the $\psi \psi \rightarrow \psi \psi$ scattering amplitude. Often the ``full theory" is referred to as the ultraviolet theory, and the effective theory as the infrared theory. The UV amplitude to order $\lambda^2$ is given by two tree-level graphs depicted in Fig.~\ref{fig:tree} and the result is 
\begin{equation}
\label{eq:amplitude}
  {\mathcal A}_{UV} = \overline{u}(p_3) u(p_1) \overline{u}(p_4) u(p_2) \, (- i \lambda)^2 \frac{i}{(p_3-p_1)^2 - M^2} - \left\{3 \leftrightarrow 4\right\}, 
\end{equation}
where $\left\{ 3 \leftrightarrow 4\right\}$ indicates interchange of $p_3$ and $p_4$ as required by the Fermi statistics. The Dirac structure is identical in the UV and IR theories, so we can concentrate on the propagator
\begin{equation}
\label{eq:expandp}
  (- i \lambda)^2 \frac{i}{(p_3-p_1)^2 - M^2} = i \frac{\lambda^2}{M^2}  \frac{1}{1-\frac{(p_3-p_1)^2 }{M^2}} 
     \approx  i \frac{\lambda^2}{M^2} \left(1 + \frac{(p_3-p_1)^2 }{M^2} + {\mathcal O}(\frac{p^4 }{M^4})\right)
\end{equation}
and neglect terms higher than second order in external field momenta. The ratio $\frac{p^2 }{M^2}$ is the expansion parameter and we can construct an effective theory to the desired order in this expansion. Since the effective theory does not include the heavy scalar of mass $M$, it is clear that the effective theory must break down when the scattering energy approaches $M$.

To the zeroth order in external momenta we can reproduce the $\psi \psi \rightarrow \psi \psi$ scattering amplitude by the four-fermion Lagrangian
\begin{equation}
   {\mathcal L}_{p^0,\lambda^2}=   i \overline{\psi} \,\slashed{\partial}  \, \psi + \frac{c}{2} \, \overline{\psi}  \psi \, \overline{\psi}  \psi,
\end{equation}
where the coefficient of the four-fermion term includes the $\frac{1}{2}$ symmetry factor that accounts for two factors of $\overline{\psi}  \psi$ in the interaction. We omit all the terms that depend on the light scalar $\varphi$ as such terms play no role here. We will restore these terms later. The amplitude calculated using the $ {\mathcal L}_{p^0,\lambda^2}$ Lagrangian is 
\begin{equation}
  {\mathcal A}_{IR} = \overline{u}(p_3) u(p_1) \overline{u}(p_4) u(p_2) \, (i c)  - \left\{3 \leftrightarrow 4\right\}.
\end{equation}
Comparing this with Eq.~(\ref{eq:expandp}) gives $c=\frac{\lambda^2}{M^2}$.

At the next order in the momentum expansion, we can write the Lagrangian as 
\begin{equation}
\label{eq:p2lam2}
   {\mathcal L}_{p^2,\lambda^2}=   i \overline{\psi} \,\slashed{\partial}  \, \psi + \frac{\lambda^2}{M^2} \,\frac{1}{2} \, \overline{\psi}  \psi \, \overline{\psi}  \psi + 
   d \, \partial_\mu \overline{\psi}  \partial^\mu \psi \, \overline{\psi}  \psi.
\end{equation}
We need to compare the scattering amplitude obtained from the effective Lagrangian $ {\mathcal L}_{p^2,\lambda^2}$ with the amplitude in Eqs.~(\ref{eq:amplitude}) and (\ref{eq:expandp}). The effective Lagrangian needs to be valid both on-shell and off-shell as we could build up more complicated diagrams from the effective interactions inserting them as parts of diagrams. For the matching procedure, we can use any choice of external momenta that is convenient. When comparing the full and effective theories, we can choose the momenta to be either on-shell or off-shell. The external particles, in this case fermions $\psi$, are identical in the full and effective theories. The choice of external momenta has nothing to do with the UV dynamics. In other words, for any small external momenta the full and effective theories must be identical, thus one is allowed to make opportunistic choices of momenta to simplify calculations. 

In this example, it is useful to assume that the momenta are on-shell that is $p_1^2=\ldots =p_4^2=0$. Therefore, the amplitude can only depend on the products of different momenta $p_i \cdot p_j$ with $i\neq j$. With this assumption, the effective theory needs to reproduce the $-2 i  \frac{\lambda^2}{M^2} \frac{p_1 \cdot p_3}{M^2}-\left\{3 \leftrightarrow 4\right\}$ part of the amplitude in Eq.~(\ref{eq:expandp}). The term proportional to $d$ in the  ${\mathcal L}_{p^2,\lambda^2}$ Lagrangian gives the amplitude
\begin{equation}
   {\mathcal A}_{IR} = i d \left( p_1\cdot p_3 + p_2 \cdot p_4 \right)   \overline{u}(p_3) u(p_1) \overline{u}(p_4) u(p_2) - \left\{3 \leftrightarrow 4\right\} .
\end{equation}
The momenta $p_1$ and $p_2$ are assumed to be incoming, thus they contribute $-i p^\mu_{1,2}$ to the amplitude, while the outgoing momenta contribute $+ i p^\mu_{3,4}$.  Conservation of momentum, $p_1+p_2=p_3+p_4$,  implies $p_1\cdot p_2 = p_3\cdot p_4$, $p_1\cdot p_3 = p_2\cdot p_4$, and $p_1\cdot p_4= p_2\cdot p_3$. Hence, $d=-\frac{\lambda^2}{M^4}$.

The derivative operator with the coefficient $d$ is not the only two-derivative term one can write with four fermions. For example, we could have included in the Lagrangian the term $(\partial^2\overline{\psi})  \psi \, \overline{\psi}  \psi + {\rm H.c.}$ or included the term $\partial_\mu \overline{\psi} \psi \, \overline{\psi}   \partial^\mu \psi$. When constructing a general effective Lagrangian it is important to consider all terms of a given order. There are four different ways to write two derivatives in the four-fermion interaction. Integration by parts implies that there is one relationship between the four possible terms. The term containing $\partial^2$ does not contribute on shell. In fact, this term can be removed from the effective Lagrangian using equations of motion~\cite{Politzer,Georgi}. We will discuss this in more detail in Sec.~\ref{sec:eoms}. Thus, there are only two independent two-derivative terms in this theory. At the tree level, only one of these terms turned out to be necessary to match  the UV theory.

\subsection{Renormalization group running}
\label{sec:RG}

So far we have focused on the fermions, but our original theory has two scalar fields. At tree level, we have obtained the effective Lagrangian
\begin{equation}
\label{eq:p2lam2phi}
   {\mathcal L}_{p^2,\lambda^2}=   i \overline{\psi} \,\slashed{\partial}  \, \psi + \frac{c}{2} \, \overline{\psi}  \psi \, \overline{\psi}  \psi + 
   d \, \partial_\mu \overline{\psi}  \partial^\mu \psi \, \overline{\psi}  \psi +  \frac{1}{2} (\partial_\mu \varphi)^2 - \frac{m^2}{2} \varphi^2 
                          - \eta \, \overline{\psi} \psi \varphi
\end{equation}
and calculated the coefficients $c$ and $d$~\footnote{The effective Lagrangian in Eq.~(\ref{eq:p2lam2phi}) is not complete to order $\lambda^2$ and $p^2$, it only contains all tree-level terms of this order. For example, the Yukawa coupling $ \overline{\psi} \psi \varphi$ receives corrections proportional to $\eta \lambda^2$ at one loop.}. Parameters do not exhibit scale dependence at tree level, but it will become clear that we calculated the effective couplings at the scale $M$ that is $c(\mu=M)=\frac{\lambda^2}{M^2}$ and $d(\mu=M)=- \frac{\lambda^2}{M^4}$.

Our next example will be computation of the $\psi \psi \rightarrow \psi \psi$ amplitude to the lowest order in the momenta and to order $\lambda^2 \eta^2$ in the UV coupling constants.  Such  contribution arises at one loop. Since loop integration generically yields factors of $\frac{1}{(4 \pi)^2}$ one expects then a correction of order $\frac{\eta^2}{(4 \pi)^2}$ compared to the tree-level amplitude. This is not an accurate estimate if there are several scales in the problem. We will assume that $m\ll M$, so the scattering amplitude could contain large $\log(\frac{M}{m})$. In fact, in an EFT one separates logarithm-enhanced contributions and contributions independent of large logs. The log-independent contributions arise from matching and the log-dependent ones are accounted for by the renormalization group (RG) evolution of parameters. By definition, while matching one compares theories with different field contents. This needs to be done using the same renormalization scale in both theories. This so-called matching scale is usually the mass of the heavy particle that is being integrated out. No large logarithms  can arise in the process since only one scale is involved. The logs of the matching scale divided by a low-energy scale must be identical in the two theories since the two theories are designed to be identical at low energies. We will illustrate loop matching in the next section. It is very useful that one can compute the matching and running contributions independently. This can be done at different orders in perturbation theory as dictated by the magnitudes of couplings and ratios of scales. 

In our effective theory described in Eq.~(\ref{eq:p2lam2phi}) we need to find the RG equation for the Lagrangian parameters. For concreteness, let us assume we want to know the amplitude at the scale $m$. Since we will be interested in the momentum independent part of the amplitude, we can neglect the term proportional to $d$. By dimensional analysis, the amplitude we are after must be proportional to $\frac{\lambda^2 \eta^2}{16 \pi^2 M^2}$. The two-derivative term will always be proportional to $\frac{1}{M^4}$, so it has to be suppressed by $\frac{m^2}{M^2}$ compared to the leading term arising from the non-derivative term. This reasoning only holds if one uses a mass-independent regulator, like dimensional regularization with minimal subtraction. In dimensional regularization, the renormalization scale $\mu$ only appears in logs. 

In less suitable regularization schemes, the two-derivative term could contribute as much as the non-derivative term as the extra power of $\frac{1}{M^2}$ could become $\frac{\Lambda^2}{M^2}$, where $\Lambda$ is the regularization scale. With the natural choice $\Lambda\approx  M$, the two-derivative term is not suppressed at all. Since the same argument holds for terms with more and more derivatives, all terms would contribute exactly the same and the momentum expansion would be pointless. This is, for example, how hard momentum cut off and Pauli-Villars regulators behave. Such regulators do their job, but they needlessly complicate power counting. From now on, we will only be using dimensional regularization. 

To calculate the RG running of the coefficient $c$, we need to obtain the relevant $Z$ factors. First, we need the fermion self energy diagram
\begin{eqnarray} 
\label{eq:selfenergy}
\begin{picture}(54,20) (137,-72)
    \SetWidth{1.5}
    \Line[arrow,arrowpos=0.5,arrowlength=5.833,arrowwidth=2.333,arrowinset=0.2](120,-68)(192,-68)
    \Arc[dash,dashsize=6,clock](156,-68)(18,-180,-360)
    \Text(126,-82)[lb]{$p$}
    \Text(145,-82)[lb]{$p+k$}
    \Text(153,-62)[lb]{$k$}
  \end{picture} 
  & = & (-i \eta)^2 \int \frac{d^d k}{(2\pi)^d} \frac{i (\slashed{k} + \slashed{p})}{(k+p)^2} \frac{i}{k^2-m^2} \nonumber 
  = \eta^2 \int \frac{d^d l}{(2\pi)^d} \int_0^1 dx\,  \frac{ \slashed{l} + (1-x) \slashed{p}}{[l^2 - \Delta^2]^2} \nonumber \\
  & = &\frac{i \eta^2}{(4\pi)^2} \frac{1}{\epsilon}\left( \int_0^1 dx(1-x) \slashed{p} \right) + \ {\rm finite}  
             =  \frac{i \eta^2 \slashed{p}}{2  (4\pi)^2}  \frac{1}{\epsilon}  + \ {\rm finite},  
\end{eqnarray}
where we used Feynman parameters to combine the denominators and shifted the loop momentum $l=k + xp$. We then used the standard result for loop integrals and expanded $d=4-2 \epsilon$. Only the $ \frac{1}{\epsilon}$ pole is kept as the finite term does not enter the RG calculation.

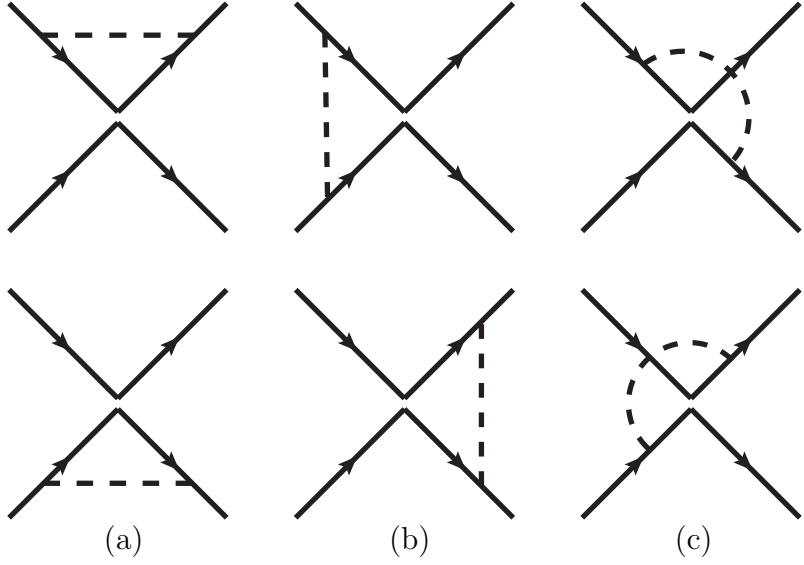
\begin{figure}[htb]
\begin{center}
  \begin{picture}(300,210) (50,-20)
    \SetWidth{2.0}
    \Line[arrow,arrowpos=0.5,arrowlength=5,arrowwidth=2,arrowinset=0.2](50,177)(91,136)
    \Line[arrow,arrowpos=0.5,arrowlength=5,arrowwidth=2,arrowinset=0.2](91,136)(132,177)
    \Line[arrow,arrowpos=0.5,arrowlength=5,arrowwidth=2,arrowinset=0.2](50,91)(91,132)
    \Line[arrow,arrowpos=0.5,arrowlength=5,arrowwidth=2,arrowinset=0.2](91,132)(132,91)
    \Line[arrow,arrowpos=0.5,arrowlength=5,arrowwidth=2,arrowinset=0.2](50,69)(91,28)
    \Line[arrow,arrowpos=0.5,arrowlength=5,arrowwidth=2,arrowinset=0.2](91,28)(132,69)
    \Line[arrow,arrowpos=0.5,arrowlength=5,arrowwidth=2,arrowinset=0.2](50,-17)(91,24)
    \Line[arrow,arrowpos=0.5,arrowlength=5,arrowwidth=2,arrowinset=0.2](91,24)(132,-17)
    \Line[arrow,arrowpos=0.5,arrowlength=5,arrowwidth=2,arrowinset=0.2](158,177)(199,136)
    \Line[arrow,arrowpos=0.5,arrowlength=5,arrowwidth=2,arrowinset=0.2](199,136)(240,177)
    \Line[arrow,arrowpos=0.5,arrowlength=5,arrowwidth=2,arrowinset=0.2](158,91)(199,132)
    \Line[arrow,arrowpos=0.5,arrowlength=5,arrowwidth=2,arrowinset=0.2](199,132)(240,91)
    \Line[arrow,arrowpos=0.5,arrowlength=5,arrowwidth=2,arrowinset=0.2](158,69)(199,28)
    \Line[arrow,arrowpos=0.5,arrowlength=5,arrowwidth=2,arrowinset=0.2](199,28)(240,69)
    \Line[arrow,arrowpos=0.5,arrowlength=5,arrowwidth=2,arrowinset=0.2](158,-17)(199,24)
    \Line[arrow,arrowpos=0.5,arrowlength=5,arrowwidth=2,arrowinset=0.2](199,24)(240,-17)
    \Line[arrow,arrowpos=0.5,arrowlength=5,arrowwidth=2,arrowinset=0.2](266,177)(307,136)
    \Line[arrow,arrowpos=0.5,arrowlength=5,arrowwidth=2,arrowinset=0.2](307,136)(348,177)
    \Line[arrow,arrowpos=0.5,arrowlength=5,arrowwidth=2,arrowinset=0.2](266,91)(307,132)
    \Line[arrow,arrowpos=0.5,arrowlength=5,arrowwidth=2,arrowinset=0.2](307,132)(348,91)
    \Line[arrow,arrowpos=0.5,arrowlength=5,arrowwidth=2,arrowinset=0.2](266,69)(307,28)
    \Line[arrow,arrowpos=0.5,arrowlength=5,arrowwidth=2,arrowinset=0.2](307,28)(348,69)
    \Line[arrow,arrowpos=0.5,arrowlength=5,arrowwidth=2,arrowinset=0.2](266,-17)(307,24)
    \Line[arrow,arrowpos=0.5,arrowlength=5,arrowwidth=2,arrowinset=0.2](307,24)(348,-17)
    \Line[dash,dashsize=6](62,165)(120,165)
    \Line[dash,dashsize=6](63,-4)(119,-4)
    \Line[dash,dashsize=6](169,166)(170,104)
    \Line[dash,dashsize=6](228,56)(228,-5)
    \Arc[dash,dashsize=6,clock](303.882,134)(24.93,126.654,-39.927)
    \Arc[dash,dashsize=6,clock](306.408,26.074)(22.998,-132.064,-315.173)
    \Text(86,-32)[lb]{(a)}
    \Text(194,-32)[lb]{(b)}
    \Text(302,-32)[lb]{(c)}
  \end{picture}
\end{center}
\caption{\label{fig:4fermion1loop} Diagrams contributing to the renormalization of the four-fermion interaction. The dashed lines represent the light scalar $\varphi$. The four-fermion vertices are represented by the kinks on the fermion lines. The fermion lines do not touch even though the interaction is point-like. This is not due to limited graphic skills of the author, but rather to illustrate the fermion number flow through the vertices.}   
\end{figure}

The second part of the calculation involves computing loop corrections to the four-fermion vertex. There are six diagrams with a scalar exchange because there are six different pairings of the external lines. The diagrams are depicted in Fig.~\ref{fig:4fermion1loop} and there are two diagrams in each of the three topologies. All of these diagrams are logarithmically divergent in the UV, so we can neglect the external momenta and masses if we are interested in the divergent parts. The divergent terms must be local and therefore be analytic in the external momenta. Extracting positive powers of momenta from a diagram reduces its degree of divergence which is apparent from dimensional analysis.  Diagrams (a) in Fig.~\ref{fig:4fermion1loop} are the most straightforward to deal with and the divergent part is easy to extract
\begin{equation}
\label{eq:cvertex}
  2 (-i\eta)^2 i c \int \frac{d^d k}{(2\pi)^d} \frac{i \slashed{k} }{k^2}  \frac{i \slashed{k} }{k^2}  \frac{i}{k^2} 
     = -2 c \eta^2  \int \frac{d^d k}{(2\pi)^d} \frac{1}{k^4} = - \frac{2 i c \eta^2}{(4\pi)^2} \frac{1}{\epsilon} + \ {\rm finite}.
\end{equation}

We did not mention the cross diagrams here, denoted $\left\{3 \leftrightarrow 4\right\}$ in the previous section, since they go along for the ride, but they participate in every step. Diagrams (b) in Fig.~\ref{fig:4fermion1loop} require more care as the loop integral involves two different fermion lines. To keep track of this we indicate the external spinors and abbreviate $u(p_i)=u_i$. The result is
\begin{equation}
  2 (-i\eta)^2 i c \int \frac{d^d k}{(2\pi)^d} \overline{u}_3 \frac{i \slashed{k} }{k^2} u_1 \,  \overline{u}_4 \frac{- i \slashed{k} }{k^2}  u_2  \frac{i}{k^2} 
  = \frac{ i c \eta^2}{2 (4\pi)^2}  \frac{1}{\epsilon} \, \overline{u}_3 \gamma^\mu u_1 \,  \overline{u}_4\gamma_\mu  u_2+ \ {\rm finite}.
\end{equation}
This divergent contribution is canceled by diagrams (c) in Fig.~\ref{fig:4fermion1loop} because one of the momentum lines carries the opposite sign
\begin{equation}
  2 (-i\eta)^2 i c \int \frac{d^d k}{(2\pi)^d} \overline{u}_3 \frac{i \slashed{k} }{k^2} u_1 \,  \overline{u}_4 \frac{ i \slashed{k} }{k^2}  u_2  \frac{i}{k^2}. 
\end{equation}
If the divergent parts of the diagrams (b) and (c) did not cancel this would lead to operator mixing which often takes place among operators with the same dimensions. We will illustrate this shortly.

To calculate the RG equations (RGEs) we can consider just the fermion part of the Lagrangian in Eq.~(\ref{eq:p2lam2phi}) and neglect the derivative term proportional to $d$. We can think of the original Lagrangian as being expressed in terms of the bare fields and bare coupling constants and rescale $\psi_0= \sqrt{Z_\psi} \psi$ and $c_0= c \mu^{2 \epsilon} Z_c$. As usual in dimensional regularization, the mass dimensions of the fields depend on the dimension of space-time.  In $d=4-2 \epsilon$, the fermion dimension is $[\psi]=\frac{3}{2}-\epsilon$ and $[{\mathcal L}]=4-2 \epsilon$. We explicitly compensate for this change from the usual 4 space-time dimensions by including the factor $\mu^{2 \epsilon}$ in the interaction term. This way, the coupling $c$ does not alter its dimension when $d=4-2 \epsilon$. The Lagrangian is then 
\begin{eqnarray}
\label{eq:counterterms}
   {\mathcal L}_{p^0,\lambda^2 \eta^2 \log}&=&    i \overline{\psi}_0 \, \slashed{\partial} \, \psi_0 + \frac{c_0}{2} \, \overline{\psi}_0 \psi_0 \, \overline{\psi}_0  \psi_0
   =   i Z_\psi \overline{\psi} \,\slashed{\partial} \, \psi + \frac{c}{2} \, Z_c Z_\psi^2 \mu^{2\epsilon} \overline{\psi}  \psi \, \overline{\psi}  \psi
     \nonumber \\
 & = &  i \overline{\psi} \,\slashed{\partial} \, \psi + \mu^{2 \epsilon} \frac{c}{2} \, \overline{\psi}  \psi \, \overline{\psi}  \psi + i (Z_\psi -1)  \overline{\psi} \,\slashed{\partial} \, \psi + \mu^{2 \epsilon}  \frac{c}{2} (Z_c Z_\psi^2 -1)  \, \overline{\psi}  \psi \, \overline{\psi}  \psi,
\end{eqnarray}
where in the last line we separated the counterterms. We can read off the counterterms from Eqs.~(\ref{eq:selfenergy}) and (\ref{eq:cvertex}) by insisting that the counterterms cancel the divergences we calculated previously.
\begin{equation}
\label{eq:Zfactors}
   Z_\psi-1=- \frac{ \eta^2}{2  (4\pi)^2} \frac{1}{\epsilon}\  \ {\rm and} \ \ c (Z_c Z_\psi^2-1) = \frac{2 c \eta^2}{ (4\pi)^2} \frac{1}{\epsilon},
\end{equation}
where we used the minimal subtraction (MS) prescription and hence retained only the $\frac{1}{\epsilon}$ poles. Comparing the two equations in (\ref{eq:Zfactors}), we obtain $Z_c=1+ \frac{3 \eta^2}{ (4\pi)^2} \frac{1}{\epsilon}$. 

The standard way of computing RGEs is to use the fact that the bare quantities do not depend on the renormalization scale
\begin{equation}
\label{eq:cbare}
  0=\mu \frac{d}{d\mu} c_0= \mu \frac{d}{d\mu} (c \mu^{2 \epsilon} Z_c)= \beta_c \mu^{2 \epsilon} Z_c + 2 \epsilon c \mu^{2 \epsilon} Z_c + c \mu^{2 \epsilon} \mu \frac{d}{d\mu}  Z_c,
\end{equation}
where $\beta_c\equiv  \mu \frac{d c}{d\mu}$. We have $\mu \frac{d}{d\mu}  Z_c = \frac{3}{(4\pi)^2} 2 \eta \beta_\eta  \frac{1}{\epsilon}$. Just like we had to compensate for the dimension of $c$, the renormalized coupling $\eta$ needs an extra factor of $\mu^\epsilon$ to remain dimensionless in the space-time where $d=4-2 \epsilon$. Repeating the same manipulations we used in Eq.~(\ref{eq:cbare}), we obtain $\beta_\eta=-\epsilon \eta -  \eta \frac{d \log Z_\eta}{d \log \mu}$. Keeping the derivative of $Z_\eta$ would give us a term that is of higher order in $\eta$ as for any $Z$ factor the scale dependence comes from the couplings. Thus, we keep only the first term, $\beta_\eta=-\epsilon \eta$, and get 
$\mu \frac{d}{d\mu}  Z_c =-  \frac{6 \eta^2}{(4\pi)^2}$. Finally,
\begin{equation}
\label{eq:betac}
  \beta_c= \frac{6 \eta^2}{(4\pi)^2} c.
\end{equation}

We can now complete our task and compute the low-energy coupling, and thus the scattering amplitude, to the leading log order 
\begin{equation}
\label{eq:clow}
  c(m)= c(M) - \frac{6 \eta^2}{(4\pi)^2} c \log\left(\frac{M}{m}\right)= \frac{\lambda^2}{M^2}\left[1 - \frac{6 \eta^2}{(4\pi)^2} \log\left(\frac{M}{m}\right)\right] .
\end{equation}
Of course, at this point it requires little extra work to re-sum the logarithms by solving the RGEs. First, one needs to solve for the running of $\eta$. We will not compute it in detail here, but $\beta_\eta=\frac{5 \eta^3}{(4\pi)^2}$. Solving this equation gives
\begin{equation}
\label{eq:etasol}
  \frac{1}{\eta^2(\mu_2)} - \frac{1}{\eta^2(\mu_1)} =
        \frac{10}{(4 \pi)^2} \log\frac{\mu_1}{\mu_2}.
\end{equation}
Putting the $\mu$ dependence of $\eta$ from Eq.~(\ref{eq:etasol}) into Eq.~(\ref{eq:betac}) and performing the integral yields
\begin{equation}
  c(m)=C(M)\left(\frac{\eta^2(m)}{\eta^2(M)}\right)^\frac{3}{5},
\end{equation}
which agrees with Eq.~(\ref{eq:clow}) to the linear order in $ \log\left(\frac{M}{m}\right)$. 

It is worth pointing out that the Yukawa interaction in the full theory, $\eta\, \overline{\psi} \psi \varphi$, receives corrections from the exchanges of both the light and the heavy scalars. Hence, the beta function $\beta_\eta$ receives contributions proportional to $\eta^3$ and $\eta \lambda^2$. In fact, $\beta_\eta=\frac{5 \eta^3}{(4\pi)^2}+\frac{3 \eta \lambda^2}{(4\pi)^2}$. The beta function has no dependence on the mass of the heavy scalar nor on the renormalization scale, so one might be tempted to use this beta function at any renormalization scale. For example, this would imply that heavy particles contribute to the running of $\eta$ at energy scales much smaller than their mass. Clearly, this is unphysical. If this was true, we could count all the electrically charged particles even as heavy as the Planck scale by measuring the charge at two energy scales, for example by scattering at the center of mass energies equal to the electron mass and equal to the  $Z$ mass.  The fact that the beta function has no dependence on the renormalization scale is characteristic of mass-independent regulators, like dimensional regularization coupled with minimal subtraction. When using dimensional regularization, heavy particles need to be removed from the theory to get physical answers for the beta function. This is yet another  reason why dimensional regularization goes hand in hand with the EFT approach. 

The contribution from the heavy scalar, proportional to $\eta \lambda^2$, is absent in the effective theory since the heavy scalar was removed from the theory. However, diagrams that reproduce the exchanges of the heavy scalar do exist in the effective theory. Such diagrams are proportional to $c \eta$ and arise from the four-fermion vertex corrections to the Yukawa interaction. Since $c$ is proportional to $\frac{1}{M^2}$, the dimensionless $\beta_\eta$ must be suppressed by $\frac{m_\psi^2}{M^2}$. We assumed that $m_\psi=0$, so the $c \eta$ contribution vanishes. Integrating out the heavy scalar changed the Yukawa $\beta_\eta$ function in a step-wise manner while going from the full theory to the effective theory. Calculations of the beta function performed using a mass-dependent regulator yield a smooth transition from one asymptotic value of the beta function to another, see for example Ref.~\cite{Aneesh}. However, the predictions for physical quantities are not regulator dependent. 

It is interesting that lack of renormalizability  of the four-fermion interaction never played any role in our calculation. Our calculation would have looked identical if we wanted to obtain the RGE for the electric charge in QED\@. The number of pertinent terms in the Lagrangian was finite since we were interested in a finite order in the momentum expansion. 

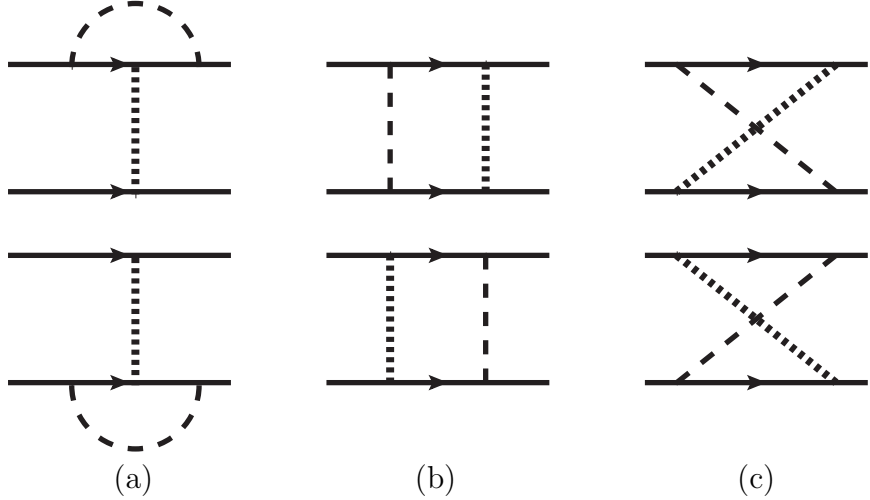
\begin{figure}[htb]
\begin{center}
\begin{picture}(326,159) (35,-10)
    \SetWidth{2.0}
    \Line[arrow,arrowpos=0.5,arrowlength=5,arrowwidth=2,arrowinset=0.2](36,129)(120,129)
    \Line[arrow,arrowpos=0.5,arrowlength=5,arrowwidth=2,arrowinset=0.2](36,81)(120,81)
    \Line[arrow,arrowpos=0.5,arrowlength=5,arrowwidth=2,arrowinset=0.2](36,57)(120,57)
    \Line[arrow,arrowpos=0.5,arrowlength=5,arrowwidth=2,arrowinset=0.2](36,9)(120,9)
    \Line[arrow,arrowpos=0.5,arrowlength=5,arrowwidth=2,arrowinset=0.2](156,129)(240,129)
    \Line[arrow,arrowpos=0.5,arrowlength=5,arrowwidth=2,arrowinset=0.2](156,81)(240,81)
    \Line[arrow,arrowpos=0.5,arrowlength=5,arrowwidth=2,arrowinset=0.2](156,57)(240,57)
    \Line[arrow,arrowpos=0.5,arrowlength=5,arrowwidth=2,arrowinset=0.2](156,9)(240,9)
    \Line[arrow,arrowpos=0.5,arrowlength=5,arrowwidth=2,arrowinset=0.2](276,129)(360,129)
    \Line[arrow,arrowpos=0.5,arrowlength=5,arrowwidth=2,arrowinset=0.2](276,81)(360,81)
    \Line[arrow,arrowpos=0.5,arrowlength=5,arrowwidth=2,arrowinset=0.2](276,57)(360,57)
    \Line[arrow,arrowpos=0.5,arrowlength=5,arrowwidth=2,arrowinset=0.2](276,9)(360,9)
    \Line[dash,dashsize=6](180,129)(180,81)
    \Line[dash,dashsize=6](216,57)(216,9)
     \Line[dash,dashsize=6](288,129)(348,81)
    \Line[dash,dashsize=6](288,9)(348,57)
     \Arc[dash,dashsize=6,clock](84,128)(24,-180,-360)
    \Arc[dash,dashsize=6](84,8)(24,-180,0)
    \SetWidth{3.0}
    \Line[dash,dashsize=2](216,129)(216,81)
    \Line[dash,dashsize=2](180,57)(180,9)
    \Line[dash,dashsize=2](288,57)(348,9)
    \Line[dash,dashsize=2](288,81)(348,129)
        \Line[dash,dashsize=2](84,56)(84,8)
    \Line[dash,dashsize=2](84,80)(84,128)
     \Text(76,-34)[lb]{(a)}
    \Text(190,-34)[lb]{(b)}
    \Text(312,-34)[lb]{(c)}
  \end{picture}
\end{center}
\caption{\label{fig:4fermionfull} The full theory analogs of the diagrams in Fig.~\protect{\ref{fig:4fermion1loop}}. The thicker dashed lines with shorter dashes represent $\Phi$, while the thinner ones with longer dashes represent $\varphi$. }
\end{figure}

The diagrams we calculated to obtain Eq.~(\ref{eq:clow}) are in a one-to-one correspondence with the diagrams in the full theory. These are depicted in Fig.~\ref{fig:4fermionfull}. Even though the EFT calculation may seem complicated, typically the EFT diagrams are simpler to compute as fewer propagators are involved. Also, computing the divergent parts of diagrams is much easier than computing the finite parts. In the full theory, one would need to calculate the finite parts of the box diagrams which can involve complicated integrals over Feynman parameters.  

There is one additional complication that is common in any field theory, not just in an EFT\@. When we integrated out the heavy scalar in the previous section, the only momentum-independent operator that is generated at tree level is $\overline{\psi}  \psi \, \overline{\psi}  \psi$. This is not the only four-fermion operator with no derivatives. There are other operators with the same field content and the same dimension, for example $\overline{\psi}  \gamma^\mu \psi \, \overline{\psi}  \gamma_\mu\psi$. Suppose that we integrated out a massive vector field with mass $M$ and that our effective theory is instead
\begin{equation}
\label{eq:p0,V}
   {\mathcal L}_{p^0,V}=   i \overline{\psi} \,\slashed{\partial} \, \psi + \frac{c_V}{2} \,\overline{\psi}  \gamma^\mu \psi \, \overline{\psi}  \gamma_\mu\psi +   \frac{1}{2} (\partial_\mu \varphi)^2 - \frac{m^2}{2} \varphi^2 
                          - \eta \, \overline{\psi} \psi \varphi.
\end{equation}
We could ask the same question about low-energy scattering in this theory, that is ask about the RG evolution of the coefficient $c_V$. The contributions from the $\varphi$ exchanges are  identical to those depicted in  Fig.~\ref{fig:4fermion1loop}. The only difference is that the four-fermion vertex contains the $\gamma^\mu$ matrices. Diagrams (a) give a divergent contribution to the $\overline{\psi}  \gamma^\mu \psi \, \overline{\psi}  \gamma_\mu\psi$ operator. However, the sum of the divergent parts of diagrams (b) and (c) is not proportional to the original operator, but instead proportional to $\overline{\psi}  \sigma^{\mu \nu} \psi \, \overline{\psi}  \sigma_{\mu \nu}\psi$, where $\sigma^{\mu \nu}=\frac{i}{2}[\gamma^\mu, \gamma^\nu]$. This means that under RG evolution these two operators mix. The two operators have the same dimensions, field content, and symmetry properties thus loop corrections can turn one operator into another. 

To put it differently, it is not consistent to just keep a single four-fermion operator in the effective Lagrangian in Eq.~(\ref{eq:p0,V}) at one loop. The theory needs to be supplemented since there needs to be an additional counterterm to absorb the divergence. At one loop it is enough to consider 
\begin{equation}
\label{eq:p0,VT}
   {\mathcal L}_{p^0,VT}=   i \overline{\psi} \,\slashed{\partial} \, \psi + \frac{c_V}{2} \,\overline{\psi}  \gamma^\mu \psi \, \overline{\psi}  \gamma_\mu\psi +   \frac{c_T}{2} \, \overline{\psi}  \sigma^{\mu \nu} \psi \, \overline{\psi}  \sigma_{\mu \nu}\psi  + \frac{1}{2} (\partial_\mu \varphi)^2 - \frac{m^2}{2} \varphi^2 
                          - \eta \, \overline{\psi} \psi \varphi ,
\end{equation}
but one expects that at higher loop orders all four-fermion operators are needed. Since we assumed that the operator proportional to $c_V$ was generated by a heavy vector field at tree level, we know that in our effective theory $c_T(\mu=M)=0$ and $c_V(\mu=M)\neq0$. At low energies, both coefficients will be nonzero. 

We do not want to provide the calculation of the beta functions for the coefficients $c_V$ and $c_T$ in great detail. This calculation is completely analogous to the one for $\beta_c$. The vector operator induces divergent contributions to itself and to the tensor operator, while the tensor operator only generates a divergent contribution for the vector operator. The coefficients of the two counterterms are
\begin{eqnarray}
\label{eq:betaVT}
  c_V (Z_V Z_\psi^2 -1 ) & = & \frac{\eta^2}{(4\pi)^2} (- c_V + 6 c_T) \frac{1}{\epsilon}, \label{eq:ZV} \\
  c_T (Z_T Z_\psi^2 -1 ) & = & \frac{\eta^2}{(4\pi)^2} c_V \frac{1}{\epsilon},  \label{eq:ZT} 
\end{eqnarray}
where we introduced separate $Z$ factors for each operator since each requires a counterterm. These $Z$ factors imply that the  beta functions are
\begin{equation}
  \beta_{c_V}  =  12 \, c_T\, \frac{\eta^2}{(4\pi)^2} \ \ {\rm and} \ \   \beta_{c_T} = 2\, ( c_T+ c_V) \frac{\eta^2}{(4\pi)^2}.
\end{equation}
This result may look surprising when compared with Eqs.~(\ref{eq:ZV}) and (\ref{eq:ZT}). The difference in the structures of the divergences and the beta functions comes from the wave function renormalization encoded in $Z_\psi$.  It is easy to solve the RGEs in Eq.~(\ref{eq:betaVT}) by treating them as one matrix equation
\begin{equation}
  \mu \frac{d}{d \mu} \left(\begin{array}{c} c_V \\ c_T \end{array} \right) = \frac{2 \eta^2}{(4\pi)^2} 
 \left(  \begin{array}{cc} 0  & 6 \\ 1 & 1 \end{array} \right) \left(\begin{array}{c} c_V \\ c_T \end{array} \right).
\end{equation}
The eigenvectors of this matrix satisfy uncoupled RGEs and they correspond to the combinations of operators that do not mix under one-loop renormalization.   

\subsection{One-loop matching}
\label{sec:1loop}

Construction of effective theories is a systematic process.  We saw how RGEs can account for each ratio of scales, and we now increase the accuracy of matching calculations. To improve our $\psi \psi \rightarrow \psi \psi$ scattering calculation we compute matching coefficients to one-loop order. As an example, we examine terms proportional to $\lambda^4$. This calculation illustrates several important points about matching calculations.

Our starting point is again the full theory with two scalars, described in Eq.~(\ref{eq:full}). Since we are only interested in the heavy scalar field, we can neglect the light scalar for the time being and consider
\begin{equation}
 {\mathcal L} = i \overline{\psi} \,\slashed{\partial} \, \psi - \sigma \overline{\psi}  \psi+ \frac{1}{2} (\partial_\mu \Phi)^2 - \frac{M^2}{2} \Phi^2     - \lambda\, \overline{\psi} \psi \Phi + {\rm terms \ that \ depend \ on}\  \varphi.
 \end{equation}  
 We added a small mass, $\sigma$, for the fermion to avoid possible IR divergences and also to be able to obtain a nonzero answer for terms proportional to $\frac{1}{M^4}$.

\begin{figure}[htb]
\begin{center}
  \begin{picture}(412,96) (39,-9)
    \SetWidth{2.0}
    \Line[arrow,arrowpos=0.5,arrowlength=5,arrowwidth=2,arrowinset=0.2](40,66)(120,66)
    \Line[arrow,arrowpos=0.5,arrowlength=5,arrowwidth=2,arrowinset=0.2](40,6)(120,6)
    \Line[arrow,arrowpos=0.5,arrowlength=5,arrowwidth=2,arrowinset=0.2](150,66)(230,66)
    \Line[arrow,arrowpos=0.5,arrowlength=5,arrowwidth=2,arrowinset=0.2](150,6)(230,6)
    \Line[arrow,arrowpos=0.5,arrowlength=5,arrowwidth=2,arrowinset=0.2](260,66)(340,66)
    \Line[arrow,arrowpos=0.5,arrowlength=5,arrowwidth=2,arrowinset=0.2](260,6)(340,6)
    \Line[arrow,arrowpos=0.5,arrowlength=5,arrowwidth=2,arrowinset=0.2](370,66)(450,66)
    \Line[arrow,arrowpos=0.5,arrowlength=5,arrowwidth=2,arrowinset=0.2](370,6)(450,6)
    \Line[dash,dashsize=5.6](60,66)(60,6)
    \Line[dash,dashsize=5.6](100,66)(100,6)
    \Line[dash,dashsize=5.6](170,66)(210,6)
    \Line[dash,dashsize=5.6](170,6)(210,66)
    \Line[dash,dashsize=5.6](420,66)(420,6)
    \Arc[dash,dashsize=5.6,clock](420,66)(20,-180,-360)
    \Line[dash,dashsize=5.6](310,46)(310,66)
    \Line[dash,dashsize=5.6](310,26)(310,6)
    \Arc[arrow,arrowpos=0.75,arrowlength=5,arrowwidth=2,arrowinset=0.2](310,36)(10,270,630)
    \Text(70,-20)[lb]{(a)}
    \Text(180,-20)[lb]{(b)}
    \Text(300,-20)[lb]{(c)}
    \Text(400,-20)[lb]{(d)}
  \end{picture}
\end{center}
\caption{\label{fig:full1loop} Diagrams in the full theory to order $\lambda^4$. Diagram (d) stands in for two diagrams that differ only by the placement of the loop.}
\end{figure}
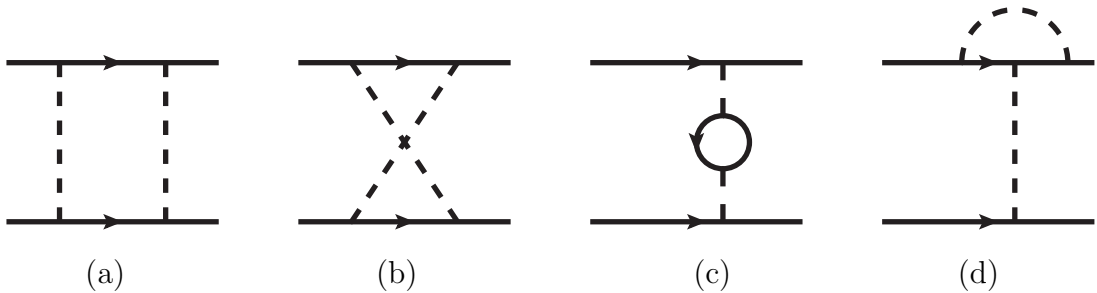

The diagrams that contribute to the scattering at one loop are illustrated in Fig.~\ref{fig:full1loop}. As we did before, we will focus on the momentum-independent part of the amplitude and we will not explicitly write the terms related by exchange of external fermions. The first diagram gives
\begin{eqnarray}
 (a) &=& (-i \lambda)^4   \int \frac{d^d k}{(2\pi)^d} \overline{u}_3  \frac{i(\slashed{k} + \sigma)}{k^2-\sigma^2} u_1 \,  \overline{u}_4 i \frac{i(-\slashed{k} + \sigma)}{k^2-\sigma^2} u_2 \frac{i^2}{(k^2-M^2)^2} \nonumber \\
  & = & \lambda^4 \left[ - \overline{u}_3 \gamma^\alpha  u_1 \,  \overline{u}_4 \gamma^\beta u_2 \int \frac{d^d k}{(2\pi)^d} \frac{k_\alpha k_\beta}{(k^2-\sigma^2)^2 (k^2-M^2)^2} \right. \nonumber \\
   && \left. + \overline{u}_3  u_1 \,  \overline{u}_4 u_2 \int \frac{d^d k}{(2\pi)^d} \frac{\sigma^2}{(k^2-\sigma^2)^2 (k^2-M^2)^2} \right].
\end{eqnarray}
The loop integrals are straightforward to evaluate using Feynman parameterization
\begin{equation}
  \frac{1}{(k^2-\sigma^2)^2 (k^2-M^2)^2}=6 \int_0^1 dx \frac{x (1-x)}{(k^2 - x M^2 - (1-x)\sigma^2)^4}.
\end{equation}  
The final result for diagram (a) is 
\begin{eqnarray}
\label{eq:aF}
\!\!\!\!\! (a)_F &=& \frac{i \lambda^4}{(4 \pi)^2}\left[   U_V \frac{1}{2}  \int_0^1 dx \frac{x (1-x)}{x M^2 + (1-x) \sigma^2} + \sigma^2 U_S \int_0^1 dx \frac{x (1-x)}{(x M^2 + (1-x) \sigma^2)^2} \right] \nonumber \\
        &=&  \frac{i \lambda^4}{(4 \pi)^2}\left[  U_V \left(\frac{1}{4 M^2} + \frac{\sigma^2}{4 M^4} (3 - 2 \log(\frac{M^2}{\sigma^2})) \right)  
              + U_S \frac{ \sigma^2}{M^4} (\log(\frac{M^2}{\sigma^2})-2) \right] +\ldots, 
\end{eqnarray}
where we abbreviated $U_S=\overline{u}_3  u_1 \,  \overline{u}_4 u_2$, $U_V=\overline{u}_3 \gamma^\alpha  u_1 \,  \overline{u}_4 \gamma_\alpha u_2$, and in the last line omitted terms of order $\frac{1}{M^6}$ and higher. The subscript $F$ stands for the full theory, We will denote the corresponding amplitudes in the effective theory with the subscript $E$. The cross box amplitude (b) is nearly identical, except for the sign of the momentum in one of the fermion propagators
 \begin{equation}
 \label{eq:bF}
 (b)_F =  \frac{i \lambda^4}{(4 \pi)^2}\left[  - U_V \left(\frac{1}{4 M^2} + \frac{\sigma^2}{4 M^4} (3 - 2 \log(\frac{M^2}{\sigma^2})) \right)  
              + U_S \frac{\sigma^2}{M^4} (\log(\frac{M^2}{\sigma^2})-2) \right] +\ldots.
\end{equation}
Diagrams (c) and (d) are even simpler to evaluate, but they are divergent.
\begin{equation}
\label{eq:cF}
 (c)_F=- 4 \frac{i \lambda^4}{(4 \pi)^2} \frac{\sigma^2}{M^4} U_S \left[ 3 \frac{1}{\overline{\epsilon}} + 3  \log(\frac{\mu^2}{\sigma^2}) + 1 \right] + \ldots,
\end{equation}
where $\frac{1}{\overline{\epsilon}} = \frac{1}{\epsilon} - \gamma + \log(4 \pi)$. $\mu$ is the regularization scale and it enters since coupling $\lambda$ carries a factor of $\mu^\epsilon$ in dimensional regularization. The four Yukawa couplings give $\lambda^4 \mu^{4\epsilon}$. However, $\mu^{2 \epsilon}$ should be factored out of the calculation to give the proper dimension of the four-fermion coupling, while the remaining $\mu^{2 \epsilon}$ is expanded for small $\epsilon$ and yields $\log(\mu^2)$.  In the following expression a factor of two is included to account for two diagrams
\begin{equation}
  (d)_F= -2 \frac{i \lambda^4}{ (4 \pi)^2 M^2} U_S \left[ \frac{1}{\overline{\epsilon}} + 1 + \log(\frac{\mu^2}{M^2}) +\frac{\sigma^2}{M^2} \left(2-3 \log(\frac{M^2}{\sigma^2}) \right) \right]+\ldots.
\end{equation}
The sum of all of these contributions is 
\begin{equation}
\label{eq:abcdF}
  (a+\ldots+d)_F=\frac{2 i \lambda^4 U_S}{(4 \pi)^2 M^2 }   \left[ -\frac{1}{\overline{\epsilon}} -1 - \log(\frac{\mu^2}{M^2})  + 
       \frac{\sigma^2}{M^2} \left(-\frac{6}{\overline{\epsilon}} - 6  \log(\frac{\mu^2}{\sigma^2}) - 6 + 4  \log(\frac{M^2}{\sigma^2}) \right) \right].
\end{equation}

\begin{figure}[htb]
\begin{center}
 \begin{picture}(450,106) (25,-7)
    \SetWidth{2.0}
  \Line[arrow,arrowpos=0.5,arrowlength=5,arrowwidth=2,arrowinset=0.2](30,78)(60,48)
    \Line[arrow,arrowpos=0.5,arrowlength=5,arrowwidth=2,arrowinset=0.2](120,48)(150,78)
    \Arc[arrow,arrowpos=0.5,arrowlength=5,arrowwidth=2,arrowinset=0.2,clock](90,48)(30,-180,-360)
    \Line[arrow,arrowpos=0.5,arrowlength=5,arrowwidth=2,arrowinset=0.2](30,13)(60,43)
    \Arc[arrow,arrowpos=0.5,arrowlength=5,arrowwidth=2,arrowinset=0.2](90,43)(30,-180,0)
    \Line[arrow,arrowpos=0.5,arrowlength=5,arrowwidth=2,arrowinset=0.2](120,43)(150,13)
    \Line[arrow,arrowpos=0.5,arrowlength=5,arrowwidth=2,arrowinset=0.2](314,94)(346,86)
    \Line[arrow,arrowpos=0.5,arrowlength=5,arrowwidth=2,arrowinset=0.2](346,86)(377,96)
    \Line[arrow,arrowpos=0.5,arrowlength=5,arrowwidth=2,arrowinset=0.2](312,7)(346,16)
    \Line[arrow,arrowpos=0.5,arrowlength=5,arrowwidth=2,arrowinset=0.2](346,16)(381,7)
    \Arc[arrow,arrowpos=0.5,arrowlength=5,arrowwidth=2,arrowinset=0.2](386,51)(50,143.13,216.87)
    \Arc[arrow,arrowpos=0.5,arrowlength=5,arrowwidth=2,arrowinset=0.2](306,51)(50,-36.87,36.87)
     \Line[arrow,arrowpos=0.5,arrowlength=5,arrowwidth=2,arrowinset=0.2](399,95)(430,84)
    \Line[arrow,arrowpos=0.5,arrowlength=5,arrowwidth=2,arrowinset=0.2](437,85)(461,95)
    \Arc[arrow,arrowpos=0.5,arrowlength=5,arrowwidth=2,arrowinset=0.2](480.198,57.11)(57.426,150.943,216.44)
    \Arc[arrow,arrowpos=0.5,arrowlength=5,arrowwidth=2,arrowinset=0.2](382.165,54.844)(60.835,-31.564,29.716)
    \Line[arrow,arrowpos=0.5,arrowlength=5,arrowwidth=2,arrowinset=0.2](400,7)(434,18)
    \Line[arrow,arrowpos=0.5,arrowlength=5,arrowwidth=2,arrowinset=0.2](434,18)(460,6)
    \Line[arrow,arrowpos=0.5,arrowlength=5,arrowwidth=2,arrowinset=0.2](170,96)(200,66)
    \Line[arrow,arrowpos=0.5,arrowlength=5,arrowwidth=2,arrowinset=0.2](260,66)(290,96)
    \Arc[arrow,arrowpos=0.5,arrowlength=5,arrowwidth=2,arrowinset=0.2,clock](230,66)(30,-180,-360)
    \Arc[arrow,arrowpos=0.5,arrowlength=5,arrowwidth=2,arrowinset=0.2,clock](229,101)(50,-53.13,-126.87)
    \Line[arrow,arrowpos=0.5,arrowlength=5,arrowwidth=2,arrowinset=0.2](169,1)(259,61)
    \Line[arrow,arrowpos=0.5,arrowlength=5,arrowwidth=2,arrowinset=0.2](199,61)(289,1)
    \Text(82,-20)[lb]{(a)}
    \Text(223,-20)[lb]{(b)}
    \Text(340,-20)[lb]{(c)}
    \Text(427,-20)[lb]{(d)}
  \end{picture}
\end{center}
\caption{\label{fig:eff1loop} Diagrams in the effective theory to order $c^2$. Diagram (d) stands in for two diagrams that are related by an upside-down reflection. As we drew in Fig.~\protect{\ref{fig:4fermion1loop}}, the four-fermion vertices are not exactly point-like, so one can follow each fermion line.}
\end{figure}
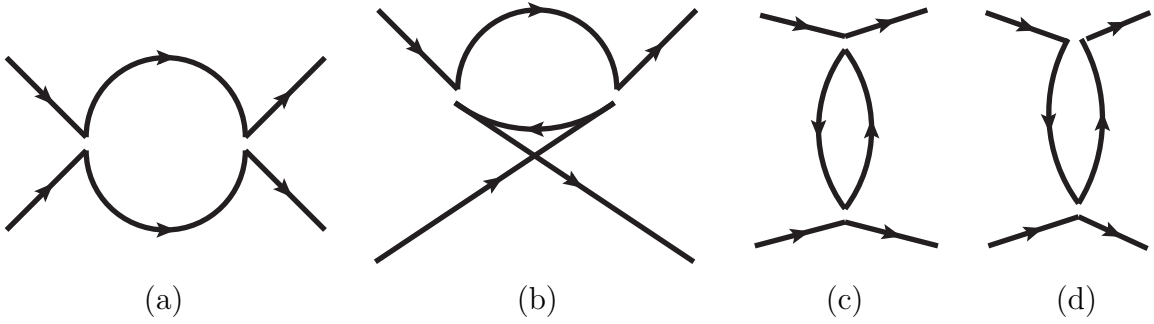

We also need the fermion two-point function in order to calculate the wave function renormalization in the effective theory. The calculation is identical to that in Eq.~(\ref{eq:selfenergy}). We need the finite part as well. The amplitude linear in momentum is  $i \slashed{p} \frac{\lambda^2}{2 (4 \pi)^2} \left( \frac{1}{\overline{\epsilon}} +  \log(\frac{\mu^2}{M^2}) + \frac{1}{2}+\ldots \right)$.

It is time to calculate in the effective theory. The effective theory has a four-fermion interaction that was induced at tree level. Again, we neglect the light scalar $\varphi$ as it does not play any role in our calculation. The effective Lagrangian is 
\begin{equation}
\label{eq:effmass}
 {\mathcal L} = i z \overline{\psi} \,\slashed{\partial} \, \psi - \sigma \overline{\psi}  \psi+ \frac{c}{2}\, \overline{\psi}  \psi\,  \overline{\psi}  \psi.
 \end{equation}  
We established that at tree level, $c=\frac{\lambda^2}{M^2}$, but do not yet want to substitute the actual value of $c$ as not to confuse the calculations in the full and effective theories. To match the amplitudes we also need to compute one-loop scattering amplitude in the effective theory. The two-point amplitude for the fermion kinetic energy vanishes in the effective theory.  The four-point diagrams in the effective theory are depicted in Fig.~\ref{fig:eff1loop}. Diagrams in an effective theory have typically higher degrees of UV divergence as they contain fewer propagators. For example, diagram $(a)_E$ is quadratically divergent, while $(a)_F$ is finite. This is not an obstacle. We simply regulate each diagram using dimensional regularization.

Exactly like in the full theory, the fermion propagators in diagrams $(a)_E$ and $(b)_E$ have opposite signs of momentum, thus the terms proportional to $U_V$ cancel. The parts proportional to $U_S$ are the same and the sum of these diagrams is
\begin{equation}
\label{eq:abE}
  (a+b)_E=2 \frac{i c^2 \sigma^2}{(4 \pi)^2} U_S  \left[ \frac{1}{\overline{\epsilon}} +  \log(\frac{\mu^2}{\sigma^2}) \right] + \ldots .
\end{equation}
If one was careless with drawing these diagrams, one might think that there is a closed fermion loop and assign an extra minus sign. However, the way of drawing the effective interactions in Fig.~\ref{fig:eff1loop} makes it clear that the fermion line goes around the loop without actually closing.  Diagram $(c)_E$ is identical to its counterpart in the full theory. Since we are after the momentum-independent part of the amplitude, the heavy scalar propagators in  $(c)_F$ were simply equal to $\frac{-i}{M^2}$. Therefore,
\begin{equation}
\label{eq:cE}
 (c)_E=- 4 \frac{i c^2\sigma^2}{(4 \pi)^2}  U_S \left[ 3 \frac{1}{\overline{\epsilon}} + 3  \log(\frac{\mu^2}{\sigma^2}) + 1 \right] + \ldots .
\end{equation}
As in the full theory, $(d)_E$ includes a factor of two for two diagrams
\begin{equation}
\label{eq:dE}
 (d)_E=2 \frac{i c^2\sigma^2}{(4 \pi)^2}  U_S \left[ 3 \frac{1}{\overline{\epsilon}} + 3  \log(\frac{\mu^2}{\sigma^2}) + 1 \right] + \ldots .
\end{equation}
The sum of these diagrams is
\begin{equation}
\label{eq:abcdE}
  (a+\ldots+d)_E=- \frac{2 i c^2\sigma^2}{(4 \pi)^2} U_S  \left[ \frac{2}{\overline{\epsilon}}  + 2  \log(\frac{\mu^2}{\sigma^2})   + 1 \right]. 
\end{equation}
Of course, we should set $c=\frac{\lambda^2}{M^2}$ at this point. 

Before we compare the results let us make two important observations. There are several logs in the amplitudes. In the full theory,  $\log(\frac{\mu^2}{M^2})$, $\log(\frac{\mu^2}{\sigma^2})$ and $\log(\frac{M^2}{\sigma^2})$ appear, while in the effective theory  only $\log(\frac{\mu^2}{\sigma^2})$ shows up. Interestingly, comparing the full and effective theories diagram by diagram, the corresponding coefficients in front of $\log(\sigma^2)$ are identical. This means that  $\log(\sigma^2)$ drops out of the difference between the full and effective theories so $\log(\sigma^2)$ never appears in the matching coefficients. It had to be this way. We already argued that the two theories are identical in the IR, so non-analytic terms depending on the light fields must be the same. This would hold for all other quantities in the low-energy theory, for instance for terms that depend on the external momenta. This correspondence between logs of low-energy quantities does not have to happen, in general, diagram by  diagram, but it has to hold for the entire calculation. This provides a useful check on matching calculations. When the full and effective theory are compared, the only log that turns up is the $\log(\frac{\mu^2}{M^2})$. This is good news as it means that there is only one scale in the matching calculation and we can minimize the logs by setting $\mu=M$.

The $\frac{1}{\overline{\epsilon}}$ poles are different in the full and effective theories as the effective theory diagrams are more divergent. We simply add appropriate counterterms in the full and the effective theories to cancel the divergences. The counterterms in the two theories are not related. We compare the renormalized, or physical, scattering amplitudes and make sure they are equal. We are going to use the $\overline{MS}$ prescription and the counterterms will cancel just the $\frac{1}{\overline{\epsilon}}$ poles. It is clear that since the counterterms differ on the two sides, the coefficients in the effective theory depend on the choice of regulator. Of course, physical quantities will not depend on the regulator.

Setting $\mu=M$, the difference between Eqs.~(\ref{eq:abcdF}) and (\ref{eq:abcdE}) gives 
\begin{equation}
\label{eq:c1loop}
  c(\mu=M) =\frac{\lambda^2}{M^2} -  \frac{2 \lambda^4}{ (4 \pi)^2 M^2}  -  \frac{10 \lambda^4 \sigma^2}{(4 \pi)^2 M^4}.
\end{equation}
To reproduce the two-point function in the full theory we set $z=1+\frac{\lambda^2}{4 (4 \pi)^2}$ in the $\overline{MS}$ prescription since there are no contributions in the effective theory. To obtain physical scattering amplitude, the fermion field needs to be canonically normalized by rescaling $\sqrt{z} \psi \rightarrow \psi_{\rm canonical}$.  This rescaling gives an additional contribution to the $\frac{\lambda^4}{ (4 \pi)^2 M^2}$ term in the scattering amplitude from the product of the tree-level contribution and the wave function renormalization factor. Without further analysis, it is not obvious that it is consistent to keep the last term in the expression for $c(\mu=M)$. One would have to examine if there are any other terms proportional to $\frac{1}{M^4}$ that were neglected. For example, the momentum-dependent operator proportional to $d$ in Eq.~(\ref{eq:p2lam2}) could give a contribution of the same order when the RG running in the effective theory is included. Such contribution would be proportional to $\frac{\lambda^2 \eta^2 \sigma^2}{(4 \pi)^2 M^4} \log(\frac{M^2}{m^2})$. There can also be contributions to the fermion two-point function arising in the full theory from the heavy scalar exchange.  We were originally interested in a  theory with massless fermions which means that $\sigma=0$. It was a useful detour to do the matching calculation including the $\frac{1}{M^4}$ terms as various logs and UV divergences do not fully show up in this example at the $\frac{1}{M^2}$ order. 

We calculated the scattering amplitudes arising from the exchanges of the heavy scalar.  In the calculation of the $\psi\psi \rightarrow \psi \psi$ scattering cross section, both amplitudes coming from the exchanges of the heavy and light scalars have to be added. These amplitudes depend on different coupling constants, but they can be difficult to disentangle experimentally since the measurements are done at low energies. The amplitude associated with the heavy scalar is measurable only if the mass and the coupling of the light scalar can be inferred. This can be accomplished, for example, if the light scalar can be produced on-shell in the $s$ channel. Near the resonance corresponding to the light scalar, the scattering amplitude is dominated by the light scalar and its mass and coupling can be determined. Once the couplings of the light scalar are established, one could deduce the amplitude associated with the heavy scalar by subtracting the amplitude with the light scalar exchange. If the heavy and light states did not have identical spins one could distinguish their contributions more easily as they would give different angular dependence of the scattering cross section. 
 
\subsection{Naturalness and quadratic divergences}
\label{sec:naturalness}

Integrating out a fermion in the Yukawa theory emphasizes several important points. We are going to study the same ``full" Lagrangian again, but this time assume that the fermion is heavy and the scalar $\varphi$ remains light
\begin{equation}
 {\mathcal L} = i \overline{\psi} \,\slashed{\partial} \, \psi - M \overline{\psi}  \psi+ \frac{1}{2} (\partial_\mu \varphi)^2 - \frac{m^2}{2} \varphi^2  
                           - \eta \, \overline{\psi} \psi \varphi,
 \end{equation}  
where $M\gg m$. We will integrate out $\psi$ and keep $\varphi$ in the effective theory.  As we did earlier, we have neglected the potential for $\varphi$ assuming that it is zero. There are no tree-level diagrams involving fermions $\psi$ in the internal lines only. We are going to examine diagrams with two scalars and four scalars for illustration purposes. The diagrams resemble those of the Coleman-Weinberg effective potential calculation, but we do not necessarily neglect external momenta. The momentum dependence could be of interest. The two point function gives
\begin{eqnarray}
\begin{picture}(70,20)(49,-22)
    \SetWidth{1.5}
    \Arc[arrow,arrowpos=0.5,arrowlength=5.833,arrowwidth=2.333,arrowinset=0.2](78,-19)(12,90,450)
    \SetWidth{1.7}
    \Line[dash,dashsize=4.4](36,-19)(66,-19)
    \Line[dash,dashsize=4.4](90,-19)(120,-19)
  \end{picture}
&=&  (-1) (-i \eta \mu^\epsilon)^2   \int \frac{d^d k}{(2\pi)^d} i^2 \frac{{\rm Tr}[(\slashed{k} + \slashed{p} +M) (\slashed{k}+M)]}{[(k+p)^2-M^2](k^2-M^2)} \nonumber \\
 &=& -\frac{4 i \eta^2}{(4\pi)^2} \left[( \frac{3}{\overline{\epsilon}} +1 +3 \log(\frac{\mu^2}{M^2})) (M^2 - \frac{p^2}{6}) +  \frac{p^2}{2} - \frac{p^4}{20 M^2} + \ldots \right], \label{eq:2ptscalar}
 \end{eqnarray}
where we truncated the momentum expansion at order $p^4$. The four-point amplitude, to the lowest order in momentum is 
\begin{equation}
\label{eq:4ptscalar}
    \begin{picture}(55,30) (75,-15)
    \SetWidth{1.5}
    \Arc[arrow,arrowpos=0.5,arrowlength=5.833,arrowwidth=2.333,arrowinset=0.2](78,-11)(12,90,450)
    \SetWidth{1.7}
    \Line[dash,dashsize=4.4](36,7)(67,-5)
   \Line[dash,dashsize=4.4](90,-5)(120,7)
    \Line[dash,dashsize=4.4](89,-16)(121,-29)
    \Line[dash,dashsize=4.4](36,-28)(68,-16)
  \end{picture}   = - \frac{8 i \eta^4}{(4\pi)^2} \left[ 3 (\frac{1}{\overline{\epsilon}} + \log(\frac{\mu^2}{M^2})) - 8 + \ldots \right].
\end{equation}

There are no logarithms involving $m^2$ or $p^2$ in Eqs.~(\ref{eq:2ptscalar}) and (\ref{eq:4ptscalar}). Our effective theory at the tree-level contains a free scalar field only, so in that effective theory there are no interactions and no loop diagrams. Thus, logarithms involving   $m^2$ or $p^2$ do not appear because they could not be reproduced in the effective theory. Setting $\mu=M$ and choosing the counterterms to cancel the $\frac{1}{\overline{\epsilon}}$ poles we can read off the matching coefficients in the scalar theory
\begin{equation}
\label{eq:LscalarE}
    {\mathcal L}=  (1- \frac{4 \eta^2}{3( 4\pi)^2})\,  \frac{(\partial_\mu \varphi)^2}{2} - (m^2+ \frac{4 \eta^2 M^2}{(4\pi)^2})\, \frac{\varphi^2}{2} + \frac{ \eta^2}{5 (4\pi)^2 M^2} \, \frac{ (\partial^2 \varphi)^2}{2} 
    + \frac{64 \eta^2}{(4 \pi)^2}\,  \frac{\varphi^4}{4!} + \ldots
\end{equation}
To obtain physical scattering amplitudes one needs to absorb the $1- \frac{4 \eta^2}{3( 4\pi)^2}$ factor in the scalar kinetic energy, so the field is canonically normalized. The scalar effective Lagrangian in Eq.~(\ref{eq:LscalarE}) is by no means a consistent approximation. For example, we did not calculate the tadpole diagram and did not calculate the  diagram with three scalar fields. Such diagrams do not vanish since the Yukawa interaction is not symmetric under $\varphi \rightarrow - \varphi$. There are no new features in those calculations so we skipped them.

The scalar mass term, $m^2+ \frac{4 \eta^2 M^2}{(4\pi)^2}$, contains a contribution from the heavy fermion. If the sum $m^2+ \frac{4 \eta^2 M^2}{(4\pi)^2}$ is small compared to $\frac{4 \eta^2 M^2}{(4\pi)^2}$ one calls the scalar ``light" compared to the heavy mass scale $M$. This requires a cancellation between $m^2$  and $\frac{4 \eta^2 M^2}{(4\pi)^2}$. Cancellation happens when the two terms are of opposite signs and close in magnitude, yet their origins are unrelated. No symmetry of the theory can relate the tree-level and the loop-level terms. If there was a symmetry that ensured the tree-level and loop contributions are equal in magnitude and opposite in sign, then small breaking of such symmetry could make make the sum $m^2+ \frac{4 \eta^2 M^2}{(4\pi)^2}$ small. But no symmetry is present in our Lagrangian. This is why light scalars require a tuning of different terms unless there is a mechanism protecting the mass term, for example the shift symmetry or supersymmetry. 

The sensitivity of the scalar mass term to the heavy scales is often referred to as the quadratic divergence of the scalar mass term. When one uses mass-dependent regulators, the mass terms for scalar fields receive corrections proportional to $\frac{\Lambda^2}{(4 \pi)^2}$. Having light scalars makes fine tuning necessary to cancel the large regulator contribution. There are no quadratic divergences in dimensional regularization, but the fine tuning of scalar masses is just the same. In dimensional regularization, the scalar mass is quadratically sensitive to heavy particle masses. This is a much more intuitive result compared to the statement about an unphysical regulator. Fine tuning of scalar masses would not be necessary in dimensional regularization if there were no heavy particles. For example, if the Standard Model (SM) was a complete theory there would be no fine tuning associated with the Higgs mass. Perhaps the SM  is a complete theory valid even beyond the grand unification scale, but there is gravity and we expect Planck-scale particles in any theory of quantum gravity. Another term used for the fine tuning of the Higgs mass in the SM is the hierarchy problem. Having a large hierarchy between the Higgs mass and other large scales requires fine tuning, unless the Higgs mass is protected by symmetry.  

 It is apparent from our calculation that radiative corrections generate all terms allowed by symmetries. Even if zero at tree level, there is no reason to assume that the potential for the scalar field vanishes. The potential is generated radiatively.   We obtained nonzero potential in the effective theory when we integrated out a heavy fermion. However, generation of terms by radiative corrections is not at all particular to effective theory. The RG evolution in the full theory would do the same. We saw another example of this in Sec.~\ref{sec:RG}, where an operator absent at one scale was generated radiatively. Therefore, having terms smaller than the sizes of radiative corrections requires fine tuning. A theory with all coefficients whose magnitudes are not substantially altered by radiative corrections is called technically natural. Technical naturalness does not require that all parameters are of the same order, it only implies that none of the parameters receives radiative corrections that significantly exceed its magnitude.  As our calculation demonstrated, a light scalar that is not protected by symmetry is not technically natural.  

Naturalness is a stronger criterion. Dirac's naturalness condition is that all dimensionless coefficients are of order one and the dimensionful parameters are of the same magnitude~\cite{Dirac}. A weaker naturalness criterion, due to 't~Hooft, is that small parameters are natural if setting a small parameter to zero enhances the symmetry of the theory~\cite{'tHooft}.  Technical naturalness is yet a weaker requirement. The relative sizes of terms are dictated by the relative sizes of radiative corrections and not necessarily by symmetries, although symmetries obviously affect the magnitudes of radiative corrections. Technical naturalness has to do with how perturbative field theory works.

\subsection{Equations of motion}
\label{sec:eoms}

After determining the light field content and power counting of an EFT one turns to enumerating higher-dimensional operators. It turns out that not all operators are independent as long as one considers $S$-matrix elements with one insertion of higher-dimensional operators. Let us consider again an effective  theory of  a single scalar field theory that we discussed in  the previous section. Suppose one is interested in the following effective Lagrangian
 \begin{equation}
\label{eq:Lscalarred}
 {\mathcal L}_\varphi = \frac{1}{2} (\partial_\mu \varphi)^2 - \frac{m^2}{2} \varphi^2  
                         -\frac{\eta}{4!} \, \varphi^4 - c_1 \varphi^6 + c_2 \varphi^3 \partial^2 \varphi,
 \end{equation}  
where both coefficients $c_1$ and $c_2$ are coefficients of operators of dimension 6. We perform a field redefinition $\varphi \rightarrow \varphi' + c_2 \varphi'^3$ in the Lagrangian in Eq.~(\ref{eq:Lscalarred}). Field redefinitions do not alter the $S$ matrix as long as $\langle \varphi_1| \varphi'|0\rangle\neq 0$, where $|\varphi_1\rangle$ is a one-particle state created by the field $\varphi$. In other words, $\varphi'$ is an interpolating field for the single-particle state $|\varphi_1\rangle$. This is guaranteed by the LSZ reduction formula which picks out the poles corresponding to the physical external states in the scattering amplitude. 

Under the  $\varphi \rightarrow \varphi' + c_2 \varphi'^3$ redefinition
\begin{eqnarray}
\label{eq:Lscalarredef}
 {\mathcal L}_\varphi \!\!& \rightarrow & \frac{(\partial_\mu \varphi')^2}{2} - c_2 \varphi'^3 \partial^2 \varphi' -  \frac{m^2}{2} \varphi'^2 - c_2 m^2 \varphi'^4  -\frac{\eta}{4!} \, \varphi'^4
                - \frac{\eta}{3!} c_2 \varphi'^6 - c_1  \varphi'^6 + c_2 \varphi'^3 \partial^2 \varphi' +\ldots  \nonumber \\
            & = &   \frac{(\partial_\mu \varphi')^2}{2} -  \frac{m^2}{2} \varphi'^2 - (\frac{\eta}{4!}+ c_2 m^2) \varphi'^4  - (c_1 + \frac{\eta c_2}{3!}) \varphi'^6+\ldots,
\end{eqnarray}
where we omitted terms quadratic in the coefficients $c_{1,2}$. This field redefinition removed the $\varphi^3 \partial^2 \varphi$ term and converted it into the $\varphi^6$ term. Field redefinitions are equivalent to using the lowest oder equations of motions to find redundancies among higher dimensional operators. The equation of motion following from the Lagrangian  in Eq.~(\ref{eq:Lscalarred}) is $\partial^2 \varphi=-m^2 \varphi - \frac{\eta}{3!} \varphi^3$. Substituting the derivative part of the $ \varphi^3 \partial^2 \varphi$ operator with the equation of motion gives
\begin{equation}
\label{eq:Lscalareoms}
 {\mathcal L}_{D>4} = - c_1 \varphi^6 + c_2 \varphi^3 \partial^2 \varphi \rightarrow - c_1 \varphi^6 + c_2 \varphi^3 (-m^2 \varphi - \frac{\eta}{3!} \varphi^3) = - (c_1+\frac{\eta c_2}{3!}) \varphi^6 - c_2 m^2 \varphi^4,
 \end{equation}  
which agrees with Eq.~(\ref{eq:Lscalarredef}). 
 
 One might worry that this a tree-level result only. Perhaps the cleanest argument showing that this is true for any amplitude can be given using path integrals, see Sec.~12 in Ref.~\cite{Politzer}
 and also Refs.~\cite{Georgi,Arzt}. One can show that given a Lagrangian containing a higher dimensional operator with a part proportional to the equations of motion
 \begin{equation}
    {\mathcal L} =  {\mathcal L}_{D\leq4} + c \, F(\varphi) \frac{\delta  {\mathcal L}_{D\leq4}}{\delta \varphi},
 \end{equation}
 all correlation functions of the form $\langle \varphi(x_1) \ldots \varphi(x_n) F(\varphi(y)) \frac{\delta  {\mathcal L}_{D\leq4}}{\delta \varphi(y)} \rangle$ vanish. 
 
 \begin{figure}[htb]
 \begin{center}
  \begin{picture}(297,77) (38,-20)
    \SetWidth{2.0}
    \Line[dash,dashsize=5](40,47)(70,17)
    \Line[dash,dashsize=5](40,-13)(70,17)
    \Line[dash,dashsize=5](130,17)(160,47)
    \Line[dash,dashsize=5](130,17)(160,-13)
    \Line[dash,dashsize=5](40,17)(160,17)
    \Vertex(70,17){3}
    \Vertex(130,17){3}
    \Line[dash,dashsize=5](190,47)(220,17)
    \Line[dash,dashsize=5](190,-13)(220,17)
    \Line[dash,dashsize=5](280,17)(310,47)
    \Line[dash,dashsize=5](280,17)(310,-13)
    \Line[dash,dashsize=5](190,17)(310,17)
    \Vertex(220,17){3}
    \Vertex(280,17){3}
    \Text(70,1)[lb]{$\eta$}
    \Text(220,1)[lb]{$\eta$}
    \Text(125,1)[lb]{$c_2$}
    \Text(275,1)[lb]{$c_2$}
    \Text(115,20)[lb]{$\partial^2$}
    \Text(300,20)[lb]{$\partial^2$}
    \Text(90,-33)[lb]{(a)}
    \Text(240,-33)[lb]{(b)}
  \end{picture}
\end{center}
 \caption{\label{fig:scalar6point} Diagrams with one non-derivative quartic interaction and one quartic interaction containing $\partial^2$. Diagrams (a) and (b) differ only by the placement of the derivative term. In diagram (a) the derivative acts on the internal line and shrinks the propagator to a point, while in diagram (b) the derivative acts on any of the external lines.}
 \end{figure}
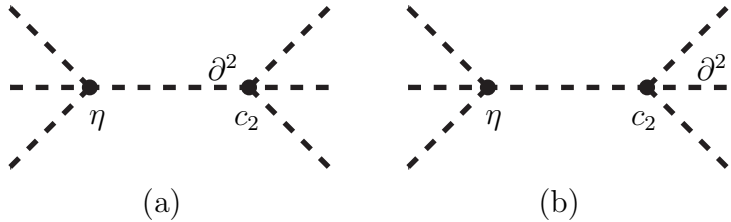
 
 These results can also be obtained diagrammatically. The diagrams in Fig.~\ref{fig:scalar6point} show the six-point amplitude arising from one insertion of $c_2 \, \varphi^3 \partial^2 \varphi$. Diagams (a) and (b) differ only by the placement of the second derivative. The derivative is associated with the internal line in diagram (a), while in diagram (b) with one of the external lines. The amplitude is 
 \begin{equation}
 \label{eq:6ptscalar}
   {\mathcal A}_{(a)} = (-i\eta) \frac{i}{k^2-m^2} (-i c_2 k^2 ) 3!  = - i \eta c_2 3!  - i \eta (-i c_2 m^2 3!) \frac{i}{k^2-m^2},
 \end{equation}
 where the momentum dependence of the interaction vertex was used to partially cancel the propagator by writing $k^2=k^2-m^2 + m^2$. The two terms on the right-hand side of Eq.~(\ref{eq:6ptscalar}) have different interpretation. The first term has no propagator, so it represents a local six-point interaction. This is a modification of the $\varphi^6$ interaction and its coefficient is the same as the one in Eq.~(\ref{eq:Lscalareoms}) even though it may not be apparent at first. When comparing the amplitudes one needs to keep track of the multiplicity factors. The $\varphi^6$ interaction comes with the $6!$ symmetry factor, while there are $\left(\begin{array}{c} 6 \\ 3 \end{array} \right)$ choices of the external lines in  Fig.~\ref{fig:scalar6point}(a).  
 
 The term with the propagator on the right-hand side of Eq.~(\ref{eq:6ptscalar}) together with diagram (b) in  Fig.~\ref{fig:scalar6point}  give the modification of the $\varphi^4$ interaction in Eq.~(\ref{eq:Lscalareoms}). Diagram (b) is associated with the $3!\cdot 3$ factor, where $3!$ comes from the permutations of lines without the derivative and $3$ comes from placing the derivative on either of the $3$ external lines. Combined with $3!$ in Eq.~(\ref{eq:6ptscalar}), we get $3! \cdot 3 + 3!=4!$ to reproduce the coefficient of the $\varphi^4$ term in Eq.~(\ref{eq:Lscalareoms}).
 
\subsection{Summary}
\label{sec:summary} 

We have constructed several effective theories so far. It is a good moment to pause and review the observations we made. To construct an EFT one needs to identify the light fields and their symmetries, and needs to establish a power counting scheme. If the full theory is known then an EFT is derived perturbatively as a chain of matching calculations interlaced by RG evolutions. Each heavy particle is integrated out and new effective theory matched to the previous one, resulting in a tower of effective field theories. Consecutive  ratios of scales are accounted for by the RG evolution.

This is a systematic procedure which can be carried out to the desired order in the loop expansion. Matching is done order by order in the loop expansion. When two theories are compared at a given loop order, the lower order results are included in the matching. For example, in Sect.~\ref{sec:1loop} we calculated loop diagrams in the effective theory including the effective interaction we obtained at the tree level. At each order in the loop expansion, the effective theory valid below a mass threshold is amended to match the results valid just above that threshold. Matching calculations do not depend on any light scales and if logs appear in the matching calculations, these have to be logs of the matching scale divided by the renormalization scale. Such logs can be easily minimized to avoid spoiling perturbative expansion. The two theories that are matched across a heavy threshold have in general different UV divergences and therefore different counterterms. 

EFTs naturally contain higher-dimensional operators and are therefore non-re\-nor\-mal\-izable. In practice, this is of no consequence since the number of operators, and therefore the number of parameters determined from experiment, is finite. To preserve power counting and maintain consistent expansion in the inverse of large mass scales one needs to employ a mass-independent regulator, for instance dimensional regularization. Consequently, the renormalization scale only appears in dimensionless ratios inside logarithms and so it does not alter power counting. Contributions from the heavy fields do not automatically decouple when using dimensional regularization, thus decoupling should be carried out explicitly by constructing effective theories. 

Large logarithms arise from the RG running only as one relates parameters of the theory at different renormalization scales. The field content of the theory does not change while its parameters are RG evolved. However, distinct operators of the same dimension can mix with one another. The RG running and matching are completely independent and can be done at unrelated orders in perturbation theory. The magnitudes of coupling constants and the ratios of scales dictate the relative sizes of different contributions and dictate to what orders in perturbation theory one needs to calculate. A commonly repeated phrase is that two-loop running requires one-loop matching. This is true when the logarithms are very large, for example in grand unified theories. The $\log(\frac{M_{\rm GUT}}{M_{\rm weak}})$ is almost as large as $(4 \pi)^2$, so the logarithm compensates the loop suppression factor. This is not the case for smaller ratios of scales. 

The contributions of the heavy particles to an effective Lagrangian appear in both renormalizable terms and in higher dimensional terms. For the renormalizable terms, the contributions from heavy fields are often unobservable as the coefficients of the renormalizable terms are determined from low-energy experiments. The contributions of the heavy fields simply redefine the coefficients that were determined from experiments instead of being predicted by the theory. The coefficients of the higher-dimensional operators are suppressed by inverse powers of the heavy masses. As one increases the masses of the heavy particles, their effects diminish. This is the observation originally made in Ref.~\cite{A-C}. This typical situation is referred to as the decoupling of heavy fields. 

Counterexamples of  ``non-decoupling'' behavior are rare and easy to understand. The suppression of higher-dimensional operators can be overcome by large dimensionless coefficients. Suppose that the coefficient of a higher-dimensional operator is proportional to $\frac{h^2}{M^2}$, where $h$ is a dimensionless coupling constant. If $h$ and $M$ are proportional to each other, then taking $M\rightarrow \infty$ does not bring $\frac{h^2}{M^2}$ to $0$. Instead, $\frac{h^2}{M^2}$ can be finite in the $M\rightarrow \infty$ limit. This happens naturally in theories with spontaneous symmetry breaking. For example, the fermion Yukawa couplings in the SM are proportional to the fermion masses divided by the Higgs vacuum expectation value.   We are going to see examples of non-decoupling in Sec~\ref{sec:ST}. The non-decoupling examples should be regarded with some degree of caution. When $M$ is large, the dimensionless coupling $h$ must be large as well. Thus the non-decoupling result, that is a nonzero limit for $\frac{h^2}{M^2}$ as $M\rightarrow \infty$, is not in the realm of perturbation theory. For masses $M$ small enough that the corresponding value of $h$ is perturbative, there is no fall off of $\frac{h^2}{M^2}$ with increasing $M$ and such results are trustworthy. 

When the high-energy theory is not known, or it is not perturbative, one still benefits from constructing an EFT\@. One can power count the operators and then enumerate the pertinent operators to the desired order. One cannot calculate the coefficients, but one can estimate them. In a perturbative theory, explicit examples tell us what magnitudes of coefficients to expect at any order of the loop expansion. In strongly coupled QCD-like theories, or in supersymmetric theories, one estimates coefficients differently, see for example Refs.~\cite{NDA,SUSYNDA}.

\section{Precision electroweak measurements}

A common task for anyone interested in extensions of the SM is making sure that the proposed hypothetical particles and their interactions are consistent with current experimental knowledge. The sheer size of the Particle Data Book~\cite{PDG} suggests that the amount of available data is vast. A small subset of accurate data, consisting of a few dozen observables on flavor diagonal processes involving the electroweak $W$ and $Z$ gauge bosons, is referred to as the precision electroweak measurements. The accuracy of the measurements in this set is at the $1\%$ level or better. We will describe the precision electroweak (PEW) measurements in Sec.~\ref{sec:measurements}.

This common task of analyzing SM extensions and comparing with experiments is in principle straightforward. One needs to calculate all the observables, including the contributions of the proposed new particles, and needs to make sure that the results agree with the experiments within errors. In practice, this can be quite tedious. When the new particles are heavy compared to the energies at which the PEW measurements were made, one can integrate the new particles out and construct an effective theory in terms of the SM fields only~\cite{Buchmuller,GW}. The PEW experiments can be used to constrain the coefficients of the effective theory. This can be, and has been, done once for all, or at least until there is new data and the bounds need to be updated.  Various SM extensions can be constrained by comparing with the bounds on the effective coefficients instead of comparing to the experimental data. The EFT approach in this case is simply a time and effort saver, as direct contact with experimental quantities can be done only once when constraining coefficients of higher-dimensional operators. Constraints on the effective operators can be used to constrain masses and couplings of proposed particles. Integrating out fields is much less time consuming than computing numerous cross sections and decay widths. 

The PEW measurements contain some low-energy data, observables at the $Z$ pole, and LEP2 data on $e^+ e^-$ scattering at various  CM  energies between the $Z$ mass and 209 GeV\@. Particles heavier than a few hundred GeV could not have been produced directly in these experiments, so we can accurately capture the effects of such particles using effective theory. The field content of the effective theory is the same as the SM field content. We know all the light fields and their symmetries, except for the sector responsible for EW symmetry breaking. We are going to assume that EW symmetry is broken by the Higgs doublet and construct the effective theory accordingly. Of course, it is possible to make a different assumption---that there is no Higgs boson and the EW symmetry is nonlinearly realized. In that case there would be no  Higgs doublet in the effective theory, but just the three eaten Goldstone bosons, see Refs.~\cite{EWCL,Wudka}. However, the logic of applying effective theory in the two cases is completely identical, so we only concentrate on one of them. It is worth noting that the SM with a light Higgs boson fits the experimental data very well, suggesting that the alternative is much less likely. 

Given a Lagrangian for an extension of the SM we want to construct the effective Lagrangian
\begin{equation}
   {\mathcal L}(\varphi_{SM};\chi_{BSM})  \longrightarrow  {\mathcal L}_{eff} 
         = {\mathcal L}_{SM}(\varphi_{SM}) + \sum_i a_i  \, O_i (\varphi_{SM}),
\end{equation}
where we collectively denoted the SM fields as $\varphi_{SM}$ and the heavy fields as $\chi_{BSM}$. All the information about the original Lagrangian and its parameters is now encoded in the coefficients $a_i$ of the higher-dimensional operators $O_i$. The operators $O_i$ are independent of any hypothetical SM extension because they are constructed from the known SM fields. We will discuss various $O_i$ that are important for PEW measurements in the following sections. 

One can find two different approaches in the literature to constructing effective theories for PEW observables. The difference between the two approaches is in the treatment of the EW gauge sector.  In one approach, an EFT is constructed in terms of the gauge boson mass eigenstates---the $\gamma$, $Z$, and $W$ bosons. In the other approach, an effective theory is expressed in terms of the $SU(2)_L\times U(1)_Y$ gauge multiplets, $A^i_\mu$ and $B_\mu$. Of course, actual calculations of any experimental quantity are done in terms of the mass eigenstates. In the EFT approach, one avoids carrying out these calculations anyway. However, when one expands around the Higgs vacuum expectation value (vev) one completely looses all information about the gauge symmetry and the constraints it imposes. For our goal, that is for constraining heavy fields with masses above the Higgs vev, using the full might of EW gauge symmetry is a much better choice. The EW symmetry is broken by the Higgs doublet at scales lower than the masses of particles that we integrate out to obtain an effective theory. The interactions in any extension of the SM must obey the  $SU(2)_L\times U(1)_Y$ gauge invariance, so we should impose this symmetry on our effective Lagrangian. 

To stress this point further, let us compare the coefficients of two similar operators written in terms of the $W$ and $Z$ bosons. 
\begin{eqnarray}
  &(A):& \ W_\mu^+ W^{- \mu} \, W_\nu^+ W^{- \nu},  \\
   &(B): & \  Z_\mu Z^\mu \, Z_\nu Z^\nu.
\end{eqnarray} 
Both operators have the same dimension and the same Lorentz structure.   Operator $(A)$ is present in the SM in the non-Abelian part of the gauge field strength $A^i_{\mu\nu} A^{i \mu \nu}$ and has a coefficient of order one. However, operator $(B)$ is absent in the SM and can only arise from a gauge-invariant operator of a very high dimension, thus its coefficient is strongly suppressed in any theory with a light Higgs. This information is simply lost when one does not use gauge invariance.\footnote{Even in theories without a light Higgs,  in which the electroweak symmetry is nonlinearly realized, there is still information about the $SU(2)_L\times U(1)_Y$ gauge invariance. Such effective theories can also be written in terms of gauge eigenstates.} If one cannot reliably estimate coefficients of operators then the effective theory is useless as it cannot be made systematic. 

From now on, all operators will be explicitly $SU(2)_L\times U(1)_Y$ gauge invariant and built out of quarks, leptons,  gauge and Higgs fields. All the operators we are going to discuss are of dimension 6. There is only one gauge invariant operator of dimension 5 consistent with gauge invariance and it gives the Majorana mass for the neutrinos. The neutrino mass is inconsequential for PEW measurements. Thus, the interesting operators start at dimension 6 and given the agreement of the SM with data we do not need operators of dimension 8, or higher. 

\subsection{The $S$ and $T$ parameters}
\label{sec:ST}

There is a special class of dimension-6 operators that arises in many extensions of the SM\@. We are going to analyze this class of operators in this section and the next one as well. These are the operators that do not contain any fermion fields. Such operators originate whenever heavy fields directly couple only to the SM gauge fields and the Higgs doublet. We are going to refer to such operators as ``universal'' because they universally affect all quarks and leptons through fermion couplings to the SM gauge fields. Sometimes such operators are referred to as ``oblique.''

It is easy to enumerate all dimension-6 operators containing the gauge and the Higgs fields only. The operator $(H^\dagger H)^3$, where $H$ denotes the Higgs doublet, is an example. This operator is not constrained by the current data, as we have not yet observed the Higgs boson. It alters the Higgs potential, but without knowing the Higgs mass and its couplings we have no information on operators like $(H^\dagger H)^3$. Here are another two operators that are not constrained by the present data: $H^\dagger H \, D_\mu H^\dagger D^\mu H $ and $H^\dagger H \, A^i_{\mu \nu} A^{i \mu \nu} $. Since there are no experiments involving Higgs particles, operators involving the Higgs doublet are sensitive to the Higgs vev only. After electroweak symmetry breaking, the two operators we just mentioned renormalize dimension-4 terms that are already present in the SM: the Higgs kinetic energy and the kinetic energies of the $W$ and $Z$ bosons, respectively.

Two important, and very tightly constrained experimentally, universal operators are 
\begin{eqnarray}
  O_S  & = & H^\dagger \sigma^i H A^i_{\mu \nu} \,B^{\mu \nu}, \label{eq:OS}  \\
  O_T  & = & \left| H^\dagger D_\mu H \right|^2, \label{eq:OT}
\end{eqnarray} 
where $\sigma^i$ are the Pauli matrices, meanwhile $B_{\mu \nu}$ and $A^i_{\mu \nu}$ are the $U(1)_Y$ and $SU(2)_L$ field strengths, respectively. The operator $O_S$ introduces kinetic mixing between $B_\mu$ and $A^3_\mu$ when the vev is substituted for $H$. The second operator, $O_T$, violates the custodial symmetry.  The custodial symmetry guarantees the tree-level relation between the $W$ and $Z$ masses, $M_W=M_Z \cos \theta_w$, where $\theta_w$ is the weak mixing angle. After substituting $\langle H \rangle$ in $O_T$, $O_T\propto Z_\mu Z^\mu$ while there is no corresponding contribution to the $W$ mass. 

The custodial symmetry can be made explicit by combining the Higgs doublet $H$ with $\tilde{H}=i \sigma_2 H^*$ into a two-by-two matrix $\Omega=\left(\tilde{H}, H\right)$, see for example Ref.~\cite{TASI-SM} for more details. The SM Higgs Lagrangian 
\begin{equation}
  {\mathcal L}_{Higgs}=\frac{1}{2} {\rm tr} \left( D_\mu \Omega^\dagger D^\mu \Omega\right) - V\left( {\rm tr}  (\Omega^\dagger \Omega)\right)
\end{equation}
is invariant under $SU(2)_L\times SU(2)_R$ transformations that act $\Omega \rightarrow L \Omega R^\dagger$. The Higgs vev breaks $SU(2)_L\times SU(2)_R$ to its diagonal subgroup which is called the custodial $SU(2)_c$. The custodial symmetry is responsible for the relation $M_W=M_Z \cos \theta_w$. The operator $O_T$ is contained in the operator ${\rm tr} (\Omega^\dagger D_\mu \Omega \sigma_3) \,  {\rm tr} (D^\mu \Omega^\dagger \Omega \sigma_3)$ that does not preserve  $SU(2)_L\times SU(2)_R$, but only preserves its $SU(2)_L\times U(1)_Y$ subgroup.

For the time being, we want to consider the SM Lagrangian amended by the two higher-dimensional operators in Eqs.~(\ref{eq:OS}) and (\ref{eq:OT}):
\begin{equation}
\label{eq:LeffST}
  {\mathcal L}= {\mathcal L}_{SM} + a_S \, O_S + a_T \, O_T.
\end{equation}
We called these operators $O_S$ and $O_T$ because there is a one-to-one correspondence between these operators and the $S$ and $T$ parameters of Peskin and Takeuchi~\cite{STU}~\footnote{There are three parameters introduced in Ref.~\cite{STU}: $S$, $T$, and $U$. The $U$ parameter corresponds to a dimension-8 operator in  a theory with a light Higgs boson. All three parameters are on equal footing in theories in which the electroweak symmetry is nonlinearly realized.}, see also Ref.~\cite{oblique} for earlier work on this topic. The $S$ and $T$ parameters are related to the coefficients $a_S$ and $a_T$ in Eq.~(\ref{eq:LeffST}) as follows
\begin{equation}
\label{eq:ST}
  S=\frac{4 s c v^2}{\alpha} \, a_S   \ \ {\rm and}  \ \ T =-\frac{v^2}{2 \alpha} \, a_T,
\end{equation}
where $v$ is the Higgs vev, $s=\sin \theta_w$, $c=\cos \theta_w$, and $\alpha$ is the fine structure constant. The coefficients $a_S$ and $a_T$ should be evaluated at the renormalization scale equal to the electroweak scale. In practice, scale dependence is often too tiny to be of any relevance. 

\begin{figure}[ht]
\begin{center} 
\vspace{-25pt}
  \includegraphics[scale=0.5,angle=-90]{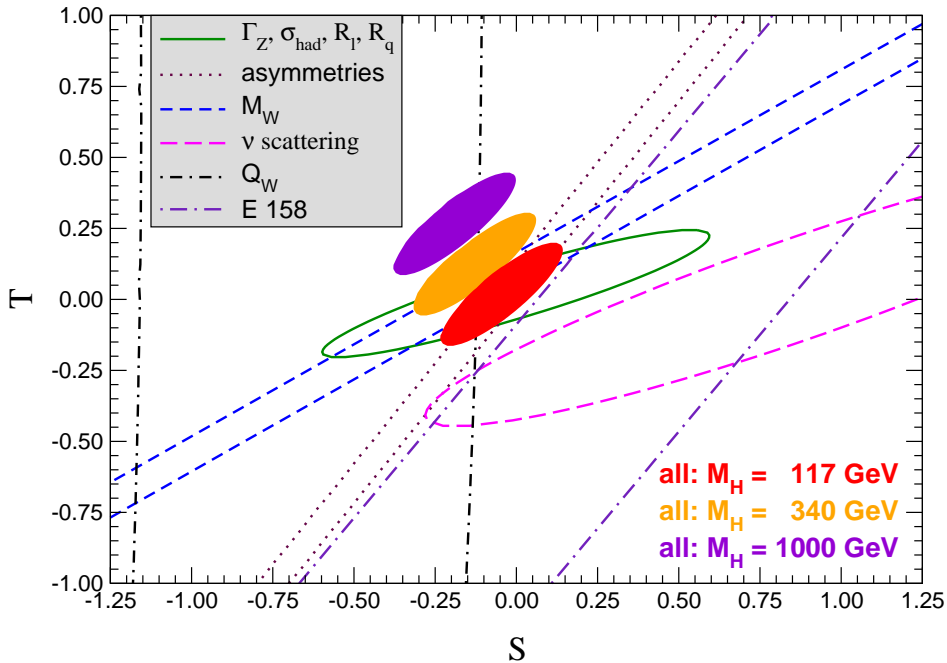}
\end{center}
\caption{\label{fig:ST} Combined constraints on the $S$ and $T$ parameters. This figure is reproduced from the review by J.~Erler and P.~Langacker in Ref.~\cite{PDG}. Different contours correspond to different assumed values of the Higgs mass and are all at the $1\sigma$ ($39\%$) confidence level. The Higgs mass dependence is discussed in Sect.~\ref{sec:sausage}.}
\end{figure}

The experimentally allowed range of the $S$ and $T$ parameters is shown in Fig.~\ref{fig:ST}. This is a key figure for understanding the EFT approach to constraints on new physics from PEW measurements. The colored regions are allowed at the $1\sigma$ confidence level. The regions indicate the values of the operator coefficients that are consistent with data. What is crucial is that  Fig.~\ref{fig:ST} incorporates all the relevant experimental data simultaneously. This is often referred to as global analysis of PEW measurements. The relevant data are combined into one statistical likelihood function from which bounds on masses an couplings of hypothetical new particles are determined.  The global analysis provides more stringent constraints than considering a few independent experiments and it also takes into account the correlations between experimental data. 

The global analysis, that includes all data and correlations, is possible using the EFT methods. All of the data is included in bounding the effective parameters $S$ and $T$. One needs to consider the two-dimensional allowed range for $S$ and $T$ instead of  the independent bounds on these parameters. When $S$ and $T$ are bounded independently, one of the parameters is varied while the other one is set to zero. This only gives bounds along the $S=0$ and $T=0$ axes of Fig.~\ref{fig:ST}, and the corresponding limits are $S=-0.04\pm0.09$ and $T=0.02\pm0.09$~\cite{PDG}. It is clear that Fig.~\ref{fig:ST} contains a lot more information. Suppose that an extension of the SM predicts nonzero values of $S$ and $T$ depending, for the sake of argument, on one free parameter. The allowed range of this free parameter depends on how $S$ and $T$ are correlated. If $S$ and $T$ happen to vary along the elongated part of Fig.~\ref{fig:ST} the allowed range could be quite large. If $S$ and $T$ happen to lie along the thin part of the allowed region, the range could be quite small. This information would not be available if one considered one effective parameter at a time by restricting the other one to be zero. Considering simultaneous bounds on $S$ and $T$ is equivalent to using the likelihood function directly from the data and the EFT provides simply an intermediate step of the calculation. In Sections~\ref{sec:YW} and \ref{sec:all6} we are going to study effective Lagrangians, very much like the one in Eq.~(\ref{eq:LeffST}), with more effective operators, but the logic of the approach will be exactly the same. 

Provided with the bounds in Fig.~\ref{fig:ST} one simply needs to match an extension of the SM to Eq.~(\ref{eq:LeffST}). We will consider here a hypothetical  fourth family of quarks as an example. We will call such new quarks $B$ and $T$ and assume that they have the same $SU(2)_L\times U(1)_Y$ quantum numbers as the ordinary quarks. The Lagrangian is
\begin{equation}
\label{eq:4thgeneration}
  {\mathcal L}_{new}=i \overline{Q}_L  \slashed{D} Q_L + i \overline{T}_R \slashed{D} T_R + i \overline{B}_R \slashed{D} B_R - \left[ y_T \, \overline{Q}_L  \tilde{H} T_R +  y_B\, \overline{Q}_L  H  B_R+ H.c. \right],
\end{equation} 
where $Q_L = \left(\begin{array}{c} T \\ B \end{array} \right)_L$ is the left-handed $SU(2)$ doublet and $y_{T,B}$ are the Yukawa couplings. Given that $\langle H \rangle = \left(\begin{array}{c} 0 \\ \frac{v}{\sqrt{2}} \end{array} \right)$, the quark masses are $M_{B,T}=\frac{v}{\sqrt{2}}y_{B,T}$. Since the new quarks, $B$ and $T$,     do not couple directly to the SM fermions, the operators induced by integrating out these quarks are necessary universal. The Yukawa part of the quark Lagrangian can be rewritten using the matrix representation of the Higgs field, $\Omega$, by combining the right-handed fields into a doublet
\begin{equation}
  y_T \overline{Q}_L  \tilde{H} T_R +  y_B\overline{Q}_L  H  B_R+ H.c. = \frac{y_T+y_B}{2} \overline{Q}_L \Omega \,Q_R + \frac{y_T - y_B}{2} \overline{Q}_L \Omega \, \sigma_3 \, Q_R + H.c.
\end{equation}
Due to the presence of $\sigma_3$ in the term proportional to $y_T - y_B$, that term violates the custodial symmetry, so we can expect contributions to the $T$ parameter whenever $y_T\neq y_B$.

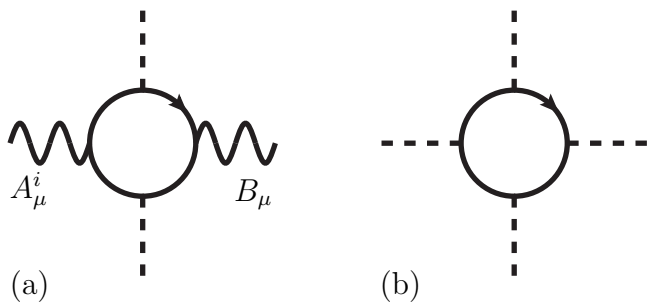
\begin{figure}[htb]
\begin{center}
  \begin{picture}(238,102) (18,-18)
    \SetWidth{2.0}
    \Arc[arrow,arrowpos=0.375,arrowlength=5,arrowwidth=2,arrowinset=0.2,flip](70,32)(20,270,630)
    \Photon(50,32)(20,32){7.5}{2}
    \Photon(90,32)(120,32){7.5}{2}
    \Arc[arrow,arrowpos=0.375,arrowlength=5,arrowwidth=2,arrowinset=0.2,flip](210,32)(20,270,630)
    \Line[dash,dashsize=5](70,52)(70,82)
    \Line[dash,dashsize=5](70,12)(70,-18)
    \Line[dash,dashsize=5](210,52)(210,82)
    \Line[dash,dashsize=5](210,12)(210,-18)
    \Line[dash,dashsize=5](190,32)(160,32)
    \Line[dash,dashsize=5](230,32)(260,32)
    \Text(20,7)[lb]{$A_\mu^i$}
    \Text(105,7)[lb]{$B_\mu$}
    \Text(20,-28)[lb]{(a)}
    \Text(160,-28)[lb]{(b)}
  \end{picture}
\end{center}
\caption{\label{fig:STdiagrams} Fermion contributions to the operators $O_S$ (a) and $O_T$ (b). The dashed lines represent the Higgs doublet.}
\end{figure}

Before we plunge into calculations we can estimate how the $S$ and $T$ parameters depend on the quark masses. The one-loop diagrams are depicted in Fig.~\ref{fig:STdiagrams}. Assuming that $M_B=M_T$ and therefore $y_B=y_T$, the contribution to the $S$ parameter can be estimated from diagram (a) in Fig.~\ref{fig:STdiagrams} to be
\begin{equation}
\label{eq:Sest}
  a_S \sim \frac{N_c}{(4\pi)^2} \frac{g g' y^2}{M^2}  \approx \frac{N_c g g' }{(4\pi)^2}\frac{y^2}{y^2 v^2} =   \frac{N_c g g' }{(4\pi)^2} \frac{1}{v^2},
\end{equation}
where $N_c=3$ is the number of colors, while $g$ and $g'$ are the $SU(2)_L$ and $U(1)_Y$ gauge couplings, respectively. The external lines consist of two gauge fields and two Higgs fields, hence the diagram is proportional to the square of the Yukawa coupling and to the $g$ and $g'$ gauge couplings. This is an example of a non-decoupling result as $a_S$ is constant for large quark mass $M$. Using Eq.~(\ref{eq:ST}), we expect that $S\sim \frac{N_c}{\pi}$ for large $M$. This is the situation we mentioned in Sec.~\ref{sec:summary} where dimensionless coefficients compensate for mass suppression. The $T$ parameter is even more interesting. Let us assume that $M_T \gg M_B$ so that only the $T$ quark runs in the loop in Fig.~\ref{fig:STdiagrams}(b). The estimate for this diagram is 
\begin{equation}
\label{eq:Test}
  a_T \sim \frac{N_c}{(4\pi)^2} \frac{y_T^4}{M_T^2} \approx \frac{N_c}{(4\pi)^2} \frac{M_T^2}{v^4 }
\end{equation}
and thus $T\sim \frac{N_c}{4 \pi} \frac{M_T^2}{v^2}$. Since four powers of the Yukawa coupling are needed to generate $O_T$, it is not surprising that the $T$ parameter grows as $M_T^2$.  If we did not take into account the full $SU(2)_L\times U(1)_Y$ symmetry, the Higgs Yukawa couplings would have been absorbed into quark masses and it would be difficult to understand Eqs.~(\ref{eq:Sest}) and (\ref{eq:Test}).

The actual calculation is easy, but there is one complication. We have been treating the quarks as massive while they only obtain masses when the theory is expanded around the Higgs vev. Chiral quarks are not truly massive fields, so  we need a small trick. We are going to match the theories in Eqs.~(\ref{eq:LeffST}) and (\ref{eq:4thgeneration}) with the Higgs background turned on. We will compare Eqs.~(\ref{eq:LeffST}) and (\ref{eq:4thgeneration}) as a function of the Higgs vev~\cite{SVVZ}. We do not need to keep any external Higgs fields and only keep the external gauge bosons. The Higgs vev will appear implicitly in the masses of the quarks. This is quite a unique complication that does not happen for fields with genuine mass terms, for example vector quarks. When the Higgs background is turned on, the calculation is very similar to the one done in the broken theory. However, we do not need to express the gauge fields in terms of the mass eigenstates. 

\begin{figure}[htb]
\begin{center}
  \begin{picture}(373,55) (30,0)
    \SetWidth{2.0}
    \Arc[arrow,arrowpos=0.5,arrowlength=5,arrowwidth=2,arrowinset=0.2,flip](65,25)(13.342,283,643)
    \Photon(51,25)(28,25){5}{2}
    \Photon(102,25)(79,25){5}{2}
    \Arc[arrow,arrowpos=0.5,arrowlength=5,arrowwidth=2,arrowinset=0.2,flip](165,25)(13.342,283,643)
    \Photon(151,25)(128,25){5}{2}
       \Photon(202,25)(179,25){5}{2}
 \Arc[arrow,arrowpos=0.5,arrowlength=5,arrowwidth=2,arrowinset=0.2,flip](265,25)(13.342,283,643)
    \Photon(251,25)(228,25){5}{2}
    \Photon(302,25)(279,25){5}{2}
 \Arc[arrow,arrowpos=0.5,arrowlength=5,arrowwidth=2,arrowinset=0.2,flip](365,25)(13.342,283,643)
    \Photon(351,25)(328,25){5}{2}
    \Photon(402,25)(379,25){5}{2}   
     \Text(61,45)[lb]{$T$}
    \Text(161,45)[lb]{$B$}
    \Text(261,45)[lb]{$T$}
    \Text(361,45)[lb]{$B$}
    \Text(61,-2)[lb]{$T$}
    \Text(161,-2)[lb]{$B$}
    \Text(261,-2)[lb]{$B$}
    \Text(361,-2)[lb]{$T$}
    \Text(30,35)[lb]{$A^3_\mu$}
    \Text(90,35)[lb]{$A^3_\nu$}
    \Text(130,35)[lb]{$A^3_\mu$}
    \Text(190,35)[lb]{$A^3_\nu$}
    \Text(230,35)[lb]{$A^1_\mu$}
    \Text(290,34)[lb]{$A^1_\nu$}
    \Text(330,35)[lb]{$A^1_\mu$}
    \Text(390,35)[lb]{$A^1_\nu$}
    \Text(111,21)[lb]{$+$}
    \Text(211,21)[lb]{$-$}
    \Text(311,21)[lb]{$-$}
  \end{picture}
\end{center}
\caption{\label{fig:Tcalc} Diagrams that contribute to $O_T$ in the Higgs background. The Higgs vev  is incorporated into the masses of the quarks in this calculation. }
\end{figure}
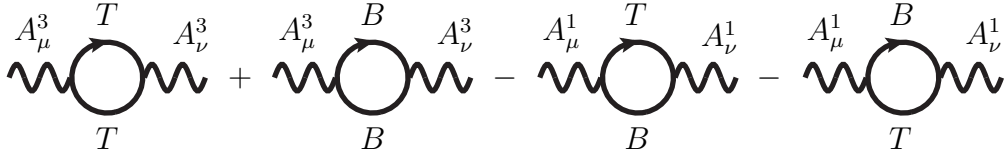

Expanding $O_T$ in Eq.~\ref{eq:OT} around the Higgs vev gives $\left| H^\dagger D_\mu H \right|^2 = \frac{v^4}{4} \frac{g^2}{4} (A^3_\mu)^2+\ldots$, where we omitted terms with the $B_\mu$ field and terms with derivatives. The relevant diagrams are shown in Fig.~\ref{fig:Tcalc} and they can be calculated at zero external momentum. We need to subtract diagrams with two external $A^1_\mu$ bosons because the diagrams with $A^3_\mu$'s contribute to both the $O_T$ operator and to an overall, custodial symmetry preserving, gauge boson mass renormalization. The operators that preserve custodial symmetry have equal coefficients of terms proportional to $(A^1_\mu)^2$ and  $(A^3_\mu)^2$.
\begin{eqnarray}
   \begin{picture}(65,35) (35,20)
    \SetWidth{2.0}
    \Arc[arrow,arrowpos=0.5,arrowlength=5,arrowwidth=2,arrowinset=0.2,flip](65,25)(13.342,283,643)
    \Photon(51,25)(28,25){5}{2}
    \Photon(102,25)(79,25){5}{2}
     \Text(61,45)[lb]{$T$}
       \Text(61,-2)[lb]{$T$}
     \Text(30,35)[lb]{$A^3_\mu$}
    \Text(90,35)[lb]{$A^3_\nu$}
  \end{picture}
& = & -N_c \left(\frac{ig \mu^\epsilon}{2}\right)^2 \int \frac{d^d k}{(2 \pi)^d} 
 \frac{i^2 {\rm tr} [\gamma^\mu P_L (\slashed{k} + M_T) \gamma^\nu P_L (\slashed{k} + M_T) ]}{(k^2-M_T^2)^2}
  \nonumber \\
&=& \frac{i N_c g^2}{2} \frac{M_T^2 g^{\mu \nu}}{(4 \pi)^2} \left( \frac{1}{\overline{\epsilon}} +  \ln(\frac{\mu^2}{M_T^2}) \right),   \label{eq:TW3}
\end{eqnarray}
where $P_L=\frac{1-\gamma^5}{2}$. The diagram with the $B$ quark in the loop gives the same answer, except for the $M_T \rightarrow M_B$ replacement.  The two diagrams with external $A^1_\mu$ bosons are identical and can be evaluated as
\begin{eqnarray} 
   \begin{picture}(65,35) (35,20)
    \SetWidth{2.0}
    \Arc[arrow,arrowpos=0.5,arrowlength=5,arrowwidth=2,arrowinset=0.2,flip](65,25)(13.342,283,643)
    \Photon(51,25)(28,25){5}{2}
    \Photon(102,25)(79,25){5}{2}
     \Text(61,45)[lb]{$T$}
       \Text(61,-2)[lb]{$B$}
     \Text(30,35)[lb]{$A^1_\mu$}
    \Text(90,35)[lb]{$A^1_\nu$}
  \end{picture}
& = & -N_c  \left(\frac{ig \mu^\epsilon}{2}\right)^2 \int \frac{d^d k}{(2 \pi)^d} 
 \frac{i^2 {\rm tr} [\gamma^\mu P_L (\slashed{k} + M_T) \gamma^\nu P_L (\slashed{k} + M_B) ]}{(k^2-M_T^2) (k^2-M_B^2)}
  \nonumber \\
&=& \frac{i N_c g^2 g^{\mu \nu}}{2 (4 \pi)^2} \int_0^1 dx (x M_T^2 + (1-x) M_B^2) \left[ \frac{1}{\overline{\epsilon}} +  \ln(\frac{\mu^2}{x M_T^2 + (1-x)M_B^2}) \right] \nonumber \\
&=&    \frac{i N_c g^2 g^{\mu \nu}}{2 (4 \pi)^2}\left[ \frac{M_T^2 + M_B^2}{2 \overline{ \epsilon}} + \frac{M_T^4 \ln (\frac{\mu^2}{M_T^2}) -    M_B^4 \ln (\frac{\mu^2}{M_B^2})  + \frac{M_T^4-M_B^4}{2}}{2 (M_T^2-M_B^2)} \right]. \label{eq:TW1}
\end{eqnarray}
When combining the four diagrams in Fig.~\ref{fig:Tcalc}, the divergent parts of Eqs.~(\ref{eq:TW3}) and (\ref{eq:TW1}) cancel. In a renormalizable theory there cannot be any divergences  for higher-dimensional operators, as divergences would indicate need for new counterterms and would spoil renormalizability. The remaining, finite, part gives the $T$ parameter when the amplitude is compared with $O_T$ expanded around the Higgs vev and the relation in Eq.~(\ref{eq:ST}) is used  
\begin{equation}
  T= - \frac{2 N_c}{v^2 \alpha (4 \pi)^2} \frac{M_T^2 M_B^2 \ln (\frac{M_T^2}{M_B^2})  -\frac{1}{2} M_T^4+\frac{1}{2} M_B^4}{M_T^2-M_B^2}.
\end{equation}
As we anticipated, for large $M_T$, $T\propto \frac{M_T^2}{v^2}$~\cite{rho}. Moreover, it is easy to check that $T\rightarrow 0$ when $M_B\rightarrow M_T$ which is consistent with the argument based on custodial symmetry.

Another example of a field that contributes to the $T$ parameter is a scalar that transforms in the three-dimensional representation of $SU(2)_L$. We postpone the discussion of triplet scalars to Appendix \ref{sec:appendix}. Integrating out the triplet at tree level is not more involved than the examples presented in this section. Obtaining one-loop results is more tedious and it would take too much space here, hence the triplet example is presented in the appendix.  

To calculate the quark doublet contribution to the $S$ parameter we expand $O_S$ around the Higgs vev,  $O_S=-\frac{v^2}{2} A^3_{\mu \nu} \,B^{\mu \nu}+ \ldots$. There are four diagrams that contribute, these are shown on the left-hand sides of Eqs.~(\ref{eq:STR}) through (\ref{eq:SBL}). We assume that the quark doublet has hypercharge $Y$ such that $D_\mu =\partial_\mu - i g \frac{\sigma^i}{2} A^i_\mu - i g' Y B_\mu$ to make our result general. For a genuine fourth generation quark doublet, $Y=\frac{1}{6}$. In order to simplify this calculation further, we calculate the diagrams mixing $A^3_\mu$ and $B_\nu$ and only keep terms proportional to $p^2 g^{\mu \nu}$. 
\begin{eqnarray}
 \begin{picture}(127,30) (20,12)
    \SetWidth{2.0}
    \Arc[arrow,arrowpos=0.5,arrowlength=5,arrowwidth=2,arrowinset=0.2,flip](84,16)(12,270,630)
    \Photon(96,16)(132,16){5}{3}
    \Photon(72,16)(42,16){5}{3}
    \Text(80,12)[lb]{$T$}
    \Text(26,8)[lb]{$A^3_\mu$}
    \Text(136,10)[lb]{$B_\nu$}
    \Text(64,-2)[lb]{$P_L$}
    \Text(95,-2)[lb]{$P_R$}
  \end{picture}
 &=& -\frac{i g g' N_c}{(4 \pi)^2} \left(Y+ \frac{1}{2}\right) \frac{p^2 g^{\mu \nu}}{6}  + \ldots , \label{eq:STR} \\
 \begin{picture}(127,30) (20,12)
    \SetWidth{2.0}
    \Arc[arrow,arrowpos=0.5,arrowlength=5,arrowwidth=2,arrowinset=0.2,flip](84,16)(12,270,630)
    \Photon(96,16)(132,16){5}{3}
    \Photon(72,16)(42,16){5}{3}
    \Text(80,12)[lb]{$B$}
    \Text(26,8)[lb]{$A^3_\mu$}
    \Text(136,10)[lb]{$B_\nu$}
    \Text(64,-2)[lb]{$P_L$}
    \Text(95,-2)[lb]{$P_R$}
  \end{picture}  &=& \frac{i g g' N_c}{(4 \pi)^2} \left(Y-  \frac{1}{2}\right) \frac{p^2 g^{\mu \nu}}{6} + \ldots , \label{SBR}\\  
  \begin{picture}(127,30) (20,12)
    \SetWidth{2.0}
    \Arc[arrow,arrowpos=0.5,arrowlength=5,arrowwidth=2,arrowinset=0.2,flip](84,16)(12,270,630)
    \Photon(96,16)(132,16){5}{3}
    \Photon(72,16)(42,16){5}{3}
    \Text(80,12)[lb]{$T$}
    \Text(26,8)[lb]{$A^3_\mu$}
    \Text(136,10)[lb]{$B_\nu$}
    \Text(64,-2)[lb]{$P_L$}
    \Text(95,-2)[lb]{$P_L$}
  \end{picture}   &=& -\frac{i g g' N_c}{(4 \pi)^2} Y \frac{p^2 g^{\mu \nu}}{3}  \left( \frac{1}{\overline{\epsilon}} +  \ln(\frac{\mu^2}{M_T^2}) - \frac{1}{2} \right)+   \ldots , \label{STL} \\
  \begin{picture}(127,30) (20,12)
    \SetWidth{2.0}
    \Arc[arrow,arrowpos=0.5,arrowlength=5,arrowwidth=2,arrowinset=0.2,flip](84,16)(12,270,630)
    \Photon(96,16)(132,16){5}{3}
    \Photon(72,16)(42,16){5}{3}
    \Text(80,12)[lb]{$B$}
    \Text(26,8)[lb]{$A^3_\mu$}
    \Text(136,10)[lb]{$B_\nu$}
    \Text(64,-2)[lb]{$P_L$}
    \Text(95,-2)[lb]{$P_L$}
  \end{picture}      &=& \frac{i g g' N_c}{(4 \pi)^2} Y \frac{p^2 g^{\mu \nu}}{3}  \left( \frac{1}{\overline{\epsilon}} +  \ln(\frac{\mu^2}{M_B^2}) - \frac{1}{2} \right)+  \ldots , \label{eq:SBL} 
\end{eqnarray}
where we omitted all terms that do not depend on the momentum as $p^2 g^{\mu\nu}$. Summing the four diagrams in Eqs.~(\ref{eq:STR}) through (\ref{eq:SBL}) and matching to the expansion of $O_S$ gives
\begin{equation}
 v^2 a_S = \frac{g g' N_c}{6 (4 \pi)^2} (1 + 2 Y \log(\frac{M_B^2}{M_T^2})).
\end{equation}
Using the conversion factor (\ref{eq:ST}) between $a_S$ and $S$ yields
\begin{equation}
  S= \frac{N_c}{6 \pi } (1 + 2 Y \log(\frac{M_B^2}{M_T^2})).
\end{equation}
Indeed, the $S$ parameter does not depend on the quark mass when $M_B=M_T$, which is an example of non-decoupling.

\subsection{More on the universal parameters: $Y$ and $W$}
\label{sec:YW}

A systematic study of all operators of dimension 6 shows that $O_S$ and $O_T$ are not the only operators that can be called universal~\cite{BPRS}. There are two more operators that can be constructed out of the gauge fields only
\begin{eqnarray} 
 O_Y & = & \frac{1}{2} (\partial_\rho B_{\mu \nu})^2 , \label{eq:OY}  \\
  O_W  & = &  \frac{1}{2} (D_\rho A^i_{\mu \nu})^2. \label{eq:OW}
\end{eqnarray} 
These operators are clearly of the same dimension as $O_S$ and $O_T$ and just as important.  It turns out that there are no more universal operators of dimension 6 that are bound by the current data. The effective Lagrangian 
\begin{equation}
\label{eq:LeffSTYW}
  {\mathcal L}= {\mathcal L}_{SM} + a_S \, O_S + a_T \, O_T + a_Y \, O_Y + a_W \, O_W 
\end{equation}
contains all the universal operators for which PEW constrains exist. 
  
It is useful to rewrite  $O_Y$ using the Bianchi identity $\partial_\rho B_{\mu \nu} + \partial_\mu B_{\nu \rho}+ \partial_\nu B_{\rho \mu}=0$
\begin{equation}
\label{eq:OYBianchi}
  O_Y =\frac{1}{2} \left(- \partial_\rho B_{\mu \nu}  \partial^\mu B^{\nu \rho} - \partial_\rho B_{\mu \nu}   \partial_\nu B_{\rho \mu}\right)= (\partial_\mu B^{\mu \nu})^2,
\end{equation}
where the last equality is obtained by integrating by parts and using the antisymmetry of the field strength. Similarly,
\begin{equation}
\label{eq:OWBianchi}
  O_W =  (D_\mu A^{i \mu \nu})^2.
\end{equation}
These forms are often more suitable for calculations.

As an example of applicability of this formalism we consider a $U(1)$ extension of the SM\@. Suppose that the SM gauge symmetry is extended to $[SU(3)_c\times SU(2)_L \times U(1)_Y ]\times U(1)'$ such that the Lagrangian is 
\begin{equation}
\label{eq:extraU1}
  {\mathcal L}= {\mathcal L}_{SM} -\frac{1}{4} (B'_{\mu \nu})^2 + \frac{\kappa}{2} B_{\mu \nu}  B'^{\mu \nu} + {\mathcal L}(\Phi),
\end{equation}
where ${\mathcal L}(\Phi)$ is a scalar field Lagrangian that spontaneously breaks the $U(1)'$ symmetry. The details of ${\mathcal L}(\Phi)$ are not relevant and we will assume that as a result of symmetry breaking the gauge field $B'_\mu$ acquires mass $M$. We have assumed that the new sector communicates with the SM only through the kinetic mixing with the hypercharge $U(1)_Y$ which would certainly be the case if SM fields do not carry any charges under $U(1)'$. There could be heavy particles that carry both the SM  and $U(1)'$ quantum numbers. Such particles would induce kinetic mixing between $B_\mu$ and $B'_\mu$. 

To constrain this new theory we need to calculate one diagram only. The Lagrangian in Eq.~(\ref{eq:extraU1}) gives a tree-level contribution to $O_Y$. 
\begin{eqnarray}
\begin{picture}(90,25) (60,0)
    \SetWidth{2.0}
    \Photon(60,3)(150,3){5}{9}
    \Vertex(90,3){4}
    \Vertex(120,3){4}
    \Text(70,13)[lb]{$B$}
    \Text(100,13)[lb]{$B'$}
    \Text(135,13)[lb]{$B$}
  \end{picture}
    &=& i \kappa (p^2 g^{\mu \alpha} - p^\mu p^\alpha) \frac{-i (g^{\alpha \beta} - \frac{p^\alpha p^\beta}{M^2})}{p^2-M^2} 
                            i \kappa (p^2 g^{\beta \nu} - p^\beta p^\nu) \nonumber \\
  &=& i \kappa^2 \frac{p^2}{p^2-M^2} (p^2 g^{\mu \nu} - p^\mu p^\nu)\approx -  \frac{i \kappa^2}{M^2} p^2  (p^2 g^{\mu \nu} - p^\mu p^\nu).
    \label{eq:Ydiagram}
 \end{eqnarray}
We compare this result with the Feynman rule for the operator $O_Y$. Writing $O_Y$ explicitly in terms of the $B_\mu$ gauge field and derivatives
\begin{equation}
   O_Y=(\partial_\mu B^{\mu \nu})^2= (\partial^2 B^\nu)^2 
           - 2 (\partial^2 B^\nu) \partial_\nu \partial^\rho B_\rho +
           (\partial_\nu \partial_\mu B^\mu) (\partial^\nu \partial_\rho B^\rho),
 \end{equation}
yields  the amplitude with one insertion of $a_Y O_Y$
 \begin{equation}
 \label{eq:Yinsertion}
    2 i a_Y  (g^{\mu \nu} p^4 - 2 p^2 p^\mu p^\nu + p^2 p^\mu p^\nu) = 2 i a_Y p^2 (p^2 g^{\mu \nu} - p^\mu p^\nu).
\end{equation}
Comparing Eqs.~(\ref{eq:Ydiagram}) and (\ref{eq:Yinsertion}) gives 
\begin{equation}
  a_Y = - \frac{\kappa^2}{2 M^2}.
\end{equation}
Ref.~\cite{BPRS} contains combined bounds on the coefficients of the four universal operators, including the bounds on $a_Y$. Obtaining the bounds on the $U(1)$ extension of the SM was certainly a straightforward exercise, yet it is not a simplified toy  example. Many extension of the SM contain extra $U(1)$ gauge symmetries and such extensions are studied in the literature, see for instance Ref.~\cite{Zprime}.

\subsection{All flavor-conserving operators}
\label{sec:all6}

There are many extensions in which the heavy fields couple directly to the SM fermions. In such extensions integrating out the heavy fields yields not only the universal operators that are included in Eq.~(\ref{eq:LeffSTYW}) but yields other operators as well. We now turn to examine a larger a class of operators that will enable us to constrain a wide range of SM extensions. 

A complete list of all baryon and lepton number conserving operators of dimension 6 in the SM is provided in Ref.~\cite{Buchmuller}. The equations of motion were used to eliminate redundant operators and there are still 80 operators listed in Ref.~\cite{Buchmuller} even with the assumption that there is only one family of quarks and leptons.  We will use the notation of Ref.~\cite{Buchmuller} for the names of the operators.

In most of this section we will follow the analysis in Ref.~\cite{HS}. There are several symmetry assumptions one can make to focus on the operators that are relevant to PEW measurements. The most important assumption is about flavor and $CP$ violation. It is likely that the flavor structure in the SM is generated at a much higher scale than the EW symmetry breaking scale. Current constraints on flavor and CP violation expressed as bounds on coefficients of dimension 6 operators point to suppression scales of order $10^3$ to $10^4$ TeV\@. Such stringent constraints can be inferred, for example, from the $K-\overline{K}$ mass difference or from the limits on the $\mu\rightarrow e \gamma$ decay. It is then reasonable to assume that that the electroweak symmetry breaking is independent of flavor physics. It is possible to lower the scale of new flavor physics, for example by assuming the minimal flavor violation structure of new physics~\cite{MFV}, but we will assume that the EW and flavor scales are well separated. We will concentrate on operators that have nothing to do with flavor, but that can be relevant for modifications of the electroweak symmetry-breaking sector. 

The SM has a large flavor symmetry when the Yukawa couplings are neglected. The kinetic energy terms for the fermions do not distinguish fields of different flavors. Thus, for three families of fermions with the same charge assignment there is a $U(3)$ symmetry. For instance, if we denote by $u$ the triplet of the right-handed up, charm, and top quarks then the kinetic terms for the right-handed up quarks are invariant under $U(3)$. Suppressing the flavor indices, we have a $U(3)^5$ symmetry under which
\begin{equation} 
q \rightarrow U_q \, q, \  \ u \rightarrow U_u\, u,  \  \ d\rightarrow U_d\, d,  \  \ l\rightarrow U_l \,l,  \  \  e\rightarrow U_e \,e, 
\end{equation}  
where $q$ denotes the left-handed quarks, $u$ and $d$ the right-handed quarks, $l$ the left-handed leptons, and $e$ the right handed leptons. We will assume that the operators of interest obey the $U(3)^5$ flavor symmetry.

Imposing the $U(3)^5$ symmetry and CP conservation on the 80 operators in Ref.~\cite{Buchmuller} reduces the list to 52 operators.\footnote{The flavor symmetry assumption can be relaxed, for example to single out the third generation, see Ref.~\cite{Zhenyu}. In some models, the third generation is integral to EW symmetry breaking.} At this step, operators that change fermion chirality are eliminated since fermions of different chiralities  transform independently under the $U(3)$ flavor symmetry.

It is only worthwhile to consider operators that are well constrained by the data, as poorly constrained operators contribute little to constraints on hypothetical new particles. The bounds on some operators are very mild. This is the case for operators that only affect  QCD processes for which experimental precision does not match that of PEW measurements. For example, four-fermion quark operators, or the $f^{abc} G^{a \mu}_\nu G^{b \nu}_\rho G^{c \rho}_\mu $ operator, are rather poorly constrained, where $G^a_{\mu \nu}$ is the gluon field strength. We therefore study operators that either contain some $SU(2)_L\times U(1)_Y$ gauge fields, or contain some leptons. This reduces the number of operators to 34. 

Of the 34 remaining operators, 6 are not observable at present, as they renormalize existing terms in the SM Lagrangian when the Higgs field is replaced by its vev. We saw a few examples of such operators in Sect.~\ref{sec:ST}. Finally, 7 operators satisfy all assumptions we made so far, but are nevertheless very poorly constrained by the available data. All 7 are operators of the form $i\overline{\psi} \gamma^\mu D^\nu \psi F_{\mu \nu}$, where $\psi$ represents SM fermions. The interference terms between such operators and the SM contributions vanish, except at the $Z$ pole. However, at the $Z$ pole the interference term is suppressed by $\frac{\Gamma_Z}{M_Z}$. Since the interference terms with the SM vanish for such operators, the amplitude square is proportional to the square of the operator coefficient which would be of the same order as an interference term of a dimension 8 operator with the SM. Thus, it would not be consistent to keep the operators of the form $i\overline{\psi} \gamma^\mu D^\nu \psi F_{\mu \nu}$ while we otherwise have truncated the expansion at dimension 6. 

We are left with 21 operators that can be divided into 4 classes. 
\begin{enumerate}
 \item Two universal operators $O_S$ and $O_T$. (These are, respectively,  called $O_{WB}$ and $O_H$ in Ref.~\cite{Buchmuller}.)
 \item 11 four-fermion operators
 \begin{eqnarray} &&
  O_{ll}^s=\frac{1}{2} (\overline{l} \gamma^\mu l) (\overline{l} \gamma_\mu l), \  \ \
  O_{ll}^t=\frac{1}{2} (\overline{l} \gamma^\mu \sigma^a l) (\overline{l} \gamma_\mu \sigma^a l),
      \nonumber \\ &&
  O_{lq}^s= (\overline{l} \gamma^\mu l) (\overline{q} \gamma_\mu q), \ \ \
  O_{lq}^t= (\overline{l} \gamma^\mu \sigma^a l) (\overline{q} \gamma_\mu \sigma^a
 q),
     \nonumber \\ &&
  O_{le}= (\overline{l} \gamma^\mu l) (\overline{e} \gamma_\mu e),  \ \ \
  O_{qe}=(\overline{q} \gamma^\mu q) (\overline{e} \gamma_\mu e),
     \nonumber  \\ &&
  O_{lu}= (\overline{l} \gamma^\mu l) (\overline{u} \gamma_\mu u),  \ \ \
  O_{ld}= (\overline{l} \gamma^\mu l) (\overline{d} \gamma_\mu d),
      \nonumber \\ &&
  O_{ee}=\frac{1}{2} (\overline{e} \gamma^\mu e) (\overline{e} \gamma_\mu e), \ \ 
\
  O_{eu}=(\overline{e} \gamma^\mu e) (\overline{u} \gamma_\mu u),  \ \ \
  O_{ed}=(\overline{e} \gamma^\mu e) (\overline{d} \gamma_\mu d).
      \nonumber
  \end{eqnarray}
 \item 7 operators that are products of the Higgs current with various fermion currents
  \begin{eqnarray} &&
   O_{Hl}^s = i (H^\dagger D^\mu H)(\overline{l} \gamma_\mu l) + {\rm H.c.}, \ \ \
   O_{Hl}^t = i (H^\dagger \sigma^a D^\mu H)(\overline{l} \gamma_\mu \sigma^a l)+ 
{\rm H.c.},
     \nonumber \\ &&
   O_{Hq}^s = i (H^\dagger D^\mu H)(\overline{q} \gamma_\mu q)+ {\rm H.c.}, \ \ \
   O_{Hq}^t = i (H^\dagger \sigma^a D^\mu H)(\overline{q} \gamma_\mu \sigma^a q)+ 
{\rm H.c.},
   \nonumber\\ &&
      O_{Hu} = i (H^\dagger D^\mu H)(\overline{u} \gamma_\mu u)+ {\rm H.c.}, \ \ \
   O_{Hd} = i (H^\dagger D^\mu H)(\overline{d} \gamma_\mu d)+ {\rm H.c.},
          \nonumber  \\ &&
   O_{He} = i (H^\dagger D^\mu H)(\overline{e} \gamma_\mu e)+ {\rm H.c.}\,.
    \nonumber
\end{eqnarray}
When the vev is substituted for the Higgs doublet, these operators modify the couplings of the $Z$ and $W$ to the fermions. 
\item One operator that alters the cubic gauge boson couplings
  \begin{displaymath}
O_W = \epsilon^{abc} \, W^{a \nu}_{\mu} W^{b\lambda}_{\nu} W^{c \mu}_{\lambda}.
\end{displaymath}
\end{enumerate}
Note that the operators $O_Y$ and $O_W$ discussed in the previous section are not on the list. Eqs.~(\ref{eq:OYBianchi})
and (\ref{eq:OWBianchi}) make it clear that these operators can be easily re-expressed using the equations of motion for the gauge fields, for example $\partial_\mu B^{\mu \nu} = j_Y^\nu$, where $j_Y$ is the hypercharge current that consists of the fermion and Higgs contributions. The square of the current can be written in terms of $O_T$, four-fermion operators, and the operators of the form $O_{H\psi}$. More details on the use of equations of motion are contained in Refs.~\cite{Buchmuller,eoms}. Of course, if the heavy fields couple to the gauge and Higgs bosons only, it is much more straightforward to deal with the set of four universal operators described in Secs.~\ref{sec:ST} and \ref{sec:YW}. If the couplings are not universal, it is better to avoid $O_Y$ and $O_W$ in favor of the operator basis presented in this section because $O_Y$ and $O_W$ are four-derivative operators. Matching is more messy when one needs to evaluate diagrams to the fourth order in external momenta. 

The effective theory we will consider now is 
\begin{equation}
\label{eq:Leffall}
  {\mathcal L}= {\mathcal L}_{SM} + \sum_{i=1}^{21} a_i \, O_i,
\end{equation}
where $O_i$ stand for the operators enumerated in this section. As we did before, to constrain an extension of the SM one matches the new theory to the effective Lagrangian (\ref{eq:Leffall}). With more than two parameters, it is difficult to visualize the experimentally allowed space of the coefficients $a_i$. We will discuss how the constraints on $a_i$ are obtained in Sec.~\ref{sec:sausage}. Briefly, each relevant observable $X_\alpha$ is computed as a function of the SM couplings, collectively denoted $g_{SM}$, and the coefficients $a_i$
\begin{equation}
\label{eq:Xalpha}
  X_\alpha(g_{SM},a_i) = X^{SM}_\alpha(g_{SM}) +  a_i X^i_\alpha  + a_i a_j X_\alpha^{ij},
 \end{equation} 
where $X_\alpha^i$ is the interference term between SM and operator $O_i$ and $X_\alpha^{ij}$ are the products of the amplitudes containing an insertion of $O_i$ and an insertion of $O_j$. As we mentioned earlier, terms quadratic in $a_i$ can be neglected because these would be equivalent, by power counting, to the interference of dimension-8 terms with the SM amplitudes. By comparing with experimental data, a $\chi^2$ distribution is constructed
\begin{equation}
\label{eq:chi2}
  \chi^2(a_i)=\sum_\alpha \frac{\left(X^{exp}_\alpha - X_\alpha(a_i)\right)^2}{\sigma_\alpha^2}
    = \chi^2_{min} + \sum_{i,j=1}^{21} (a_i-\hat{a}_i) {\mathcal M}_{ij} (a_j- \hat{a}_j),
\end{equation}
where the last equation follows because $\chi^2(a_i)$ is quadratic in $a_i$. This is because we only kept the linear terms in $a_i$ in Eq.~(\ref{eq:Xalpha}). The sum over $\alpha$ runs over all different observable quantities, $X^{exp}_\alpha$ are the measured values of the observable, while $\sigma_\alpha$ are the corresponding errors. In practice, one also needs to take into account correlations between measurements, but this does not change the fact that $\chi^2(a_i)$ is quadratic in $a_i$.

It is worth stressing that the matrix ${\mathcal M}_{ij}$ in  Eq.~(\ref{eq:chi2}) and the coefficients $\hat{a}_i$, for which $\chi^2$ is minimized, are constants determined from experiments. The allowed region in the space of coefficients is a 21-dimensional ellipsoid centered at $\hat{a}_i$ whose axes are determined by the matrix ${\mathcal M}_{ij}$. Eq.~(\ref{eq:chi2}) is an analog of the $S-T$ plot in Fig.~\ref{fig:ST}. The $S-T$ plot is obtained when all coefficients, except $a_S$ and $a_T$, are set to zero.

\begin{table}
\begin{center}
\begin{tabular}{|r||c|c|c|}
\hline
  & 68.27\% & 95.45\% & 99.73\% \\  \hline 
1  & 1& 4 & 9 \\
2 & 2.29 & 6.18 & 11.8 \\
3 &  3.53 & 8.02 & 14.2 \\
5 & 5.89 & 11.3 & 18.2 \\
10 & 11.5 & 18.6 & 26.9 \\
21 & 23.5 & 33.1 & 43.5 \\ \hline 
\end{tabular}
\end{center}
\caption{\label{tabledelta} Increments of the $\chi^2$ distribution depending on the number of free parameters and on confidence levels. Confidence levels are listed in the top row, while the number of degrees of freedom in the leftmost column. The ``allowed'' values of $\chi^2$ are those for which  $\chi^2 \leq \chi^2({\rm ``best\ fit"}) + \Delta$. }
\end{table}

By matching, the operator coefficients $a_i$ are calculated in terms of the masses and couplings of the heavy fields. The allowed range of the parameters is then determined by finding the minimum of $\chi^2$ and accepting the values of the underlying parameters for which $\chi^2 \leq \chi^2({\rm ``best\ fit"}) + \Delta$, where $\Delta$ is determined by the desired confidence level and by the number of free parameters. Table~\ref{tabledelta} shows the values of  $\Delta$ for several confidence levels and several numbers of free parameters. In general, $\chi^2({\rm ``best\ fit"}) \geq \chi^2_{min}$, but $\chi^2({\rm ``best\ fit"})$ is less than the SM value $\chi^2(a_i=0)$. 

Eq.~(\ref{eq:chi2}) allows one to constrain arbitrary linear combinations of operators $O_i$ instead of just constraining each coefficient independently one at a time. As we already discussed in Sec.~\ref{sec:ST}, this is necessary for implementing a global analysis in the EFT approach. Once the heavy fields are integrated out, the operator coefficients $a_i$ are given in terms of the underlying parameters. The coefficients $a_i$ are determined by the same couplings and masses of the heavy states in the full theory, so these coefficients are typically not independent.

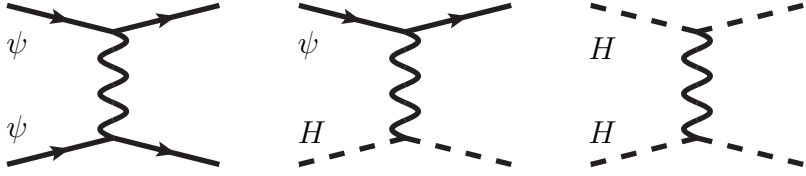
\begin{figure}[htb]
 \begin{center}
  \begin{picture}(300,52) (38,-8)
    \SetWidth{2.0}
    \Line[arrow,arrowpos=0.5,arrowlength=5,arrowwidth=2,arrowinset=0.2](40,-16)(80,-6)
    \Line[arrow,arrowpos=0.5,arrowlength=5,arrowwidth=2,arrowinset=0.2](80,-6)(120,-16)
    \Line[arrow,arrowpos=0.5,arrowlength=5,arrowwidth=2,arrowinset=0.2](40,44)(80,34)
    \Line[arrow,arrowpos=0.5,arrowlength=5,arrowwidth=2,arrowinset=0.2](80,34)(120,44)
    \Line[arrow,arrowpos=0.5,arrowlength=5,arrowwidth=2,arrowinset=0.2](150,44)(190,34)
    \Line[arrow,arrowpos=0.5,arrowlength=5,arrowwidth=2,arrowinset=0.2](190,34)(230,44)
    \Line[dash,dashsize=6](260,44)(300,34)
    \Line[dash,dashsize=6](300,34)(340,44)
    \Photon(80,34)(80,-6){5}{3}
    \Line[dash,dashsize=6](150,-16)(190,-6)
    \Line[dash,dashsize=6](190,-6)(230,-16)
    \Line[dash,dashsize=6](260,-16)(300,-6)
    \Line[dash,dashsize=6](300,-6)(340,-16)
    \Photon(190,34)(190,-6){5}{3}
    \Photon(300,34)(300,-6){5}{3}
    \Text(40,24)[lb]{$\psi$}
    \Text(40,-8)[lb]{$\psi$}
    \Text(150,24)[lb]{$\psi$}
    \Text(150,-8)[lb]{$H$}
    \Text(260,24)[lb]{$H$}
    \Text(260,-8)[lb]{$H$}
  \end{picture}
\end{center}
\caption{\label{fig:WH} Diagrams with exchanges of heavy vector bosons that give products of the fermion and Higgs currents.   }
\end{figure}

In the remainder of this section we are going to consider two sample extensions of the SM and integrate out the heavy fields to further illustrate how one obtains the coefficients $a_i$ and thus how one constrains new theories. As the first example, suppose that the EW sector of the SM is extended to have the $SU(2)_1 \times SU(2)_2 \times U(1)_Y$ gauge group. The $SU(2)_1 \times SU(2)_2$ group is spontaneously broken to its diagonal subgroup, that is to $SU(2)_L$. Moreover, we assume that the SM fermions are charged under the $SU(2)_1$ group, while the SM Higgs boson is charged under $SU(2)_2$ so that the couplings are
\begin{equation}
  {\mathcal L}=g_1 A_{1\mu}^i \, j_\psi^{i\mu} + g_2 A_{2\mu}^i \,j_H^{i\mu},
\end{equation}
where $j_\psi^{i\mu} = \overline{q} \frac{\sigma^i}{2} \gamma^\mu q +  \overline{l} \frac{\sigma^i}{2}  \gamma^\mu l$ is the $SU(2)$ fermion current, while $j_H^{i\mu} = i H^\dagger \frac{\sigma^i}{2}  D_\mu H - i (D_\mu H^\dagger)  \frac{\sigma^i}{2}   H$ is the $SU(2)$ Higgs current. When the product $SU(2)_1 \times SU(2)_2$ group is broken to the diagonal $SU(2)_L$, the $SU(2)_L$ coupling constant is given by
\begin{equation}
 g=\frac{g_1 g_2}{\sqrt{g_1^2+g_2^2}} \ \ {\rm and} \ \ g= g_1 s_H = g_2 c_H,
 \end{equation}
 where we introduced the sine and cosine of the mixing angle between the gauge couplings, denoted $s_H$ and $c_H$, respectively. One linear combination of the vector bosons, $W_H^i=c_H A_1^i - s_H A_2^i$, becomes massive, while the orthogonal combination gives the $A^i$ bosons of the $SU(2)_L$.  
The $SU(2)_1 \times SU(2)_2$ gauge bosons can be expressed as $A_1^i=c_H W_H^i + s_H A^i$ and $A_2^i=c_H A^i - s_H W_H^i$. Diagrams representing tree-level exchanges of $W_H$ are shown in Fig.~\ref{fig:WH}. Integrating out $W_H^i$ gives
\begin{equation}
  {\mathcal L}=- \frac{g_1^2 c_H^2}{M^2} (j_\psi^{i\mu})^2  
      + \frac{g_1 g_2 s_H c_H}{M^2} j_{\psi \mu}^i \,  j_H^{i \mu} - \frac{g_2^2 s_H^2}{M^2} (j_H^{i \mu})^2,
\end{equation}   
where $M$ is the mass of the heavy vector bosons. Since the operator $(j_H^{i \mu})^2$ does not break the custodial symmetry, it does not contain a piece proportional to $O_T$.  $O_T$ is the only operator on our list with the same field content as $(j_H^{i \mu})^2$  that is containing just the Higgs fields and derivatives. If there is no $O_T$ in $(j_H^{i \mu})^2$ we can neglect this term because $(j_H^{i \mu})^2$ must correspond to unobservable, or poorly constrained, operators. This can be checked by an explicit calculation. The other two products of currents give
\begin{equation}
  a^t_{lq}=a^t_{ll} = - \frac{g^2 c_H^2}{2 s_H^2 M^2} \ \ {\rm and} \ \  a^t_{Hl}=a^t_{Hq}=\frac{g^2}{4 M^2}.
\end{equation}

\begin{figure}[htb]
\begin{center}
  \begin{picture}(240,62) (64,-10)
    \SetWidth{2.0}
    \Photon(250,-3)(280,37){5}{4}
    \Line[arrow,arrowpos=0.5,arrowlength=5,arrowwidth=2,arrowinset=0.2](40,-3)(160,-3)
    \Line[dash,dashsize=6](40,37)(80,-3)
    \Line[dash,dashsize=6](120,-3)(160,37)
    \Line[arrow,arrowpos=0.5,arrowlength=5,arrowwidth=2,arrowinset=0.2](200,-3)(320,-3)
    \Line[dash,dashsize=6](200,37)(240,-3)
    \Line[dash,dashsize=6](280,-3)(320,37)
    \Text(95,-25)[lb]{(a)}
    \Text(255,-25)[lb]{(b)}
   \Text(41,15)[lb]{$H$}
     \Text(41,1)[lb]{$d,u$}   
      \Text(95,1)[lb]{$Q$}
  \end{picture}
\end{center}
\caption{\label{fig:QH}Heavy doublet contributions to $O_{Hd}$ and $O_{Hu}$.}
\end{figure}
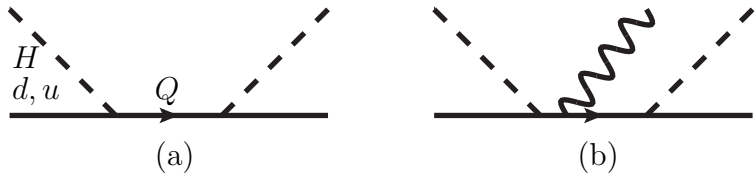

Our next example is an additional vector-like doublet of quarks. We choose the left-handed doublet $Q$ to have the same hypercharge as the SM quark doublets, so that the quarks can mix. The Lagrangian for the heavy quarks is 
\begin{equation}
\label{eq:QH}
  {\mathcal L} = - M \overline{Q} Q - ( \lambda_d \overline{Q} d H + \lambda_u \overline{Q} u \tilde{H} + H.c.),
\end{equation}
where the mass term is allowed because both the right- and left-handed components of $Q$ have the same quantum numbers.  The relevant diagram is shown in Fig.~\ref{fig:QH}(a). Since this diagram will match to the $O_{Hd}$ operator we need to extract the amplitude proportional to one power of the external momentum.  The corresponding amplitude for the external $d$ quarks is 
\begin{equation}
\label{eq:QHintout}
  {\mathcal A}=  (-i \lambda_d)^2 \, \overline{u}(p_4) P_L \frac{i (\slashed{p} + M)}{p^2 - M^2}  P_R u(p_3), 
\end{equation}
where the $d$ quarks are by assumption right-handed, so the projection operators pick out the $\slashed{p}$ part of the propagator in Eq.~(\ref{eq:QHintout}). The momentum flowing through the internal line is $p=p_1+p_3=p_2+p_4=\frac{p_1+p_2+p_3+p_4}{2}$. However, the external quarks are massless, so $\slashed{p}_3 u=0$ and $\overline{u} \slashed{p}_4=0$. Comparing this result with the amplitude from an insertion of $O_{Hd}$ we obtain
\begin{equation}
\label{eq:aHd}
  a_{Hd}=\frac{\lambda_d^2}{2 M^2}.
\end{equation}
Obtaining the amplitude with external $u$ quarks is just as simple, but one needs to convert the current written in terms of $\tilde{H}$ to the current written in terms of $H$. This results in an extra minus sign compared to Eq.~(\ref{eq:aHd})
\begin{equation}
\label{eq:aHu}
  a_{Hu}=-\frac{\lambda_u^2}{2 M^2}.
\end{equation}

It is worth pointing out that when we matched the UV amplitude to the operators $O_{Hd}$ and $O_{Hu}$ we only took into account the partial derivative part of the Higgs current. These operators also have a part proportional to the gauge fields. This part arises from the diagram in Fig.~\ref{fig:QH}(b). We could have calculated either diagram (a) or (b), but since the two are related by gauge invariance it was enough to calculate one of them. Extracting the amplitude with the gauge fields allows one to neglect all external momenta
\begin{equation}
  {\mathcal A}= (-i \lambda_d)^2 \overline{u}_d \frac{i (\slashed{p} + M)}{p^2 - M^2} (i \frac{g}{2} \sigma^i \slashed{A}^i + i \frac{g'}{6} \slashed{B})   \frac{i (\slashed{p} + M)}{p^2 - M^2} u_d\approx \frac{\lambda_d^2}{M^2}  \overline{u}_d  (i \frac{g}{2} \sigma^i \slashed{A}^i + i \frac{g'}{6} \slashed{B}) u_d, 
\end{equation}
which is even simpler than the previous calculation. This agrees with Eq.~(\ref{eq:aHd}) when either the amplitude with an external $B_\mu$ or with an external $A^i_\mu$ are compared with the corresponding terms in $O_{Hd}$.

An interesting exercise is checking the results in Eqs.~(\ref{eq:aHd}) and (\ref{eq:aHu}) directly by diagonalizing the quark mass terms in Eq.~(\ref{eq:QH}). The light mass eigenstates are mixtures of the ``original'' SM right-handed quarks $d$ and $u$ with the right-handed part of $Q$. Since Q has different quantum numbers than $u$ and $d$, the light quarks couple differently to the $Z$ boson compared to the ordinary SM quarks. The modifications of the $Z$ couplings can be compared, and have to agree, with those given by the operators $O_{Hd}$ and $O_{Hu}$. Refs.~\cite{examples} contain several further examples of various applications of this formalism for constraining interesting extensions of the SM\@. 

\subsection{Measurements}
\label{sec:measurements}

The PEW constraints arise from data gathered by many different experiments. For the purpose of this discussion we divide the data into four categories. We briefly review the types of data in this section and discuss which operators are sensitive to different measurements. The four types of measurements are
\begin{enumerate} 
\item $Z$-pole observables gathered by the experiments at LEP1 and at SLAC\@. These include the $Z$ mass $M_Z$, the $Z$ width $\Gamma_Z$, branching ratios of the $Z$ into quarks and leptons, forward-backward asymmetries, and left-right asymmetries depending on the beam polarizations. The $Z$-pole measurements achieved very high statistics and typically these measurements are the most relevant for PEW constraints. However, not all operators can be constrained by the $Z$-pole data.
\item $W$ mass. We single out this measurement because of its high accuracy and also because it is obtained by both the Tevatron and LEP2 experiments. Due to its accuracy, this measurement puts very stringent constraints on several operators. 
\item LEP2 measurements. These include measurements of $e^+ e^- \rightarrow \overline{\psi} \psi$ scattering at the center of mass energies above the $Z$ mass as well as $e^+ e^- \rightarrow W^+ W^-$ scattering. The data that is used includes a combination of total cross sections, asymmetries, and differential scattering cross sections in a few channels. 
\item Low-energy observables. This class encompasses many diverse experiments. The two most precisely measured quantities are the QED fine structure constant $\alpha$ and the Fermi coupling $G_F$. There is a lot of data on neutrino scattering, both deep inelastic scattering of $\nu_\mu$ on nucleons, and neutrino-electron scattering. Measurements of atomic parity violation constitute the next set of measurements. These are usually reported in terms of an effective weak charge of the nucleus, for example $Q_W(Cs)$ or $Q_W(Tl)$. The nuclei in which the highest precision has been achieved are cesium and thallium, but there are also measurements of atomic parity violation in lead and bismuth. Other experiments include Moller, that is $e^- e^-\rightarrow e^- e^-$, scattering and the measurements of the muon anomalous magnetic moment.  
\end{enumerate}
No data from hadron colliders, other than the $W$ mass measurement are included in this list. There are many processes which would be useful for constraining effective operators. For example, jet production cross sections probe quark four-fermion interactions. The accuracy of such measurements, due to poor knowledge of the parton distribution functions and limited precision of hadronic measurements, is much smaller than the accuracy of the measurements that are considered  PEW observables. 

In the electroweak sector, the SM has three undetermined parameters  that is the gauge coupling constants $g$ and $g'$ and the electroweak vev $v$. Three most precisely measured quantities, $\alpha$, $G_F$, and $M_Z$, are used to determine the parameters of the SM\@. These three measurements cannot be therefore used to constrain new physics. As we will discuss in the next section, the precision of the measurements requires one-loop electroweak calculations in the SM that depend on the top quark mass. Even though the top quark mass is known, it has not been measured as accurately as other parameters of the SM\@. The uncertainty in $m_{top}$ is sometimes important for comparisons of the SM with experiment and needs to be included in the estimates of errors. 

Looking back at Eq.~(\ref{eq:chi2}), it is clear that experimental uncertainties determine the size of the allowed region in the space of coefficients $a_i$ that is encoded in the matrix ${\mathcal M}_{ij}$. The quadratic dependence on $a_i$ is solely determined by the uncertainties. The central values of the coefficients, denoted $\hat{a}_i$ in Eq.~(\ref{eq:chi2}), are determined by the differences between the central values of measurements and the SM predictions. It is important that many measurements are correlated instead of being independent. The expression for $\chi^2$ in Eq.~(\ref{eq:chi2}) assumes that the experimental quantities are independent, so Eq.~(\ref{eq:chi2}) needs to be modified to include correlations
\begin{equation}
\label{eq:chi2corr}
  \chi^2(a_i)=\sum_{\alpha,\beta} \left(X^{exp}_\alpha - X_\alpha(a_i)\right) \left(\sigma_{\alpha\beta}^2\right)^{-1}  \left(X^{exp}_\alpha - X_\alpha(a_i)\right),
\end{equation}
where the error matrix $\sigma_{\alpha\beta}^2$ can be expressed in terms of the correlation matrix $\rho_{\alpha \beta}$ and the standard deviations as  $\sigma_{\alpha\beta}^2= \sigma_\alpha \rho_{\alpha \beta} \sigma_\beta$. Correlations are particularly prominent among the $Z$-pole measurements~\cite{LEP-EW} and among LEP2 measurements. The differential cross sections at LEP2 measured at different energies are correlated. 

\subsection{How the sausage is made}
\label{sec:sausage}

As the title suggests, not everyone may be interested in reading this section. In a way, that is the point of the EFT approach. The bounds on the coefficients of operators have already been extracted and the details how it was done are not that important. One can constrain their favorite model without ever being concerned with the actual experimental data. 

Foremost, to constrain new physics one needs accurate SM calculations. This is a topic that we will not discuss in these notes. The precision of measurements generally requires one-loop electroweak corrections and often higher-oder QCD corrections. The electroweak corrections depend on the masses of the SM particles, including the unknown Higgs mass. Thus, the predictions are always shown with a chosen reference value for the Higgs mass, as illustrated in Fig.~\ref{fig:ST}. Since the couplings of the Higgs to the light fermions are tiny, only the universal parameters are sensitive to the Higgs mass. The leading dependence of $S$ and $T$ on the Higgs mass is logarithmic~\cite{STU}
\begin{equation}
  \Delta  S\approx \frac{1}{12 \pi} \log\left(\frac{M_h^2}{M^2_{h,ref}}\right) \ \ {\rm and} \ \  \Delta T \approx \frac{- 3}{16 \pi c^2} \log\left(\frac{M_h^2}{M^2_{h,ref}}\right).
\end{equation}
It is this dependence that gives indirect estimates of the Higgs mass in the SM\@.

To constrain the coefficients of operators we use the interference terms between the SM and the effective operators. The experimental accuracy of PEW measurements is comparable to the one-loop electroweak corrections. Thus, the suppression of higher dimensional operators is of the same order.  When computing the interference terms electroweak loop corrections can be neglected, as the product of suppression of higher-dimensional terms with the electroweak loop suppression is much smaller than the experimental accuracy.

\begin{table}[hbt]
\begin{center}
\begin{tabular}{|c|c|c|c|c|c|c|c|}
\hline
 Operator(s)&shift&$M_W$&Z-pole&$\nu$&$ APV $&$\psi \overline \psi$ &$W^+W^-$\\
 \hline
 $O_S$&$\alpha, \, M_Z$& &$\surd$&$\surd$&$\surd$&$\surd$&$\surd$\\
 \hline
 $O_T$&$M_Z$&&&&&&\\
 \hline
 $O_{ll}^t$&$G_F$&&&$\surd$&&$\surd$&\\
 \hline
 $O_{ll}^s$, $O_{le}$ &&&&$\surd$&&$\surd$&\\ \hline
 $O_{ee}$ & & & & & & $\surd$ & \\ \hline
 $O_{lq}^s,O_{lq}^t,O_{lu},O_{ld}$&&&&$\surd$&$\surd$&$\surd$&\\ \hline
 $O_{eq},O_{eu},O_{ed}$&&&&&$\surd$&$\surd$&\\
 \hline
 $O_{hl}^t$&$G_F$&&$\surd$&$\surd$&$\surd$&$\surd$&$\surd$\\
 \hline
 $O_{hl}^s,O_{he}$&&&$\surd$&$\surd$&$\surd$&$\surd$&$\surd$\\
 \hline
 $O_{hu},O_{hd},O_{hq}^s,O_{hq}^t$&&&$\surd$&$\surd$&$\surd$&$\surd$&\\
 \hline
 $O_W$&&&&&&&$\surd$\\
\hline
\end{tabular}
\end{center}
\caption{ \label{table}
Measurements affected by different operators. The abbreviations used for the types of measurements are: $\nu$ for neutrino scattering experiments, APV for atomic parity violation, $\psi \overline \psi$ for $e^+ e^- \rightarrow \psi \overline \psi$ at LEP2,
and $W^+W^-$ for $e^+ e^- \rightarrow W^+W^-$  at LEP2. The check marks, $\surd$,  indicate  ``direct" corrections only. The operators that shift input parameters are marked in the ``shift'' column by indicating the affected input quantity. }
\end{table}

We now examine two examples of how constraints on the coefficients of effective operators are obtained. We consider the operators $O_T$ and $O_{ee}$. These examples illustrate two distinct possibilities. The operator $O_T$ does not directly contribute to any observables used for constraining new physics. There are no diagrams with an insertion of $O_T$ that give rise to scattering or decay widths, etc. Instead, $O_T$ contributes to the $Z$ mass. Since the $Z$ mass determines the SM input parameters all SM predictions will be altered when $O_T$ is present. We calculate the ``shifts'' in the values of the input parameters to the linear order in $a_T$ because we are interested in the interference terms only. The operators $O_S$, $O^t_{ll}$, and $O^t_{Hl}$ also shift the SM input parameters, but these operators also contribute directly to some of the observables. That is when either $O_S$, $O^t_{ll}$, or $O^t_{Hl}$ are present, the SM input parameters need to be shifted and insertions of these operators considered in the scattering amplitudes. A lucid explanation of how to account for the shifts in the SM inputs is contained in Ref.~\cite{Canada}. Table~\ref{table}, adopted from Ref.~\cite{HS}, shows which operators  contribute to different experiments, or contribute to the shifts of the input parameters, and in turn which operators are constrained by which measurements. 

In this section we deal with cross sections and decay rates, thus we need to use the gauge boson mass eigenstates. Expanding $O_T$ around the Higgs vev we obtain $O_T=\frac{m_Z^2}{2} \frac{v^2}{2} Z_\mu^2 + \ldots$. Hence if $a_T O_T$ is present in the  effective Lagrangian there is an extra contribution to the $Z$ mass. Below the EW symmetry breaking scale
\begin{equation}
\label{eq:SMwithOT}
  {\mathcal L}_{SM} + a_T O_T  \supset -\frac{1}{4} A_{\mu \nu} A^{\mu \nu} -  \frac{1}{4} Z_{\mu \nu} Z^{\mu \nu} + \frac{\hat{m}_Z^2 (1 + \gamma)}{2} Z_\mu Z^\mu - e A_\mu j_{em}^\mu - \frac{e}{s c} Z_\mu j^\nu_{NC},
\end{equation}
where we only included the photon and the $Z$ kinetic terms and their couplings to the currents, while $\gamma=\frac{a_T v^2}{2}$. We are going to consider the electric charge, the $Z$ mass, and the weak mixing angle as the input parameters, and we abbreviate $s=\sin \theta_w$, $c=\cos\theta_w$. These parameters are equivalent to $g$, $g'$, and $v$. In the absence of higher dimensional operators one would extract the values $e$, $s$, and $m_Z$ from the measurements of $\alpha$, $M_Z$, and $G_F$. When $a_T O_T$ is added to the Lagrangian, one deduces instead the values $\hat{e}$, $\hat{s}$, and $\hat{m}_Z$. We use the lower-case $m_Z$ for the Lagrangian parameter, while the physical value of the $Z$ mass is denoted $M_Z$.  

Reading off from Eq.~(\ref{eq:SMwithOT}) we obtain
\begin{equation}
  e=\hat{e}, \ \ m_Z^2 = \hat{m}_Z^2 (1 + \gamma), \ \ {\rm and} \ \ \frac{8 G_F}{\sqrt{2}}=\frac{e^2}{m_Z^2 s^2 c^2}=\frac{\hat{e}^2}{\hat{m}_Z^2 \hat{s}^2 \hat{c}^2},
\end{equation}
where the expression for $G_F$ can be taken as our definition of $s$ and $c$. Solving these equations to the linear order in $\gamma$ gives
\begin{equation}
\label{eq:shifts}
  \hat{e}=e, \ \ \hat{m}_Z^2=m_Z^2(1-\gamma), \ \ \hat{s}^2=s^2 (1+ \frac{\gamma c^2}{c^2-s^2}),  \ \  {\rm and} \ \ \hat{c}^2=c^2 (1- \frac{\gamma s^2}{c^2-s^2}).
\end{equation}
For every observable, for example the $Z$ width into fermions $\psi$ $\Gamma(Z\rightarrow \psi \overline{\psi})$, one can take the corresponding tree-level expression in terms of the input parameters and calculate its change due to the shift in the input parameters in Eq.~(\ref{eq:shifts}). At the tree level,
\begin{equation}
\label{eq:gammaZ}
  \Gamma(Z\rightarrow \psi \overline{\psi})=\frac{M_Z}{12 \pi} \left(g_V^2 + g_A^2\right),
\end{equation}
where $g_V=\frac{\hat{e}}{\hat{s}\hat{c}} \left(T_3 - Q \hat{s}^2\right)$ and $g_A=- \frac{\hat{e}}{\hat{s}\hat{c}} T_3$. Meanwhile, $T_3$ denotes the third component of the $SU(2)_L$ generator, that is $\pm\frac{1}{2}$, and  $Q$ the electric charge of the fermion. Combining Eqs.~(\ref{eq:shifts}) and (\ref{eq:gammaZ}) gives the change of $\Gamma$ due to $O_T$:
\begin{equation}
 \delta  \Gamma(Z\rightarrow \psi \overline{\psi})= - \frac{a_T v^2}{4} \frac{M_Z}{6 \pi} \frac{e^2}{s^2 c^2} \left[ (T_3 - Q s^2)(T_3 + Q \frac{s^2}{c^2-s^2}) + T_3^2 \right].
 \end{equation} 
Of course, such a calculation needs to be repeated for every observable before $\chi^2$ in Eq.~(\ref{eq:chi2corr}) can be calculated. For instance,
\begin{equation}
  M_W^2= \hat{m}_Z^2 \hat{c}^2 = m_z^2 (1-\gamma) c^2 (1-\gamma \frac{s^2}{c^2-s^2})= m_z^2 c^2  (1-\gamma \frac{c^2}{c^2-s^2}) + {\mathcal O}(\gamma^2)
\end{equation}
so that $\delta M_W^2 = - \frac{a_T v^2}{2}  \frac{c^4 M_Z^2}{c^2-s^2}$. In the equation for the predicted change of the $W$ mass, denoted $\delta M_W^2$, we replaced $m_Z$ with $M_Z$. This is justified because the difference between $m_Z$ with $M_Z$ is given by loop effects. Loop corrections can be neglected when multiplied by the small parameter $a_T v^2$. The four operators that shift the input parameters, $O_S$, $O_T$, $O^t_{ll}$, and $O^t_{Hl}$ have the most stringent bounds on their coefficients among all the operators considered here. This happens because shifts of the input parameters affect all observables, so all measurements are statistically combined when obtaining bounds. 

\begin{figure}
    \begin{center}
  \begin{picture}(150,80) (18,-5)
    \SetWidth{2.0}
    \Line[arrow,arrowpos=0.5,arrowlength=5,arrowwidth=2,arrowinset=0.2,flip](20,64)(40,24)
    \Line[arrow,arrowpos=0.5,arrowlength=5,arrowwidth=2,arrowinset=0.2,flip](40,24)(20,-16)
    \Line[arrow,arrowpos=0.5,arrowlength=5,arrowwidth=2,arrowinset=0.2](80,24)(100,64)
    \Line[arrow,arrowpos=0.5,arrowlength=5,arrowwidth=2,arrowinset=0.2](100,-16)(80,24)
    \Line[arrow,arrowpos=0.5,arrowlength=5,arrowwidth=2,arrowinset=0.2,flip](130,64)(150,24)
    \Line[arrow,arrowpos=0.5,arrowlength=5,arrowwidth=2,arrowinset=0.2,flip](150,24)(130,-16)
    \Line[arrow,arrowpos=0.5,arrowlength=5,arrowwidth=2,arrowinset=0.2](150,24)(170,64)
    \Line[arrow,arrowpos=0.5,arrowlength=5,arrowwidth=2,arrowinset=0.2](170,-16)(150,24)
    \Photon(40,24)(80,24){5}{4}
    \Vertex(150,24){4}
    \Text(50,34)[lb]{$\gamma,Z$}
  \end{picture}
\end{center}
\caption{\label{fig:eemumu}  The SM diagram and four-fermion contribution to $e^+ e^- \rightarrow \mu^+ \mu^-$ .}
\end{figure}
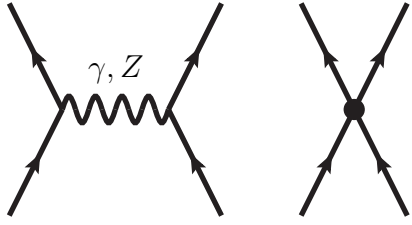

Let us briefly examine the operator $O_{ee}$ that contributes directly to some observables and does not shift the input parameters.
\begin{equation}
  O_{ee}=\frac{1}{2}  (\overline{e} \gamma^\mu e) (\overline{e} \gamma_\mu e)
     = \frac{1}{2} (\overline{e} \gamma^\mu e + \overline{\mu} \gamma^\mu \mu 
          + \overline{\tau} \gamma^\mu \tau)^2,
\end{equation}     
where $e$ denotes at first a $U(3)$ triplet of the right-handed leptons and then denotes just the electron right-handed field. Hopefully, this abuse of notation will not be confusing. All the fields are right-handed, so there are implicit chirality projectors in the equation above. Suppose we are interested in the $e^+ e^- \rightarrow \mu^+ \mu^-$ scattering. The operator $O_{ee}$ has a very simple structure and one needs to calculate the interference between the SM graph and the four-fermion interaction. The Feynman diagrams are depicted in Fig.~\ref{fig:eemumu}. The amplitude for the $ O_{ee}$ operator is simply
\begin{equation}
\label{eq:eemumua}
  {\mathcal A}_{ee}= i a_{ee}\, \overline{u}  \gamma^\mu u \, \overline{u} \gamma_\mu u,
\end{equation}
where $a_{ee}$ is the coefficient of $O_{ee}$ and $u$'s indicate Dirac spinors for the external electrons and muons. The $Z$ boson exchange amplitude is proportional to
\begin{equation}
\label{eq:eemumuZ}
  {\mathcal A}_Z \propto \left(\frac{i g}{c}\right)^2 \frac{-i}{k^2 -M_Z^2 + i \Gamma_Z M_Z} \left(g^{\mu \nu} - \frac{k^\mu k^\nu}{M_Z^2}\right) \overline{u} \gamma_\mu u \, \overline{u} \gamma_\nu u.
\end{equation}
We are not going to do this straightforward calculation in detail, but want to point something out. At the $Z$ pole, the factor multiplying the spinors in Eq.~(\ref{eq:eemumuZ}) is real. However, the analogous factor in  Eq.~(\ref{eq:eemumua}) is imaginary, so the interference of the two amplitudes vanishes. This is general: four-fermion operators are not significantly constrained by the Z-pole measurements. Of course, there is a photon exchange diagram as well, but it is suppressed by the photon propagator and therefore small. The four-fermion operators are constrained by the low-energy observables and by LEP2 data.

\appendix
\section{Scalar triplet contributions to the $T$ parameter}
\label{sec:appendix}
Scalars that transform in the triplet representation of $SU(2)_L$ are a common ingredient of many extensions of the SM\@. Triplet scalars contribute to the $T$ parameter because they violate the custodial symmetry if they acquire a vev. In this section we will integrate out scalar triplets at the tree and one-loop levels. One of the reasons for the discussion at the one-loop order are claims in the literature that the effects of triplets on the $T$ parameter do not decouple when the triplet mass is very large~\cite{triplets1,triplets2}. This is difficult to understand based on power counting. We discuss the power counting at the end of Sec.~\ref{sec:triplettree}  and again at the beginning of Sec.~\ref{sec:tripletloop} before we describe the loop calculations. However, we do not have an answer as to why the result obtained here and the results in Refs.~\cite{triplets1,triplets2} disagree qualitatively. 

We first calculate the tree-level contribution of triplets to the $T$ parameter. We obtain the coefficient of the $T$ operator in several different ways in Sec.~\ref{sec:triplettree}. The method that may seem the least straightforward at tree level will turn out to be useful in loop calculations. In Sec.~\ref{sec:tripletloop}  we calculate one-loop matching coefficients, but do not calculate one-loop running of the $T$ operator in the effective theory. While the RG contributions can be numerically significant, it is clear that such contributions cannot alter decoupling. The one-loop RG logs multiply the tree-level contribution, so the decoupling of the tree-level result implies decoupling of the RG-corrected contribution.

\subsection{Tree level}
\label{sec:triplettree}

Scalar triplets, like any other scalars that are not in the doublet representation of the $SU(2)_L$,  violate the custodial symmetry if they acquire a vev. Thus, we are interested in the triplet contributions to $O_T$. The triplet can only obtain a vev if its hypercharge is either $0$ or $\pm1$, otherwise we would have spontaneous breaking of $U(1)_{EM}$. We will use $\varphi^a$ to denote the triplet with hypercharge 0 and $\phi^a$ to denote the one with hypercharge -1. The corresponding Lagrangians, including the couplings to the SM Higgs, are
\begin{eqnarray}
   {\mathcal L}_0 &=& \frac{1}{2} D_\mu \varphi^a D^\mu \varphi^a - \frac{M^2}{2} (\varphi^a)^2 + \kappa \, H^\dagger \sigma^a H \varphi^a,  \label{eq:Ltriplet0} \\
    {\mathcal L}_{\pm 1} &=&  (D_\mu \phi^a)^* D^\mu \phi^a - M^2 |\phi^a|^2 + \kappa \left(\tilde{H}^\dagger \sigma^a H \phi^a+ H.c.\right), \label{eq:Ltriplet1}
\end{eqnarray}
where all other couplings not explicitly written in these Lagrangians are not relevant for our calculation. The covariant derivatives acting on the the triplets are $D_\mu \varphi^a = \partial_\mu \varphi^a + g \epsilon^{abc} A^b_\mu \varphi^c$ and $D_\mu \phi^a = \partial_\mu \phi^a + g \epsilon^{abc} A^b_\mu \phi^c + i g' B_\mu \phi^a $. The coupling constant $\kappa$ has mass dimension 1 since it is the coefficient of a cubic scalar interaction. When $H$ obtains a vev, the cubic terms proportional to $\kappa$ become linear terms for the triplet thus forcing a triplet vev. In the UV theory, one should not be concerned with what happens at low-energies that is with a vev for a light field. One simply integrates out the triplet which induces the operator $O_T$. $O_T$ reproduces the custodial symmetry breaking effects of either $\langle \varphi^a \rangle$ or  $\langle \phi^a \rangle$.

\begin{figure}[htb]
\begin{center}
  \begin{picture}(190,90) (28,-10)
    \SetWidth{2.0}
    \Line[dash,dashsize=6](30,67)(40,27)
    \Line[dash,dashsize=6](40,27)(30,-13)
    \Line[dash,dashsize=6](110,27)(120,67)
    \Line[dash,dashsize=6](110,27)(120,-13)
     \Text(35,-15)[lb]{$p_2$}
     \Text(35,62)[lb]{$p_1$}
     \Text(107,-15)[lb]{$p_4$}
     \Text(107,62)[lb]{$p_3$}
    \SetWidth{3.0}
    \Line[dash,dashsize=2](40,27)(110,27)
    \SetWidth{2.0}
    \Line[dash,dashsize=6](150,67)(160,27)
    \Line[dash,dashsize=6](160,27)(150,-13)
    \Line[dash,dashsize=6](230,27)(240,67)
    \Line[dash,dashsize=6](230,27)(240,-13)
    \SetWidth{3.0}
    \Line[dash,dashsize=2](160,27)(230,27)
    \SetWidth{2.0}
    \Photon(200,27)(220,67){5}{3.4}
    \Photon(180,67)(200,27){5}{3.4}
    \Text(70,-23)[lb]{(a)}
    \Text(190,-23)[lb]{(b)}
  \end{picture}
\end{center}
\caption{\label{fig:triplet}  Triplet contributions to $O_T$. The external dashed lines represent the Higgs doublet, while the internal dashed line represents the heavy triplet.}
\end{figure}
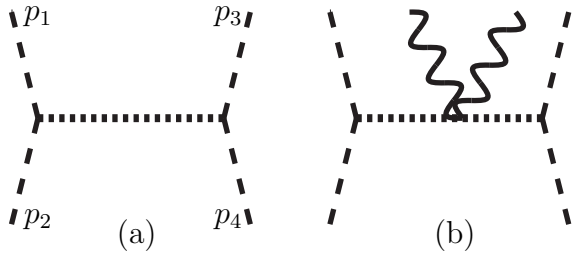

Fig.~\ref{fig:triplet}(a) depicts tree-level triplet exchange that gives $O_T$. This amplitude needs to be evaluated to the second order in the external momenta since there are no interesting terms without derivatives. The kinematic part of the amplitude arising from the $\varphi^a$ exchange, neglecting for the moment the $\sigma^a$ matrices, is
\begin{equation}
\label{eq:triplet}
 {\mathcal A}= (i \kappa)^2 \frac{i}{p^2-M^2}\approx \frac{i \kappa^2}{M^2} \left( 1 + \frac{p^2}{M^2}\right) =  
    \frac{i \kappa^2}{M^2} \left( 1 + \frac{p_1p_2 + p_3 p_4}{M^2}\right),
\end{equation}
where the last equality follows from $p=p_1+p_2=-p_3-p_4$ and from assuming that $p_1^2=\ldots=p_4^2=0$. All of the external momenta are assumed to be incoming. The momentum-dependent part of amplitude in  Eq.~(\ref{eq:triplet})  corresponds to the amplitude obtained from $D_\mu H^\dagger \sigma^a D^\mu H \, H^\dagger \sigma^a H$ which can be rewritten using the completeness relation for the Pauli matrices to produce $O_T$ and other uninteresting operators of dimension 6. Finally,
\begin{equation}
\label{eq:aT0}
  a_T^{(0)} = - \frac{2 \kappa^2}{M^4}.
\end{equation}
Integrating out $\phi^a$ does not give the same result because the amplitude in Eq.~(\ref{eq:triplet}) corresponds to the operator $D_\mu \tilde{H}^\dagger \sigma^a D^\mu H \, H^\dagger \sigma^a \tilde{H} + H.c.$, which gives
\begin{equation}
\label{eq:aT1}
  a_T^{(\pm1)} =  \frac{4 \kappa^2}{M^4}.
\end{equation}

\begin{figure}[htb]
\begin{center}
  \begin{picture}(140,125) (130,-10)
    \SetWidth{2.0}
    \Line[dash,dashsize=6](60,96)(110,46)
    \Line[dash,dashsize=6](110,46)(60,-4)
    \Line[dash,dashsize=6](110,46)(160,96)
    \Line[dash,dashsize=6](110,46)(160,-4)
    \Photon(110,46)(130,96){4.5}{4}
    \Photon(90,96)(110,46){4.5}{4}
    \Text(165,-14)[lb]{$H_1^*$}
    \Text(45,-14)[lb]{$H_1$}
    \Text(45,92)[lb]{$H_1$}
    \Text(165,92)[lb]{$H_1^*$}
    \Text(85,100)[lb]{$B_\mu$}
    \Text(125,100)[lb]{$B_\nu$}
    \Text(90,-24)[lb]{$i g^{\mu \nu} g'^2 c_B$}
    \Line[dash,dashsize=6](230,96)(280,46)
    \Line[dash,dashsize=6](280,46)(230,-4)
    \Line[dash,dashsize=6](280,46)(330,96)
    \Line[dash,dashsize=6](280,46)(330,-4)
    \Photon(280,46)(300,96){4.5}{4}
    \Photon(260,96)(280,46){4.5}{4}
   \Text(335,-14)[lb]{$H_1^*$}
    \Text(215,-14)[lb]{$H_1$}
    \Text(215,92)[lb]{$H_1$}
    \Text(335,92)[lb]{$H_1^*$}
    \Text(255,100)[lb]{$A^1_\mu$}
    \Text(295,100)[lb]{$A^1_\nu$}
    \Text(260,-24)[lb]{$i g^{\mu \nu} g^2 c_A$}
  \end{picture}
\end{center}
\caption{\label{fig:4H2g}  The amplitudes that define the coefficients $c_B$ and $c_A$.}
\end{figure}
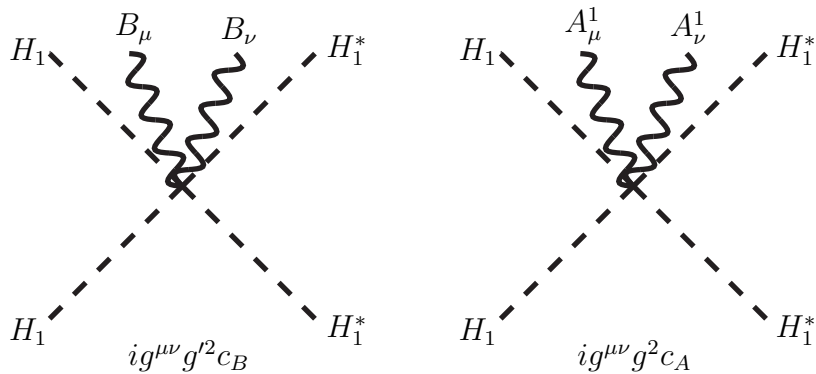

As we observed before, gauge invariance ensures that diagram (b) in Fig.~\ref{fig:triplet} reproduces the gauge field dependent part of $O_T$ even though we only matched the part without any external gauge fields. We can also use that amplitude  to derive $a_T$. This way of matching the effective theory will turn out to be very useful in the next section. Expanding the covariant derivatives in $O_T$ gives
\begin{equation}
  O_T= \left| H^\dagger \partial_\mu H \right|^2 + \frac{g'^2 B_\mu^2}{4}  (H^\dagger H)^2  + \frac{g^2 A^i_\mu A^{j \mu}}{4}  H^\dagger \sigma^i H H^\dagger \sigma^j H    
             + \ldots,
\end{equation}
where we omitted terms linear in the gauge fields. Expressing the Higgs doublet in components $H=\left(\begin{array}{c} H_1 \\ H_2 \end{array} \right)$
\begin{equation}
  O_T=  |H_1\partial_\mu H_1|^2 +  |H_1 \partial_\mu H_1|^2 + \frac{g'^2 B_\mu^2}{4} \left( |H_1|^4 + \ldots \right) + \frac{g^2 (A^1_\mu)^2 }{4} (H_1^* H_2 + H_1 H_2^*)^2 + \ldots 
\end{equation} 
we notice that $B_\mu^2$ couples to $ |H_1|^4$ while $(A^1_\mu)^2$ does not. It is not enough to extract the coefficient of the term $B_\mu^2 |H_1|^4$ to obtain $O_T$ since there are other operators of dimension 6 that contain this term, for example $D_\mu H^\dagger D^\mu H H^\dagger H$. However, all operators containing four Higgs and two gauge fields that do not violate the custodial symmetry have equal coefficients for the terms $B_\mu^2 |H_1|^4$ and $(A^1_\mu)^2 |H_1|^4$. Thus, we will extract the difference between the amplitudes depicted in Fig.~\ref{fig:4H2g}. This difference is proportional to the $T$ parameter 
\begin{equation}  
\label{eq:cBminuscA}
  a_T = \frac{1}{2} ( c_B - c_A),
\end{equation}
where appropriate powers of the gauge couplings and $i g^{\mu \nu}$ have been absorbed into the definitions of $c_B$ and $c_A$, as described in Fig.~\ref{fig:4H2g}. Let us test this method on the tree-level triplet contributions. The diagram in Fig.~\ref{fig:triplet}(b) gives for the hypercharge 0 triplet 
\begin{equation}
\label{eq:cA}
  c_A^{(0)} = 4 \frac{\kappa^2}{M^4}.
\end{equation}
Since $\varphi^a$ has no hypercharge, $c_B^{(0)} = 0$ and Eq.~(\ref{eq:cA}) agrees with Eq.~(\ref{eq:aT0}). Analogous computation for the charged triplet yields Eq.~(\ref{eq:aT1}). 

Yet another way of obtaining Eqs.~(\ref{eq:aT0}) and (\ref{eq:aT1}) is by matching the coefficient of $O_T$ in the background of the Higgs field. Expanding the Lagrangians (\ref{eq:Ltriplet0}) and (\ref{eq:Ltriplet1}) around the Higgs vev gives a linear term for the triplet field. The linear term forces a vev for the triplet, which in turn gives extra contributions to the masses of the gauge bosons. One needs to compare the mass terms for the gauge bosons with the gauge boson masses arising from $O_T$ in the Higgs background discriminating against contributions from other operators of dimension 6 that do not violate the custodial symmetry. This can be done, for example, by calculating the difference between the mass terms for $A^3_\mu$ and $A^1_\mu$, which we used in Sec.~\ref{sec:ST}. 

Note that Eqs.~(\ref{eq:aT0}) and (\ref{eq:aT1}) exhibit decoupling even if $\kappa \propto M$. It is certainly natural to assume that parameters of mass dimension 1 scale proportionately to large masses in the theory.  Here, one can assume that $\kappa \propto M$. Even with such a scaling, one does not expect non-decoupling effects of higher-dimensional operators similar to the non-decoupling we observed when dimensionless quantities scale proportionately to large masses. In perturbation theory, the amplitudes depend on positive powers of the couplings. Thus, whenever coupling constants have positive mass dimensions, the coefficients of higher-dimensional operators must be suppressed by a power of the heavy particle masses larger than the dimension of the coupling constants. Obviously, this argument has nothing to do with tree-level perturbation theory. In the next section, we are going to examine two types of one-loop contributions to the $T$ parameter. One contribution is proportional to $\frac{\kappa^2}{M^4}$ and another one proportional to $\frac{\kappa^4}{M^6}$. Both contributions vanish in the limit $\kappa \propto M \rightarrow \infty$.

\subsection{One-loop level}
\label{sec:tripletloop}

We now turn to the one-loop contributions of the scalar triplets. We are going to discuss the effects of the hypercharge-neutral triplet only, but there is no qualitative distinction between the charged and the neutral cases. We will not present a complete analysis of all one-loop effects. We will calculate certain classes of diagrams chosen such that it is clear that in the effective theory the triplet contributions to the $T$ parameter decouple. 

As usual in an effective theory, log-enhanced contributions come from RG running and terms without large logs arise from matching. We matched the theory with the triplet to the SM and found that $a_T = - \frac{2 \kappa^2}{M^4}$ at tree level. We will omit the superscript $(0)$ for $a_T$ since we will only deal with the neutral triplet in this section. There are two types of diagrams that correct $O_T$ at one loop: gauge boson exchanges and Higgs quartic interactions. Schematically, these give either $a_T\sim \frac{g^2}{(4 \pi)^2} \frac{\kappa^2}{M^4} \log(\frac{m_h}{M})$ or $a_T\sim \frac{\lambda}{(4 \pi)^2} \frac{\kappa^2}{M^4} \log(\frac{m_h}{M})$, where $m_h$ is the Higgs mass and $\lambda$ is the Higgs quartic coupling constant. Neglecting the masses of the SM fields, compared to $M$, the dimensionless couplings in the SM cannot alter the proportionality of $a_T$ to $\frac{\kappa^2}{M^4}$ through the RG running.  Hence, it is clear that the log-enhanced terms decouple in the limit $\kappa \propto M \rightarrow \infty$. Moreover, there is no contribution to the running of $O_T$ from two insertions of $O_T$ when the masses of the $SM$ fields are neglected. Two insertions of $O_T$ in the effective theory yield a coefficient proportional to $\frac{\kappa^4}{M^8}$, which could give $O_T$ only when multiplied by the mass squared of a SM field, for example it could give $a_T \sim m_h^2 \frac{\kappa^4}{M^8}$. This term is additionally suppressed by $\frac{m_h^2}{M^2}$ compared to the terms we will consider next.~\footnote{There is also an RG contribution of order $a_T\sim \frac{1}{(4 \pi)^2} \frac{\kappa^4}{M^6} \log(\frac{m_h}{M})$ arising from one insertion of $O_T$ and one insertion of $\frac{\kappa^2}{2 M^2} (H^\dagger H)^2$ that one also obtains from tree-level matching. This contribution is not distinguishable at low energies from $a_T\sim\frac{\lambda}{(4 \pi)^2} \frac{\kappa^2}{M^4} \log(\frac{M}{m_h})$ since both terms arise from the same Higgs quartic coupling.} 

The tree-level result is also modified by the Higgs wave function renormalization due to the triplet exchange. Straightforward calculation gives $(1+\frac{3 \kappa^2}{2 (4 \pi)^2 M^2}) D^\mu H^\dagger D_\mu H$ for the Higgs kinetic energy in the effective theory. This gives another contribution of order $\frac{\kappa^4}{M^6}$ without any log enhancement. 

\begin{figure}[htb]
\begin{center}
\begin{picture}(422,163) (59,-28)
    \SetWidth{2.0}
    \Line[dash,dashsize=6](180,33)(230,3)
    \Line[dash,dashsize=6](230,3)(180,-27)
    \SetWidth{3.0}
    \Line[dash,dashsize=2](200,-16)(200,22)
    \SetWidth{2.0}
    \Line[dash,dashsize=6](270,3)(300,33)
    \Line[dash,dashsize=6](270,3)(300,-27)
    \SetWidth{3.0}
    \Line[dash,dashsize=2](230,3)(270,3)
    \SetWidth{2.0}
    \Line[dash,dashsize=6](220,103)(250,133)
    \Line[dash,dashsize=6](220,103)(250,73)
    \Line[dash,dashsize=6](170,133)(220,103)
    \Line[dash,dashsize=6](220,103)(170,73)
    \SetWidth{3.0}
    \Line[dash,dashsize=2](190,84)(190,122)
    \SetWidth{2.0}
    \Line[dash,dashsize=6](60,33)(160,33)
    \Line[dash,dashsize=6](60,-27)(160,-27)
    \SetWidth{3.0}
    \Line[dash,dashsize=2](80,33)(80,-27)
    \Line[dash,dashsize=2](140,33)(140,-27)
    \SetWidth{2.0}
    \Arc[dash,dashsize=6](400,3)(20,180,540)
    \SetWidth{3.0}
    \Line[dash,dashsize=2](380,3)(348,3)
    \SetWidth{2.0}
    \Line[dash,dashsize=6](350,3)(320,33)
    \Line[dash,dashsize=6](350,3)(320,-27)
    \Line[dash,dashsize=6](450,3)(480,33)
    \Line[dash,dashsize=6](450,3)(480,-27)
    \SetWidth{3.0}
    \Line[dash,dashsize=2](420,3)(450,3)
    \SetWidth{2.0}
    \Line[dash,dashsize=6](368,102)(398,132)
    \Line[dash,dashsize=6](368,102)(398,72)
    \Arc[dash,dashsize=6](348,103)(20,180,540)
    \SetWidth{3.0}
    \Line[dash,dashsize=2](328,103)(299,103)
    \SetWidth{2.0}
    \Line[dash,dashsize=6](300,103)(270,133)
    \Line[dash,dashsize=6](300,103)(270,73)
        \Text(210,60)[lb]{$(1)$}
        \Text(310,60)[lb]{$(2)$}
        \Text(105,-44)[lb]{$(3)$} 
         \Text(245,-44)[lb]{$(4)$} 
          \Text(395,-44)[lb]{$(5)$} 
  \end{picture}
\end{center}
\caption{\label{fig:4Hloop}  Diagrams that contribute to the $T$ parameter at orders $\frac{\kappa^2}{M^4} \lambda$ (top) and $\frac{\kappa^4}{M^6}$ (bottom). The short dashed lines represent massive triplet fields, while the lines with long dashes represent the Higgs doublet.}
\end{figure}
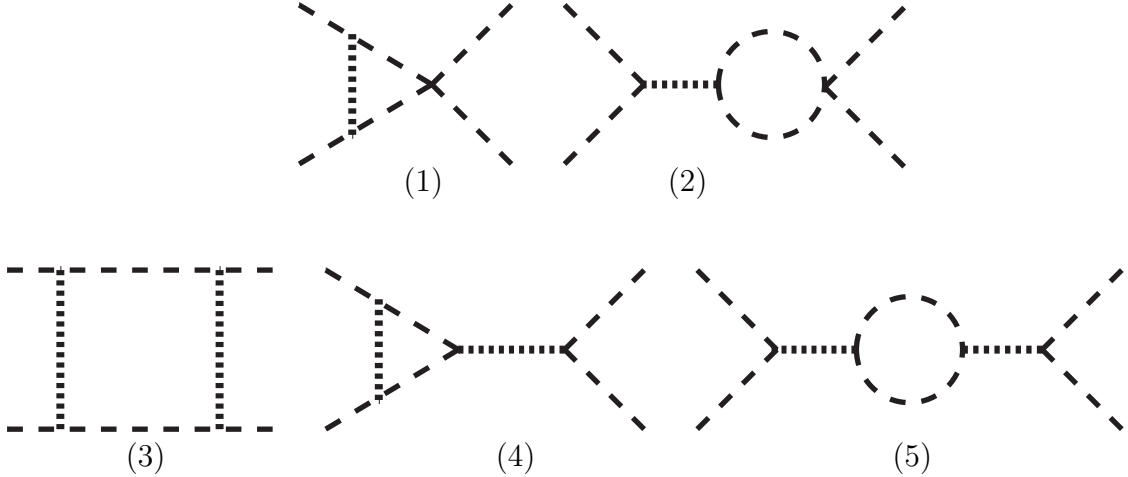

We will now discuss two cases of matching contributions. To gain experience with less complex calculations first, we will start with diagrams that give $a_T\sim \frac{\lambda}{(4 \pi)^2} \frac{\kappa^2}{M^4}$. Then we compute terms proportional to $\frac{1}{(4 \pi)^2} \frac{\kappa^4}{M^6}$. The corresponding diagrams are shown in Fig.~\ref{fig:4Hloop}.
For discussion of decoupling, diagrams (3), (4), and (5) in Fig.~\ref{fig:4Hloop} are the most interesting. These diagrams have the highest power of the cubic coupling $\kappa$ one can get at one loop, so one expects that these are the most important when $\kappa$ is large. However, since the $T$ parameter corresponds to an operator of dimension 6, these diagrams are proportional to $\frac{\kappa^4}{M^6}$. Of course, this dimensional argument is not particular to the one-loop approximation.

To perform one-loop matching we will not work directly with the diagrams in Fig.~\ref{fig:4Hloop}, but instead extract the coefficients of the terms $B_\mu^2 |H_1|^4$ and $(A^1_\mu)^2 |H_1|^4$. This is the second method of calculating the $T$ parameter we used in Sec.~\ref{sec:triplettree}. Extracting the coefficient of $\left|H^\dagger \partial_\mu H\right|^2$ is actually more difficult because it depends on the momenta of the external states. Keeping external momenta makes loop calculations more complicated. An additional complication is that all the diagrams in Fig.~\ref{fig:4Hloop} are IR divergent. This means that one cannot simply expand the propagators around the zero values of external momenta and then retain terms quadratic in those momenta. 

\begin{figure}[htb]
\begin{center}
  \begin{picture}(404,227) (18,-8)
    \SetWidth{2.0}
    \Line[dash,dashsize=6](40,187)(140,187)
    \Line[dash,dashsize=6](40,127)(140,127)
    \SetWidth{3.0}
    \Line[dash,dashsize=2](60,187)(60,127)
    \Line[dash,dashsize=2](120,187)(120,127)
    \Text(80,107)[lb]{(a)}
    \SetWidth{2.0}
    \Photon(90,187)(60,217){4.5}{4}
    \Photon(90,187)(120,217){4.5}{4}
    \Line[dash,dashsize=6](170,187)(270,187)
    \Line[dash,dashsize=6](170,127)(270,127)
    \SetWidth{3.0}
    \Line[dash,dashsize=2](190,187)(190,127)
    \Line[dash,dashsize=2](250,187)(250,127)
    \Text(210,107)[lb]{(b)}
    \SetWidth{2.0}
    \Photon(232,187)(262,217){4.5}{4}
    \Photon(210,187)(180,217){4.5}{4}
    \SetWidth{2.0}
    \Line[dash,dashsize=6](40,67)(140,67)
    \Line[dash,dashsize=6](40,7)(140,7)
    \SetWidth{3.0}
    \Line[dash,dashsize=2](60,67)(60,7)
    \Line[dash,dashsize=2](120,67)(120,7)
    \Text(80,-13)[lb]{(d)}
    \SetWidth{2.0}
    \Line[dash,dashsize=6](300,187)(400,187)
    \Line[dash,dashsize=6](300,127)(400,127)
    \SetWidth{3.0}
    \Line[dash,dashsize=2](320,187)(320,127)
    \Line[dash,dashsize=2](380,187)(380,127)
    \Text(340,107)[lb]{(c)}
    \SetWidth{2.0}
    \Photon(320,157)(290,177){4.5}{4}
    \Photon(320,157)(290,137){4.5}{4}
    \SetWidth{2.0}
    \Photon(90,67)(120,97){4.5}{4}
    \Photon(60,37)(20,37){4.5}{4}
    \SetWidth{2.0}
    \Line[dash,dashsize=6](170,67)(270,67)
    \Line[dash,dashsize=6](170,7)(270,7)
    \SetWidth{3.0}
    \Line[dash,dashsize=2](190,67)(190,7)
    \Line[dash,dashsize=2](250,67)(250,7)
    \Text(210,-13)[lb]{(e)}
    \SetWidth{2.0}
    \Photon(190,27)(150,27){4.5}{4}
    \SetWidth{2.0}
    \Photon(190,47)(150,47){4.5}{4}
    \Line[dash,dashsize=6](300,67)(400,67)
    \Line[dash,dashsize=6](300,7)(400,7)
    \SetWidth{3.0}
    \Line[dash,dashsize=2](320,67)(320,7)
    \Line[dash,dashsize=2](380,67)(380,7)
    \Text(340,-13)[lb]{(f)}
    \SetWidth{2.0}
    \Photon(320,37)(280,37){4.5}{4}
     \SetWidth{2.0}
    \Photon(420,37)(380,37){4.5}{4}
  \end{picture}
\end{center}
\caption{\label{fig:4H2gloop} Diagrams with two gauge bosons obtained from diagram $(3)$ in Fig.~\protect{\ref{fig:4Hloop}}. }
\end{figure}
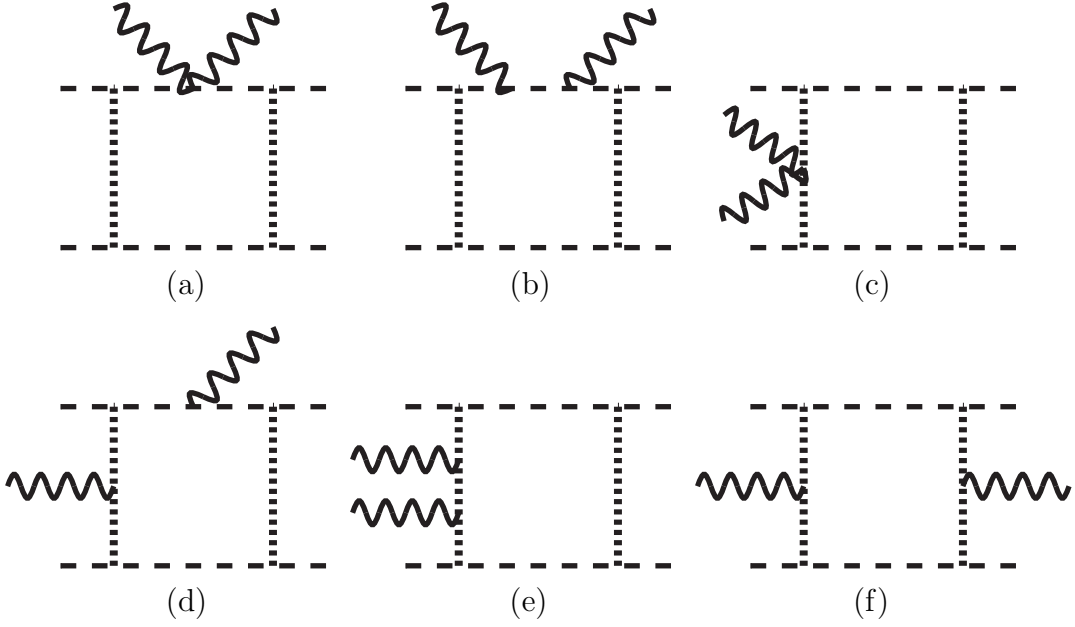

To extract the coefficients of the terms $B_\mu^2 |H_1|^4$ and $(A^1_\mu)^2 |H_1|^4$ we attach two gauge bosons in all possible ways to the internal lines of the diagrams in  Fig.~\ref{fig:4Hloop} and set all the external momenta to zero. The loop integrals are much simpler to compute, but the price of this approach is proliferation of diagrams. The diagrams with different ways of attaching the gauge bosons are depicted in Fig.~\ref{fig:4H2gloop}.
The diagrams in Fig.~\ref{fig:4H2gloop} correspond to different ways of attaching gauge bosons to diagram $(3)$ in Fig.~\ref{fig:4Hloop}. Of course, we consider all possible ways of attaching two external gauge bosons to the remaining diagrams in Fig.~\ref{fig:4Hloop}. These are completely analogous to the ones drawn in Fig.~\ref{fig:4H2gloop}, except that for diagrams $(1)$ and $(2)$ in Fig.~\ref{fig:4Hloop} there is no corresponding diagram (f) because the Higgs quartic vertex contains no gauge bosons. Fig.~\ref{fig:4H2gloop} does not show all possible permutations of attaching photons, but representative diagrams. For example, diagram (a) represents two diagrams where a pair of gauge bosons is attached to either of the two internal Higgs lines. Diagram (b) represents three diagrams in which two gauge bosons are attached to either of the two Higgs lines or one gauge boson is attached to each line, etc.   

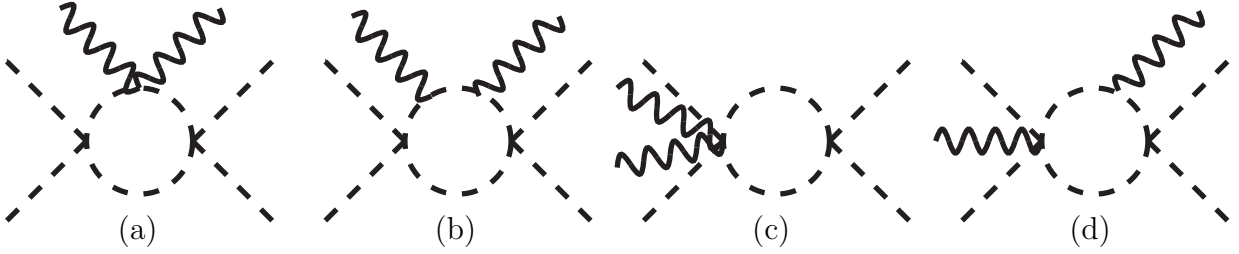
\begin{figure}[htb]
\begin{center}
\begin{picture}(463,88)(32,-7)
    \SetWidth{2.0}
    \SetColor{Black}
    \Line[dash,dashsize=6](30,58)(60,28)
    \Line[dash,dashsize=6](60,28)(30,-2)
    \Arc[dash,dashsize=6](80,28)(20,0,360)
    \Line[dash,dashsize=6](100,28)(130,58)
    \Line[dash,dashsize=6](100,28)(130,-2)
    \Photon(80,48)(110,78){4.5}{4}
    \Photon(80,48)(50,78){4.5}{4}
    \Line[dash,dashsize=6](150,58)(180,28)
    \Line[dash,dashsize=6](180,28)(150,-2)
    \Arc[dash,dashsize=6](200,28)(20,0,360)
    \Line[dash,dashsize=6](220,28)(250,58)
    \Line[dash,dashsize=6](220,28)(250,-2)
    \Line[dash,dashsize=6](270,58)(300,28)
    \Line[dash,dashsize=6](300,28)(270,-2)
    \Arc[dash,dashsize=6](320,28)(20,0,360)
    \Line[dash,dashsize=6](340,28)(370,58)
    \Line[dash,dashsize=6](340,28)(370,-2)
    \Line[dash,dashsize=6](390,58)(420,28)
    \Line[dash,dashsize=6](420,28)(390,-2)
    \Arc[dash,dashsize=6](440,28)(20,0,360)
    \Line[dash,dashsize=6](460,28)(490,58)
    \Line[dash,dashsize=6](460,28)(490,-2)
    \Photon(208,45)(238,75){4.5}{4}
    \Photon(190,45)(160,75){4.5}{4}
    \Photon(300,28)(260,48){4.5}{4}
    \Photon(300,28)(260,18){4.5}{4}
    \Photon(449,47)(479,77){4.5}{4}
    \Photon(420,28)(380,28){4.5}{4}
    \Text(72,-12)[lb]{(a)}
    \Text(192,-12)[lb]{(b)}
    \Text(312,-12)[lb]{(c)}
    \Text(432,-12)[lb]{(d)}
  \end{picture}
\end{center}
\caption{\label{fig:4H2gEff}  Diagrams in the effective theory. The dashed lines represent the Higgs doublet.}
\end{figure}

The diagrams in Fig.~\ref{fig:4H2gloop} are still IR divergent. The IR divergences must be matched by the loop diagrams in the effective theory using the matching coefficients obtained at tree level. The effective theory diagrams are shown in Fig.~\ref{fig:4H2gEff}. We will compare diagrams in the full theory with the corresponding diagrams in the effective theory to make sure that the IR divergences match. Diagrams (a) through (d) in the full theory correspond to diagrams (a) through (d) in the effective theory, respectively. The full theory diagrams (e) and (f) are finite in the IR. The cancellations of IR divergences happens diagram by diagram, so we check this in every case. Both diagrams (c) and (e) in the full theory correspond to diagram (c) in the effective theory, but since diagram (e) is finite we evaluate it it separately. 

The full theory diagrams involve integrals of the form
\begin{eqnarray}
\label{eq:Inm}
  I_{n,m} & = & \int \frac{d^d k}{(2\pi)^d} \frac{1}{(k^2)^n (k^2-M^2)^m} 
                   = \int \frac{d^d k}{(2\pi)^d} \int dx \frac{\Gamma(n+m) (1-x)^{n-1} x^{m-1}}{\Gamma(n) \Gamma(m) (k^2 - x M^2)^{n+m}} \nonumber \\
 &=& \frac{i (-1)^{n+m}}{(4 \pi)^{2 - \epsilon} (M^2)^{n+m-2+\epsilon}} \frac{\Gamma(n+m-2 +\epsilon)}{\Gamma(n)\Gamma(m)} \int dx \frac{(1-x)^{n-1} x^{m-1}}{x^{n+m-2+\epsilon}}  \nonumber \\
  &=&  \frac{i (-1)^{n+m}}{(4 \pi)^{2 - \epsilon} (M^2)^{n+m-2+\epsilon}} \frac{\Gamma(n+m-2 +\epsilon)\Gamma(2-n-\epsilon)}{\Gamma(m)\Gamma(2-\epsilon)},
\end{eqnarray} 
where in the last line we performed the Feynman parameter integral in $d=4 - 2 \epsilon$ dimensions using the standard Euler beta function integral. The IR divergences of the integrals with $n\geq2$ manifest as poles of the $\Gamma$ function. As $\epsilon\rightarrow 0$, $\Gamma(2-n-\epsilon)$ is divergent for $n\geq2$. The poles of  $\Gamma(2-n-\epsilon)=\Gamma(d/2-n)$ occur in $d=2n, 2n-2, 2n-4, \ldots$, which is characteristic of an IR divergence. We will also need
\begin{equation}
\label{eq:Inmkk}
 I^{\mu\nu}_{n,m} = \int \frac{d^d k}{(2\pi)^d} \frac{k^\mu k^\nu}{(k^2)^n (k^2-M^2)^m} =  \frac{i (-1)^{n+m-1} g^{\mu \nu}}{(4 \pi)^{2 - \epsilon} (M^2)^{n+m-3+\epsilon}} \frac{\Gamma(n+m-3 +\epsilon)\Gamma(3-n-\epsilon)}{2 \Gamma(m)\Gamma(3-\epsilon)},
 \end{equation}
which is IR divergent when $n\geq3$. The expressions in Eqs.~(\ref{eq:Inm}) and (\ref{eq:Inmkk}) apply only when $m>0$. 

The integrals $I_{n,0}$ and $ I^{\mu\nu}_{n,0}$ vanish in dimensional regularization since there is no mass scale to make up for the dimension of the integral. However, $I_{2,0}$  and $ I^{\mu\nu}_{3,0}$ appear in the full and effective theories and these integrals contain both the IR and UV divergences. Since both the IR and UV divergences manifest as $\frac{1}{\epsilon}$ poles the two divergences cancel for $I_{2,0}$  and $ I^{\mu\nu}_{3,0}$ in dimensional regularization. To show explicitly that the IR divergences are identical in the full and effective theories we rewrite 
\begin{equation}
\label{eq:I20}
 I_{2,0}= \int \frac{d^d k}{(2\pi)^d} \frac{k^2-M^2}{(k^2)^2 (k^2-M^2)} = I_{1,1} - M^2 I_{2,1}\
   = \frac{i \, \left[\Gamma(\epsilon) \Gamma(1-\epsilon)+ \Gamma(-\epsilon) \Gamma(1+\epsilon)\right]}{(4 \pi)^{2 - \epsilon} (M^2)^{\epsilon} \, \Gamma(2-\epsilon)}.
\end{equation}
Of course,  $\Gamma(\epsilon) \Gamma(1-\epsilon)+ \Gamma(-\epsilon) \Gamma(1+\epsilon)=0$ which can be shown by multiplying by $\epsilon$ and using $z \Gamma(z)=\Gamma(z+1)$. However, by rewriting the integral we separated the UV and IR divergences which are encoded in $\Gamma(\epsilon)$ and  $\Gamma(-\epsilon)$, respectively. Similarly, we can rewrite
\begin{equation}
\label{eq:I30kk}
 I_{3,0}^{\mu\nu}= I_{2,1}^{\mu\nu} - M^2 I_{3,1}^{\mu\nu}
   = \frac{i g^{\mu \nu}}{(4 \pi)^{2 - \epsilon} (M^2)^{\epsilon}} \frac{\Gamma(\epsilon) \Gamma(1-\epsilon)+ \Gamma(-\epsilon) \Gamma(1+\epsilon)}{2 \Gamma(3-\epsilon)}.
\end{equation}
For $n>2$, rewriting $I_{n,0}$ using the trick described above does not yield anything useful because the integrals are UV convergent. Thus, dimensional regularization sets the IR divergence to zero. 

We are almost ready to do the calculation, except that in the effective theory we need all terms of order $\frac{\kappa^2}{M^2}$ and $\frac{\kappa^2}{M^4}$. We have calculated the coefficient of  $O_T$ in the previous section, but neglected all other operators. Integrating out $\varphi^a$ at tree level gives
\begin{equation}
\label{eq:Lc1c2}
  {\mathcal L}_{eff} = \frac{c_1}{4} (H^\dagger H)^2 + c_2 \left[ |H^\dagger D_\mu H|^2 + \frac{1}{4} D^2 H^\dagger H H^\dagger H + \frac{1}{4} H^\dagger D^2 H H^\dagger H - \frac{1}{2} D_\mu H^\dagger D^\mu H H^\dagger H \right],
\end{equation}
where each derivative acts only on the field immediately next to it and not on all the fields to the right of the derivative. The coefficients are $c_1=2 \frac{\kappa^2}{M^2}$ and $c_2=-2 \frac{\kappa^2}{M^4}$. The first term in Eq.~(\ref{eq:Lc1c2}) is of the same form as the ordinary Higgs quartic coupling in the SM\@. The coefficient of the quartic term in the effective theory is $-\lambda + c_1$, where $\lambda$ is the quartic coupling in the full theory above $M$. Our convention for the quartic coupling is such that, at tree level, $V(H)=\frac{\lambda}{4} (H^\dagger H - \frac{v^2}{2})^2$. It may seem odd that in the effective theory we care about terms that do not violate the custodial symmetry, for example $D_\mu H^\dagger D^\mu H H^\dagger H$. In the following calculation we will be extracting coefficients of all operators with four Higgs fields and two gauge bosons, and not just the coefficient of $O_T$. If we only cared about the cancellation of IR divergences for $O_T$ we may not need to keep all of the operators in the effective theory. However, it is a very useful consistency check on the calculation to be able to show cancellation of IR divergences for individual diagrams. 

With the integrals in Eqs.~(\ref{eq:Inm}) through (\ref{eq:I30kk}) at hand, the problem is reduced to combinatorics. We will show a couple of examples in detail and then present the results. To provide further checks we calculate separately the amplitudes depending on the flow of the scalar field number. Since the Higgs field is complex, we can assign arrows indicating the direction of the flow of the scalar field. We will separate diagrams in which the arrows on the Higgs lines in Fig.~\ref{fig:4H2gloop} are in the same direction from the ones in which the arrows are in the opposite directions. We will denote the amplitudes in the full theory by $F$ and in the effective theory by $E$ adding the superscripts ${\overrightarrow{\rightarrow}}$ and ${\overrightarrow{\leftarrow}}$ to indicate the arrow directions. The subscripts will indicate the topology of the diagram, as shown in Figs.~\ref{fig:4H2gloop} and \ref{fig:4H2gEff}, and the type of the gauge fields: either $B_\mu$ or $A^1_\mu$.

As our first example, we compute diagram $(1)$ in  Fig.~\ref{fig:4Hloop} with the Higgs lines in the same direction and  two $B_\mu$ fields coupling at the same point to the Higgs line, as represented in diagram (a) in  Fig.~\ref{fig:4H2gloop}.
\begin{equation}
  F_{B(a)}^{\overrightarrow{\rightarrow}} = 4 (i\kappa)^2 (-i \lambda) g'^2 \frac{i g^{\mu \nu}}{2} \int \frac{d^d k}{(2\pi)^d} \frac{i^3}{(k^2)^3} \frac{i}{k^2-M^2} 
                  = -2 \kappa^2 \lambda g'^2 g^{\mu \nu} I_{3,1},
\end{equation}
where the factor of four comes from exchanging the two external lines on the left due to Bose statistics and from two possible directions for the arrows. The reversal of the arrow directions corresponds to exchanging the external $H_1$ fields with the $H_1^*$'s.  The remaining factors are the coupling constants for the vertices and the propagators, where we set all the external momenta to zero. Note that  diagram $(2)$ in Fig.~\ref{fig:4H2gloop} is identically zero when the arrow directions are parallel because $\varphi^a$ couples to $H$ and $H^\dagger$. In the effective theory, the corresponding diagram gives
\begin{equation}
 E_{B(a)}^{\overrightarrow{\rightarrow}} = 2 (i c_2) (-i\lambda) g'^2 \frac{i g^{\mu \nu}}{2}   \int \frac{d^d k}{(2\pi)^d} \frac{i^3}{(k^2)^3} (-k^2) = - c_2 \lambda g'^2 g^{\mu \nu} I_{2,0},
\end{equation}
where the factor of two is due to the reversal of arrow directions, or equivalently due to exchanging the $c_2$ and $\lambda$ interaction vertices. The factor $-k^2$ arises from the two-derivative terms in Eq.~(\ref{eq:Lc1c2}). The IR divergent part of the difference $F_{B(a)}^{\overrightarrow{\rightarrow}} - E_{B(a)}^{\overrightarrow{\rightarrow}}$ is proportional to 
\begin{eqnarray}
  \Gamma(2+\epsilon) \Gamma(-1-\epsilon) + \Gamma(1+\epsilon) \Gamma(-\epsilon)&=&
       (1+\epsilon) \Gamma(1+\epsilon) \Gamma(-1-\epsilon) + \Gamma(1+\epsilon) \Gamma(-\epsilon) \nonumber \\
       &= & -  \Gamma(1+\epsilon) \Gamma(-\epsilon) + \Gamma(1+\epsilon) \Gamma(-\epsilon) = 0. \nonumber
\end{eqnarray}
In this case the IR divergent terms cancel exactly, but in some cases the difference between the diagrams is finite. Since $B_\mu$ does not couple to $\varphi^a$, diagrams (c) through (f) are absent. These diagrams vanish in the effective theory because there is no term proportional to $B_\mu$ in the effective Lagrangian in  Eq.~(\ref{eq:Lc1c2}). This is expected since the effective Lagrangian comes from integrating out $\varphi^a$, but is not apparent as the covariant derivatives in (\ref{eq:Lc1c2}) contain the $B_\mu$ field.

As the second detailed example, we compute $F_{A(c)}^{\overrightarrow{\leftarrow}}$ and $F_{A(c)}^{\overrightarrow{\leftarrow}}$ for the $\frac{\kappa^2}{M^4}\lambda$ contributions that is diagrams $(1)$ and $(2)$ in Fig.~\ref{fig:4Hloop}. Both diagrams in the full theory contribute, and since diagram $(2)$ is IR divergent it needs to be accounted for to ensure cancellation of divergences. 
\begin{equation} 
  F_{A(c)}^{\overrightarrow{\leftarrow}} = - 4 (2 + 1)  \kappa^2 \lambda g^2 g^{\mu \nu} I_{2,2} - 4 (2-1) \frac{\kappa^2 \lambda g^2 g^{\mu \nu}}{M^4} I_{2,0},
\end{equation}
where the factors of four are from exchanges of the external $H_1$ lines and exchanges of  $H_1^*$'s. The remaining factors come from the couplings. In diagram $(1)$, the triplet components can be either $\varphi^2$ or $\varphi^3$, which is responsible for the $2 + 1$ factor. In diagram $(2)$, the factor of $2-1$ comes from $H_1$'s and $H_2$'s exchanged in the loop. In each of these diagrams, $2\pm1$ originates from the relative factor of two between the $|H_1|^4$ and the $| H_1 H_2|^2$ couplings in the Higgs quartic term. In the effective theory there is only one diagram. To calculate it one needs to extract the coefficient of $|H_1|^4 (A^1_\mu)^2$ in the effective Lagrangian (\ref{eq:Lc1c2}), which then gives
\begin{equation} 
  E_{A(c)}^{\overrightarrow{\leftarrow}} =8 c_2 \lambda g^2 g^{\mu \nu} I_{2,0}.
\end{equation}
The difference between the IR divergent parts of the full and effective theory amplitudes is proportional to
\begin{eqnarray}
 -3 \Gamma(1+\epsilon) \Gamma(-\epsilon) - \Gamma(2+\epsilon) \Gamma(-\epsilon) + 4 \Gamma(1+\epsilon) \Gamma(-\epsilon)&= &
  -  \epsilon \Gamma(1+\epsilon) \Gamma(-\epsilon) \nonumber \\
  &= &  \Gamma(1+\epsilon) \Gamma(1-\epsilon),
 \end{eqnarray}
 which is finite when $\epsilon \rightarrow 0$.
 
 \begin{table}[thb!]
\vspace{-10pt}
\begin{displaymath}
\begin{array}{|r|c|c|c|c|}
\hline
   & F^{\overrightarrow{\rightarrow}}  & E^{\overrightarrow{\rightarrow}} &  F^{\overrightarrow{\leftarrow}}   &  E^{\overrightarrow{\leftarrow}}  \\ \hline \hline
 B(a) & - 2 \kappa^2 \lambda g^{\mu \nu} I_{3,1}  &  -  c_2 \lambda g^{\mu \nu}  I_{2,0} & - 8 \kappa^2 \lambda g^{\mu \nu} I_{3,1} &  -  4 c_2 \lambda g^{\mu \nu}  I_{2,0} \\[1pt]
 B(b) & 4 \kappa^2 \lambda I_{4,1}^{\mu \nu}  & 2 c_2 \lambda I_{3,0}^{\mu \nu}& 48 \kappa^2 \lambda I_{4,1}^{\mu \nu}  & 24 c_2 \lambda I_{3,0}^{\mu \nu}\\[1pt]
 A(a)& - 2 \kappa^2 \lambda g^{\mu \nu} I_{3,1}  &  -  c_2 \lambda g^{\mu \nu}  I_{2,0}  & - 8 \kappa^2 \lambda g^{\mu \nu} I_{3,1} &  -  4 c_2 \lambda g^{\mu \nu}  I_{2,0}  \\[1pt]
 A(b)& 8 \kappa^2 \lambda I_{4,1}^{\mu \nu}  & 4  c_2 \lambda I_{3,0}^{\mu \nu}& 52 \kappa^2 \lambda I_{4,1}^{\mu \nu}  & 26 c_2 \lambda I_{3,0}^{\mu \nu} \\[1pt]
 A(c)& - 4 \kappa^2 \lambda g^{\mu \nu} I_{2,2}  &  2 c_2 \lambda g^{\mu \nu}  I_{2,0}  &-4  \kappa^2\lambda g^{\mu\nu} (3 I_{2,2}+\frac{1}{M^4} I_{2,0} ) & 8 c_2 \lambda I_{2,0} \\[1pt]
 A(d)&  16 \kappa^2 \lambda I_{3,2}^{\mu \nu}  & -8  c_2 \lambda I_{3,0}^{\mu \nu} & 48 \kappa^2 \lambda I^{\mu\nu}_{3,2} & - 24 c_2 \lambda I_{3,0}^{\mu \nu}  \\[1pt]
 A(e)& 16  \kappa^2 \lambda I_{2,3}^{\mu \nu}& - & 48 \kappa^2 \lambda I_{2,3}^{\mu \nu} & - \\[1pt]  \hline \hline
 B(a) & - 2 \kappa^4 g^{\mu \nu} I_{3,2}  & c_1 c_2  g^{\mu \nu} I_{2,0}  &- 2 \kappa^4 g^{\mu\nu} (5 I_{3,2} + \frac{2}{M^2} I_{3,1}) & 4 c_1 c_2 I_{2,0} \\[1pt]
 B(b) & 4 \kappa^4 I_{4,2}^{\mu \nu} & -2 c_1 c_2  I_{3,0}^{\mu \nu}  & 12 \kappa^4 (5 I_{4,2}^{\mu\nu} + \frac{2}{M^2} I_{4,1}^{\mu\nu}) & -24 c_1 c_2  I_{3,0}^{\mu \nu} \\[1pt]
 A(a)&- 2 \kappa^4 g^{\mu \nu} I_{3,2}  & c_1 c_2  g^{\mu \nu} I_{2,0}  &  - 2 \kappa^4 g^{\mu\nu} (5 I_{3,2} + \frac{2}{M^2} I_{3,1}) & 4 c_1 c_2 I_{2,0} \\[1pt]
 A(b)& +8 \kappa^4 I_{4,2}^{\mu \nu} & - 4 c_1 c_2  I_{3,0}^{\mu \nu}  & 8 \kappa^4  (7 I_{4,2}^{\mu \nu}+ \frac{1}{M^2} I_{4,1}^{\mu \nu}) & -26 c_1 c_2  I_{3,0}^{\mu \nu}  \\[1pt]
 A(c)&  -8 \kappa^4 g^{\mu \nu} I_{2,3} &-2  c_1 c_2  g^{\mu \nu} I_{2,0} & 8 \kappa^4 g^{\mu \nu} (- 3 I_{2,3} + \frac{1}{M^4} I_{2,1} + \frac{2}{M^6} I_{2,0}) & -8 c_1 c_2  g^{\mu \nu} I_{2,0}  \\[1pt]
 A(d)& 32 \kappa^4  I_{3,3}^{\mu \nu}  &  8 c_1 c_2  I_{3,0}^{\mu \nu}  &  96  \kappa^4  I_{3,3}^{\mu \nu} &  24 c_1 c_2  I_{3,0}^{\mu \nu}\\[1pt]
 A(e)& 32  \kappa^4 I_{2,4}^{\mu \nu} & - & 96   \kappa^4 I_{2,4}^{\mu \nu} & - \\[1pt]
 A(f)&  32  \kappa^4 I_{2,4}^{\mu \nu} & - &  32  \kappa^4 I_{2,4}^{\mu \nu} & - \\[1pt]
 \hline
 \end{array}
\end{displaymath}
\caption{\label{table:coeff} The amplitudes corresponding to the diagrams in Figs.~\protect{\ref{fig:4H2gloop}} and \protect{\ref{fig:4H2gEff}}.  The rows correspond to different ways of attaching gauge boson lines as shown in the figures. $B$ and $A$ indicate the external gauge fields: either $B_\mu$ or $A^1_\mu$, respectively.  To save space, the gauge couplings are omitted. The diagrams with the $B_\mu$ fields are proportional to $g'^2$, while the ones with $A^1_\mu$ to $g^2$. The top part of this table lists the amplitudes proportional to $\frac{\kappa^2 \lambda}{M^4}$, while the bottom part proportional to $\frac{\kappa^4}{M^6}$. The columns give the full and effective theory amplitudes with either parallel or antiparallel Higgs lines.} 
\end{table}
The complete answer for all diagrams is presented in Table~\ref{table:coeff}. The IR divergences cancel in each row of the table between the two corresponding amplitudes, as we already described in the previous examples. Altogether, there are 24 cancellations of IR divergences that provide consistency checks on this calculation.  The full theory diagrams (e) and (f) are indeed IR finite, and there are no corresponding effective theory diagrams.

We can now extract the matching coefficients by calculating the differences between the full and effective theories. The coefficients $c_B$ and $c_A$ defined in Fig.~\ref{fig:4H2g} are 
\begin{eqnarray}
  c_B & = & \frac{\kappa^2 \lambda}{ (4 \pi)^2 M^4} \left(\frac{3}{\overline{\epsilon}} + \frac{19}{2} \right) +  \frac{\kappa^4}{ (4 \pi)^2 M^6} \left(-\frac{6}{\overline{\epsilon}} -23 \right),\\
  c_A & = & \frac{\kappa^2 \lambda}{(4 \pi)^2 M^4 } \left(\frac{5}{\overline{\epsilon}} + \frac{25}{2} \right) +  \frac{\kappa^4}{ (4 \pi)^2 M^6} \left(-\frac{2}{\overline{\epsilon}} -21 \right),
\end{eqnarray}
which finally gives 
\begin{equation}
\label{eq:T1loop}
  a_T= -\frac{3}{2}\frac{\kappa^2 \lambda}{ (4 \pi)^2 M^4}-   \frac{\kappa^4}{ (4 \pi)^2 M^6}+  \frac{6 \kappa^4}{ (4 \pi)^2 M^6},
\end{equation}
where the last term comes from the wave function renormalization of the tree-level term. In obtaining Eq.~(\ref{eq:T1loop}) we absorbed the $\frac{1}{\overline{\epsilon}}$ poles into counterterms using the $\overline{MS}$ prescription. These poles can be used to calculate the running of the $T$ operator in the effective theory. The renormalization scale has been set to $M$, so that the logarithms of $\frac{\mu}{M}$ are absent.

The numerical coefficients in Eq.~(\ref{eq:T1loop}) are not crucial for us. This calculation provided a thorough illustration of the methods we discussed in these notes. What is interesting is that the one-loop result exhibits decoupling in the limit $\kappa \propto M \rightarrow \infty$. There was no other possibility in the effective theory since this is guaranteed by power counting even without doing an explicit calculation. One might wonder if the effective theory reproduces properly the full theory. The cancellation of the IR divergences among various terms in Table~\ref{table:coeff} provides convincing evidence that it does. The results in Refs.~\cite{triplets1,triplets2} that motivated this calculation were obtained in the EW broken phase without using EFT methods. It is unlikely that the non-decoupling observed in Refs.~\cite{triplets1,triplets2} is a result of an algebraic error. One plausible reason may be the triplet correction to the Higgs mass term, which is proportional to $\frac{\kappa^2}{(4 \pi)^2}$. (This is another example of the quadratic sensitivity of the Higgs mass to the heavy scales, even though the diagram with the triplet exchange is only logarithmically divergent.) This contribution might creep into the Higgs vev calculation, but should be cancelled when the calculation is expressed in terms of the physical Higgs mass. Unfortunately, we have no firm argument as to why the two approaches disagree.  

\section*{Acknowledgments}
These notes are based on five lectures given at TASI during the summer of 2009. One of these lectures reviewed the Standard Model and since this topic is covered almost every summer, see for example Ref.~\cite{TASI-SM}, it is omitted here. I am grateful to the TASI organizers, especially Csaba Cs\'aki, Tom DeGrand,  and K.T.~Mahantappa, for a well designed and smoothly run program. I very much enjoyed lively reception of these lectures by the TASI participants. 

I am indebted to Walter Goldberger for discussions and comments on the manuscript, and to Zuhair Khandker for carefully inspecting the calculations and comments on the manuscript. This work was supported in part by the US Department of Energy under grant DE-FG02-92ER-40704.


\newpage

\addcontentsline{toc}{section}{References}
\thebibliography{99}

\bibitem{Weinberg}
 S.~Weinberg,
  Physica A {\bf 96}, 327 (1979).
    
\bibitem{WilsonKogut}
 K.~G.~Wilson and J.~B.~Kogut,
  Phys.\ Rept.\  {\bf 12} (1974) 75.
  
 \bibitem{A-C}
  T.~Appelquist and J.~Carazzone,
  Phys.\ Rev.\  D {\bf 11}, 2856 (1975).

\bibitem{gminus2}
J.~S.~Schwinger,
  Phys.\ Rev.\  {\bf 73}, 416 (1948);
P.~Kusch and H.~M.~Foley,
  Phys.\ Rev.\  {\bf 74}, 250 (1948).

\bibitem{EW}
 A.~Sirlin,
  Phys.\ Rev.\  D {\bf 22}, 971 (1980);
G.~Passarino and M.~J.~G.~Veltman,
  Nucl.\ Phys.\  B {\bf 160}, 151 (1979);
W.~F.~L.~Hollik,
  Fortsch.\ Phys.\  {\bf 38}, 165 (1990);
  J. Erler and P. Langacker in C.~Amsler {\it et al.}  [Particle Data Group],
  Phys.\ Lett.\  B {\bf 667}, 1 (2008), and references therein. 

 \bibitem{Aneesh}
A.~V.~Manohar,
   ``Effective field theories,''
  arXiv:hep-ph/9606222.
  
\bibitem{Ira}
I.~Z.~Rothstein,
  ``TASI lectures on effective field theories,''
  arXiv:hep-ph/0308266.

\bibitem{DBKaplan}
D.~B.~Kaplan,
  ``Effective field theories,''
  arXiv:nucl-th/9506035.

\bibitem{Walter}
W.~D.~Goldberger,
  ``Les Houches lectures on effective field theories and gravitational radiation,''
  arXiv:hep-ph/0701129;
W.~D.~Goldberger and I.~Z.~Rothstein,
  Phys.\ Rev.\  D {\bf 73}, 104029 (2006)
  [arXiv:hep-th/0409156].

\bibitem{Politzer}
  H.~D.~Politzer,
  Nucl.\ Phys.\  B {\bf 172}, 349 (1980).
  
  \bibitem{Georgi}
  H.~Georgi,
  Nucl.\ Phys.\  B {\bf 361}, 339 (1991).
  
  \bibitem{Dirac}P.~A.~M.~Dirac,
  Nature {\bf 139}, 323 (1937);
  Proc.\ Roy.\ Soc.\ Lond.\  A {\bf 165}, 199 (1938).
  
   \bibitem{'tHooft}
  G.~'t Hooft,
  ``Naturalness, Chiral Symmetry, And Spontaneous Chiral Symmetry Breaking,''
  NATO Adv.\ Study Inst.\ Ser.\ B Phys.\  {\bf 59}, 135 (1980).
    
\bibitem{Arzt}
  C.~Arzt,
  Phys.\ Lett.\  B {\bf 342}, 189 (1995)
  [arXiv:hep-ph/9304230].
 
\bibitem{NDA}
 A.~Manohar and H.~Georgi,
  Nucl.\ Phys.\  B {\bf 234}, 189 (1984).
  
\bibitem{SUSYNDA}
A.~G.~Cohen, D.~B.~Kaplan and A.~E.~Nelson,
  Phys.\ Lett.\  B {\bf 412}, 301 (1997)
  [arXiv:hep-ph/9706275].
  
\bibitem{PDG}
C.~Amsler {\it et al.}  [Particle Data Group],
  Phys.\ Lett.\  B {\bf 667}, 1 (2008).
  
 \bibitem{Buchmuller}
W.~Buchmuller and D.~Wyler,
Nucl.\ Phys.\ B {\bf 268}, 621 (1986).

\bibitem{GW}
 B.~Grinstein and M.~B.~Wise,
  Phys.\ Lett.\  B {\bf 265}, 326 (1991).
 
 \bibitem{EWCL}
T.~Appelquist and C.~W.~Bernard,
Phys.\ Rev.\ D {\bf 22}, 200 (1980);
A.~C.~Longhitano,
Phys.\ Rev.\ D {\bf 22}, 1166 (1980);
Nucl.\ Phys.\ B {\bf 188}, 118 (1981).

\bibitem{Wudka}
 J.~Wudka,
  Int.\ J.\ Mod.\ Phys.\  A {\bf 9}, 2301 (1994)
  [arXiv:hep-ph/9406205].

\bibitem{TASI-SM} 
S.~Willenbrock,
  ``Symmetries of the standard model,''
  arXiv:hep-ph/0410370.

\bibitem{STU}
M.~E.~Peskin and T.~Takeuchi,
Phys.\ Rev.\ D {\bf 46}, 381 (1992).

\bibitem{oblique}
M.~Golden and L.~Randall,
Nucl.\ Phys.\ B {\bf 361}, 3 (1991);
B.~Holdom and J.~Terning,
Phys.\ Lett.\ B {\bf 247}, 88 (1990);
M.~E.~Peskin and T.~Takeuchi,
Phys.\ Rev.\ Lett.\  {\bf 65}, 964 (1990);
G.~Altarelli and R.~Barbieri,
Phys.\ Lett.\ B {\bf 253}, 161 (1991).

\bibitem{SVVZ}
  M.~A.~Shifman, A.~I.~Vainshtein, M.~B.~Voloshin and V.~I.~Zakharov,
  Sov.\ J.\ Nucl.\ Phys.\  {\bf 30}, 711 (1979)
  [Yad.\ Fiz.\  {\bf 30}, 1368 (1979)].

\bibitem{rho}
A.~G.~Cohen, H.~Georgi and B.~Grinstein,
  Nucl.\ Phys.\  B {\bf 232}, 61 (1984);
M.~B.~Einhorn, D.~R.~T.~Jones and M.~J.~G.~Veltman,
  Nucl.\ Phys.\  B {\bf 191}, 146 (1981).

\bibitem{BPRS}
R.~Barbieri, A.~Pomarol, R.~Rattazzi and A.~Strumia,
  Nucl.\ Phys.\  B {\bf 703}, 127 (2004)
  [arXiv:hep-ph/0405040].
  
  \bibitem{Zprime}
  K.~S.~Babu, C.~F.~Kolda and J.~March-Russell,
  Phys.\ Rev.\  D {\bf 57}, 6788 (1998)
  [arXiv:hep-ph/9710441].
  
\bibitem{HS}
Z.~Han and W.~Skiba,
  Phys.\ Rev.\  D {\bf 71}, 075009 (2005)
  [arXiv:hep-ph/0412166].

\bibitem{MFV}
G.~D'Ambrosio, G.~F.~Giudice, G.~Isidori and A.~Strumia,
  Nucl.\ Phys.\  B {\bf 645}, 155 (2002)
  [arXiv:hep-ph/0207036].
  
\bibitem{Zhenyu}
  Z.~Han,
  Phys.\ Rev.\  D {\bf 73}, 015005 (2006)
  [arXiv:hep-ph/0510125];
  Z.~Han,
  AIP Conf.\ Proc.\  {\bf 903}, 435 (2007)
  [arXiv:hep-ph/0610302].

  \bibitem{eoms}
  C.~Grojean, W.~Skiba and J.~Terning,
  Phys.\ Rev.\  D {\bf 73}, 075008 (2006)
  [arXiv:hep-ph/0602154].
  
 \bibitem{examples}
 Z.~Han and W.~Skiba,
  Phys.\ Rev.\  D {\bf 72}, 035005 (2005)
  [arXiv:hep-ph/0506206];
M.~S.~Carena, E.~Ponton, J.~Santiago and C.~E.~M.~Wagner,
  Phys.\ Rev.\  D {\bf 76}, 035006 (2007)
  [arXiv:hep-ph/0701055];
S.~Mert Aybat and J.~Santiago,
  Phys.\ Rev.\  D {\bf 80}, 035005 (2009)
  [arXiv:0905.3032 [hep-ph]].
 
\bibitem{LEP-EW}
  [LEP Collaboration and \ldots],
  ``A Combination of preliminary electroweak measurements and constraints on
  the standard model,''
  arXiv:hep-ex/0312023.
   
\bibitem{Canada}
  C.~P.~Burgess, S.~Godfrey, H.~Konig, D.~London and I.~Maksymyk,
  Phys.\ Rev.\  D {\bf 49}, 6115 (1994)
  [arXiv:hep-ph/9312291].
  
\bibitem{triplets1}
M.~C.~Chen and S.~Dawson,
  Phys.\ Rev.\  D {\bf 70}, 015003 (2004)
  [arXiv:hep-ph/0311032];
M.~C.~Chen, S.~Dawson and T.~Krupovnickas,
  Int.\ J.\ Mod.\ Phys.\  A {\bf 21}, 4045 (2006)
  [arXiv:hep-ph/0504286],
  Phys.\ Rev.\  D {\bf 74}, 035001 (2006)
  [arXiv:hep-ph/0604102].

\bibitem{triplets2}
P.~H.~Chankowski, S.~Pokorski and J.~Wagner,
  Eur.\ Phys.\ J.\  C {\bf 50}, 919 (2007)
  [arXiv:hep-ph/0605302].
\end{document}